\documentclass[twocolumn, astrosymb]{aastex631}
\usepackage{makecell}
\usepackage{amssymb, amsmath}
\let\tablenum\relax
\usepackage[usenames,dvipsnames]{xcolor}
\usepackage{siunitx}
\usepackage{CJKutf8}
\usepackage{bm}
\usepackage{fontawesome5}
\usepackage{academicons}
\usepackage{graphicx}
\usepackage{float}
\usepackage{multirow}

\submitjournal{AAS Journal}

\shorttitle{Mass Density and Stellar Population Profiles}
\shortauthors{Zhang et al.}

\graphicspath{{./}{figures/}}
\hypersetup{colorlinks=true,
            citecolor=MidnightBlue,
            linkcolor=MidnightBlue,
            filecolor=magenta,
            urlcolor=MidnightBlue}
\def\apjs{{ApJS}}

\def\h{\hskip -3 mm}

\def\asec{$^{\prime\prime}$}

\newcommand{\lt}{<}
\newcommand{\gt}{>}

\def\snratio{{$\mathrm{S}/\mathrm{N}$}}

\def\ser{{S\'{e}rsic}}
\def\re{$R_{\rm e}$}

\def\msun{$M_\odot$}
\def\mstar{{$M_{*}$}}

\def\mhalo{{$M_{\rm h}$}}

\def\logmin{$\mathrm{log_{10}} (M_{\star,R>10\ \rm kpc}/M_{\odot})$}
\def\logmout{$\mathrm{log_{10}} (M_{\star,R>20\ \rm kpc}/M_{\odot})$}
\def\min{$M_{\star,R<10\ \rm kpc}$}
\def\mout{$M_{\star,R>20\ \rm kpc}$}
\def\logage{$\mathrm{log\ Age(Gyr)}$}

\def\mgfe{$\mathrm{[Mg/Fe]}$}
\def\ofe{$\mathrm{[O/Fe]}$}

\def\alphafe{$\mathrm{[\alpha/Fe]}$}

\def\feh{$\mathrm{[Fe/H]}$}
\def\mgh{$\mathrm{[Mg/H]}$}
\def\mgfex{$\mathrm{[MgFe]'}$}
\def\mgbfex{$\mathrm{Mg_b/\langle Fe\rangle}$}
\def\mgb{$\mathrm{Mg_b}$}
\def\alph{$\mathrm{\alpha}$}
\def\sigmacen{$\sigma_{\mathrm{\star,cen}}$}

\def\hsig{$\mathrm{High}\ \sigma_{*}$}
\def\lsig{$\mathrm{Low}\ \sigma_{*}$}
\def\sig{$\mathrm{\sigma_*}$}
\def\imfi{\texttt{imf1}}
\def\imfii{\texttt{imf2}}
\def\fmode{\texttt{full mode}}
\def\smode{\texttt{simple mode}}
\def\mh200c{{$M_{\mathrm{200c}}$}}
\def\delt{$\Delta$}

\def\insitu{{\textit{in-situ}}}
\def\exsitu{{\textit{ex-situ}}}

\def\alf{{\texttt{alf}}}

\definecolor{LightGray}{gray}{0.85}
\definecolor{Tab1}{RGB}{114, 158, 206}
\definecolor{Tab2}{RGB}{255, 158,  74}
\definecolor{Tab3}{RGB}{103, 191,  92}
\definecolor{Tab4}{RGB}{174, 199, 232}
\definecolor{Tab5}{RGB}{255, 187, 120}
\definecolor{Tab6}{RGB}{152, 223, 138}
\definecolor{Tab7}{RGB}{255, 152, 150}
\definecolor{Tab8}{RGB}{197, 176, 213}
\definecolor{hpurple}{HTML}{7E16DF}

\begin{document}
\begin{CJK*}{UTF8}{gbsn}

\title{A MaNGA about the Legacy I: Connecting the Assembly of Stellar Halo with the Average Star Formation History in Low-Redshift Massive Galaxies}

\correspondingauthor{Xiao-Ya Zhang (张筱雅), Song Huang (黄崧)}
\email{zhangxiaoya97@gmail.com, shuang@tsinghua.edu.cn}

\author[0009-0003-2586-2694]{Xiaoya Zhang (张筱雅)}
\affiliation{Department of Astronomy, Tsinghua University, Beijing 100084, China}

\author[0000-0003-1385-7591]{Song Huang (黄崧)}
\affiliation{Department of Astronomy, Tsinghua University, Beijing 100084, China}

% To be corrected
\author[0000-0002-4267-9344]{Meng Gu (顾梦)}
\affiliation{Department of Astronomy, Tsinghua University, Beijing 100084, China}

% -------------------------------------------------------------------------------------------- %
% Abstract & Keywords
% -------------------------------------------------------------------------------------------- %
\begin{abstract}

    We investigate the connection between stellar mass distribution, assembly history, and star formation timescales in low-redshift massive early-type galaxies (ETGs) by combining deep LegacySurvey imaging with MaNGA’s spatially resolved spectroscopy. Focusing on stellar population properties, especially the \mgfe{} abundance ratio, we analyze stacked spectra using both absorption line indices and full-spectrum fitting. We find that, among massive ETGs with identical average stellar mass distributions beyond 20 kpc, those with higher central velocity dispersion (\sigmacen{}) are older and more \alph{}-enhanced, suggesting a connection between the \insitu{} star formation in the past and the central gravitational potential today for massive ETGs with a similar stellar accretion history. Conversely, \emph{at fixed \sigmacen{} and total stellar mass}, galaxies with more extended stellar halos show lower \feh{}, higher \mgfe{}, and older ages, indicating an intriguing link between early starburst and quenching and later \exsitu{} assembly. These results demonstrate that the evolution of massive galaxies cannot be fully described by simple scaling relations alone, as the interplay between in-situ star formation and \exsitu{} accretion leaves distinct imprints in both their inner and outer stellar populations. Our findings highlight the importance of extending stellar population studies to large radii and underscore the scientific potential of next-generation IFU surveys and deep, high-resolution spectroscopy for probing the galaxy–halo connection. 
    
\end{abstract}

% https://astrothesaurus.org

% -------------------------------------------------------------------------------------------- %
% Introduction
% -------------------------------------------------------------------------------------------- %

\section{Introduction} 
    \label{sec:intro}

    The two-phase process has been a widely accepted theory for describing the formation of massive early-type galaxies (ETGs; e.g., \citealt{OserAPJ2010, JohanssonAPJ2012}). In the first phase, cold gas collapses in radial flows, forming the bulk of the \insitu{} stars. From $z\sim2-3$, major and minor mergers increase the \exsitu{} component. This picture has been successful in explaining observations, including the significant size growth compared with the relatively mild mass growth of ETGs since $z\sim 2$ \citep[e.g.,][]{NaabAPJL2009, OserAPJ2012}. 

    However, direct observational support is challenging as the light profiles of massive ETGs are very steep. Hence, it is difficult to obtain reliable observations of the faint outskirts, and this will affect the measurement of galaxy size \citep[e.g.,][]{NewmanAPJ2012,vanderWelAPJ2014} as well as total stellar mass (\mstar) \citep[e.g.,][]{BlantonAJ2011, BernardiMNRAS2013, FischerMNRAS2017, BernardiMNRAS2017}. Nonetheless, efforts have been made in recent years. \citet{HuangAPJL2013} found that nearby massive galaxies can be decomposed into a compact inner region and an extended stellar envelope, which can be explained as the result of accretion through minor mergers. In simulations, \citet{HirschmannMNRAS2015} found differences in the \insitu{} and the \exsitu{} components, where accreted stars tend to be older and more metal-poor. However, this is extremely difficult to confirm in observations. As mentioned before, the low surface-brightness outskirt means ground-based spectra suffer heavily from telluric lines, so most spectroscopic observations on galaxies focus on regions within the effective radius (\re), where the stellar population properties are mainly governed by internal processes \citep[e.g.,][]{PengAPJ2012, GreeneAPJ2015, BluckMNRAS2020}. 

    Fortunately, with the development of IFU surveys such as The Calar Alto Legacy Integral Field Area \citep[CALIFA,][]{SanchezAAP2012, HusemannAAP2013} SAMI (Sydney-Australian-Astronomical-Observatory Multi-object Integral-Field Spectrograph) Galaxy Survey \citep{CroomMNRAS2012, BryantMNRAS2015} and Mapping Nearby Galaxies at Apache Point Observatory \citep[MaNGA]{BundyAPJ2015}, we are now one step closer to the actual radial profiles of physical parameters in these massive ETGs. For example, \citet{FerrerasMNRAS2019} utilized SAMI data and measured the radial profiles of 522 ETGs out to a median value of 2.2\re{}. Their sample covers a wide stellar mass range ($10^{9.5}M_{\odot}<M_*<10^{11.7}M_{\odot}$) and environments (field/group and cluster). They find a weak trend suggesting that ETGs in a cluster environment are more sensitive to velocity dispersion, which impacts both the central stellar populations and their radial gradients. Another example is \citet{OyarzunAPJ2023}, where they focused MaNGA ETGs ($10^{9}M_{\odot}<M_*<10^{12}M_{\odot}$) and studied the radial profiles out to 1.5 \re. They found that massive satellites ($\rm M_*>10^{11.5}M_\odot$) are more \alph{}-enhanced and metal-poor in the outskirts($\rm R>R_e$) than the centrals with similar \mstar{}, as well as satellites with smaller Dark Matter Halo Mass (\mhalo{}). These results support a picture in which the outskirts of satellites are quenched by interactions with host halos ('outside-in' quenching).
    %\citet{WatsonMNRAS2022} used SAMI data and found that galaxies in higher-density regions tend to be more \alph-enhanced, meaning they have a shorter star-formation time scale. \citet{OyarzunAPJ2023} studied MaNGA ETGs and found that galaxies in more massive DM halos have more \alph-enhanced satellites. All these indicate that the central galaxy's environment and \revise{dark matter halo mass (\mhalo{})} play an essential role in quenching lower-mass galaxies. 

    At this point, it is tempting to connect the accretion history of these massive ETGs to their parent dark matter halo properties, but we can not yet directly measure the \mhalo. \citet{HuangMNRAS2018} utilized the high-quality data from Hyper Suprime-Cam \citep[HSC,][]{AiharaPASJ2018} and found that the outskirt ($R>50$ kpc) mass of massive ETGs ($M_*>10^{11.6}M_\odot$) serves as an excellent proxy for \mhalo. One explanation is that after $z\ge 2$, minor mergers begin to play a more dominant role in the mass assembly of massive galaxies and they tend to deposit \exsitu{} stars in the outer (R$>$\re{}) regions \citep[e.g.,][]{OogiMNRAS2013, BedorfMNRAS2013} and the minor merger rates can increase with \mhalo{} if the dynamical friction timescale is short \citep{NewmanAPJ2012}. These could result in a dependence between the outskirts mass ($R>50$ kpc) and \mhalo{}.
    %One explanation is that with the higher fraction of the \exsitu{} component, the outskirts would be better at reflecting the current gravitational potential and not be affected by the internal evolution of the host galaxy. 
    This means we can peek into the dark matter (DM) properties through the galactic outskirts. With IFU observations that allow us to go beyond \re where the \exsitu{} stars from minor mergers begin to play a more prominent role\citep[e.g.,][]{OogiMNRAS2013}, we can build a bridge between stellar population observations and their dark matter halo. 

    In our work, we set out to study stellar population gradients in galaxies with different central velocity dispersions (\sigmacen{}) as well as stellar mass profiles, by combining deep imaging from the DESI legacy survey \citep{DeyAJ2019} and the recently completed IFU survey MaNGA. The former provides high-quality photometry, which can help us obtain more reliable measurements of galaxy sizes and \mstar{} profiles. Our primary goal in this paper is to bridge the host DM halo of massive ETGs with their observable spectral features. 

    The paper is organized as follows: Section \ref{sec:data} describes the photometry and spectroscopy data used in this work, as well as the \mstar{} estimation used for later analysis. Section \ref{sec:methods} details the sample selection and spectral stacking method. Section \ref{sec:results} shows the results from both spectral index measurements and full-spectrum fitting. We discuss how the observations inform the general formation picture and compare with previous works in Section \ref{sec:discussions}. Finally, the summary is in section \ref{sec:summary}.

    Throughout this work, we adopt the $\Lambda$CDM model with parameters ~$H_0=70$~km~s$^{-1}$ Mpc$^{-1}$, ~$\Omega_b=0.046$ and $\Omega_m=0.3$. All magnitudes in this work are in the AB magnitude system (\cite{Oke1983}).

% -------------------------------------------------------------------------------------------- %
% Data
% -------------------------------------------------------------------------------------------- %

\section{Data} 
    \label{sec:data}
    
    This section introduces the data used in this work, including the MaNGA survey, the DESI Legacy Imaging Survey, and the Siena Galaxy Atlas (SGA). We also describe the sample selection and the method used to estimate the stellar mass and effective radius of the galaxies in our sample.
    
% -------------------------------------------------------------------------------------------- %
\subsection{The MaNGA Survey} 
    \label{ssec:manga}

    Mapping Nearby Galaxies at Apache Point Observatory \citep[MaNGA,][]{BundyAPJ2015, YanAJ2016} is part of the fourth-generation Sloan Digital Sky Survey \citep[SDSS-IV,][]{YorkAJ2000, GunnAJ2006, BlantonAJ2017, AguadoAPJS2019}. MaNGA is the most comprehensive spatially resolved spectroscopic survey of nearby galaxies so far, and it covers more than 10,000 $0.01\lesssim z\lesssim0.15$ galaxies with around $R\sim 2000$ spectral resolution within the rest frame wavelength range of $3600-10300 $\r{A}. The SDSS Collaboration has released the whole MaNGA data set, which has become a valuable resource for studying many topics in galaxy evolution, including the kinematics, stellar population, and assembly history of low-$z$ massive early-type galaxies \citep[e.g.,][]{LiuMNRAS2020, LiMNRAS2018, BevacquaMNRAS2022}
    
    MaNGA drew its galaxy sample from the NASA Sloan Atlas catalog \citep[NSA][]{BlantonAJ2005}. The selected galaxies have a uniform distribution of $\log M_{\star}$ \citep{WakeAJ2017}. MaNGA observes these targets using optimized fiber bundle sizes to spatially cover at least $1.5\times$ \re{} for the `Primary' and `Color-enhanced' samples and $2.5\times$ \re{} for the `Secondary' sample. For more details on MaNGA's sample selection, please refer to \citep{WakeAJ2017}.

    In this work, we use the \href{https://sdss-mangadap.readthedocs.io/en/latest/datamodel.html#dap-model-logcube-file}{\texttt{LOGCUBE}} data produced by MaNGA's Data Reduction Pipeline \citep[DRP,][]{LawAJ2016}, which performs flux calibration and sky background subtraction on the raw data cubes. \texttt{LOGCUBE} provides the flux, noise, bitmasks, and spectral resolution of each spaxel, whose spatial size is 0.5\asec{} $\times$ 0.5\asec{} and the wavelength is logarithmically binned.
    
    Meanwhile, MaNGA’s Data Analysis Pipeline \citep[DAP,][]{WakeAJ2017} fits both the stellar continuum and emission lines in each spaxel and provides their stellar \& gas kinematics. We use the Voronoi-binned stellar kinematics from the DAP maps for sample selection and the latter data reduction (see Section \ref{ssec:selection}). For more details on the MaNGA observing strategy and spectrophotometry calibration, we refer the readers to \citet{LawAJ2016} and \citet{YanAJ2016a}, respectively.
    
% -------------------------------------------------------------------------------------------- %
\subsection{The DESI Legacy Imaging Survey and SGA} 
    \label{ssec:legacy}
    
    Although MaNGA's spatially resolved spectroscopy is critical to this work, it does not provide the radial coverage needed to include the extended stellar halo, which is crucial for understanding the assembly history of massive ETGs. The SDSS optical images that MaNGA relied on also lack the necessary depth to reach the faint outskirts of these galaxies. Therefore, we turn to the DESI Legacy Imaging Surveys \citep[LegacySurvey,][]{SchlegelAAS2021, DeyAJ2019} for high-quality photometry. The LegacySurvey is a combination of the DECam Legacy Survey (DECaLS) using the Dark Energy Camera (DECam; \cite{FlaugherAJ2015}) in the $grz$-band, the Beijing-Arizona Sky Survey (BASS) using the Bok 2.3m telescope in the $gr$-band \citep{ZouPASP2017}, and the Mayall z-band Legacy Survey (MzLS) using the Mayall 4m telescope in the $z$-band \citep{DESICollaborationAJ2022, DESICollaborationARXIVEPRINTS2016}. In total, it covers $\sim 14,000$ deg$^2$ of the extragalactic sky that matches the footprint of the MaNGA survey.

    At $z<0.15$, massive galaxies in MaNGA have a large angular size that makes it challenging to extract their stellar halos' surface brightness distribution due to systematics such as the background subtraction (e.g., \citealt{HuangMNRAS2018, LiMNRAS2022}). Fortunately, the Siena Galaxy Atlas 2020 dataset \citep[SGA-2020,][]{MoustakasAPJS2023} provides high-quality $grz$-band surface brightness profiles of 383,620 bright galaxies from the DESI Legacy Imaging Surveys DR9\footnote{https://www.legacysurvey.org/dr9/} with large angular sizes on image mosaics generated by a customized version of the LegacySurvey pipeline (\href{https://github.com/legacysurvey/legacypipe}{\texttt{legacypipe}}) that handles the background subtraction around bright \& extended objects carefully and also prevents excessive shredding of the large galaxies. These SGA-2020 surface brightness profiles typically extend to $\mu_{r} \sim 26\ mag\cdot arcsec^{-2}$, which is perfect for characterizing the stellar halos of massive galaxies in MaNGA from $\sim 40$ to $>80$ kpc. SGA-2020 also groups nearby galaxies by projected distance and conducts a joint analysis within each group when their light distributions overlap. Since low-$z$ massive galaxies in MaNGA are often large and highly blended, the pipeline sometimes mistake the alignment of multiple objects as the position angle of the target galaxy, the estimation of ellipticity can be affect due to the same reason. Therefore, we favor the geometric measurements in SGA-2020 over those in MaNGA's input NSA catalog and use the SGA-2020 profiles to derive stellar mass distributions based on optical colors, estimating total, aperture, and outskirts stellar masses for this work (see Section \ref{ssec:selection}). For more details about the SGA-2020 dataset, we refer the readers to \citet{MoustakasAPJS2023} and the SGA website\footnote{https://www.legacysurvey.org/sga/sga2020/}.

% -------------------------------------------------------------------------------------------- %
\subsection{Stellar Mass Measurements and Sample Selection} 
    \label{ssec:selection}
    
    \begin{figure}[!th]
    \centering
        \includegraphics[width=\columnwidth]{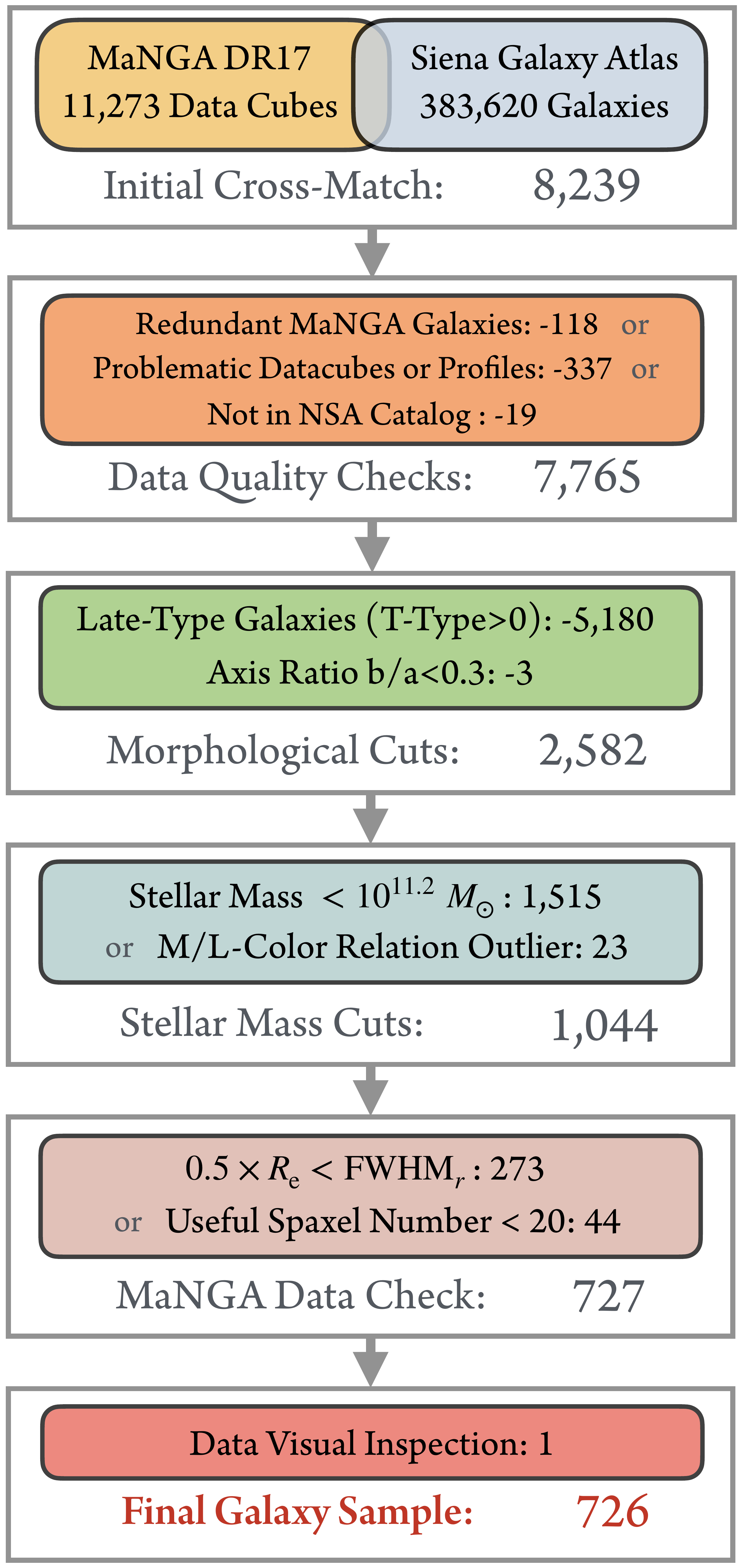}
        \caption{
            A flowchart representing the sample selection procedures used in this work. Each box represents a selection stage, and the exact criteria are listed inside. 
            }
        \label{fig:flowchart}
    \end{figure}

    \begin{figure}[!h]
        \centering
        \includegraphics[width=1\linewidth]{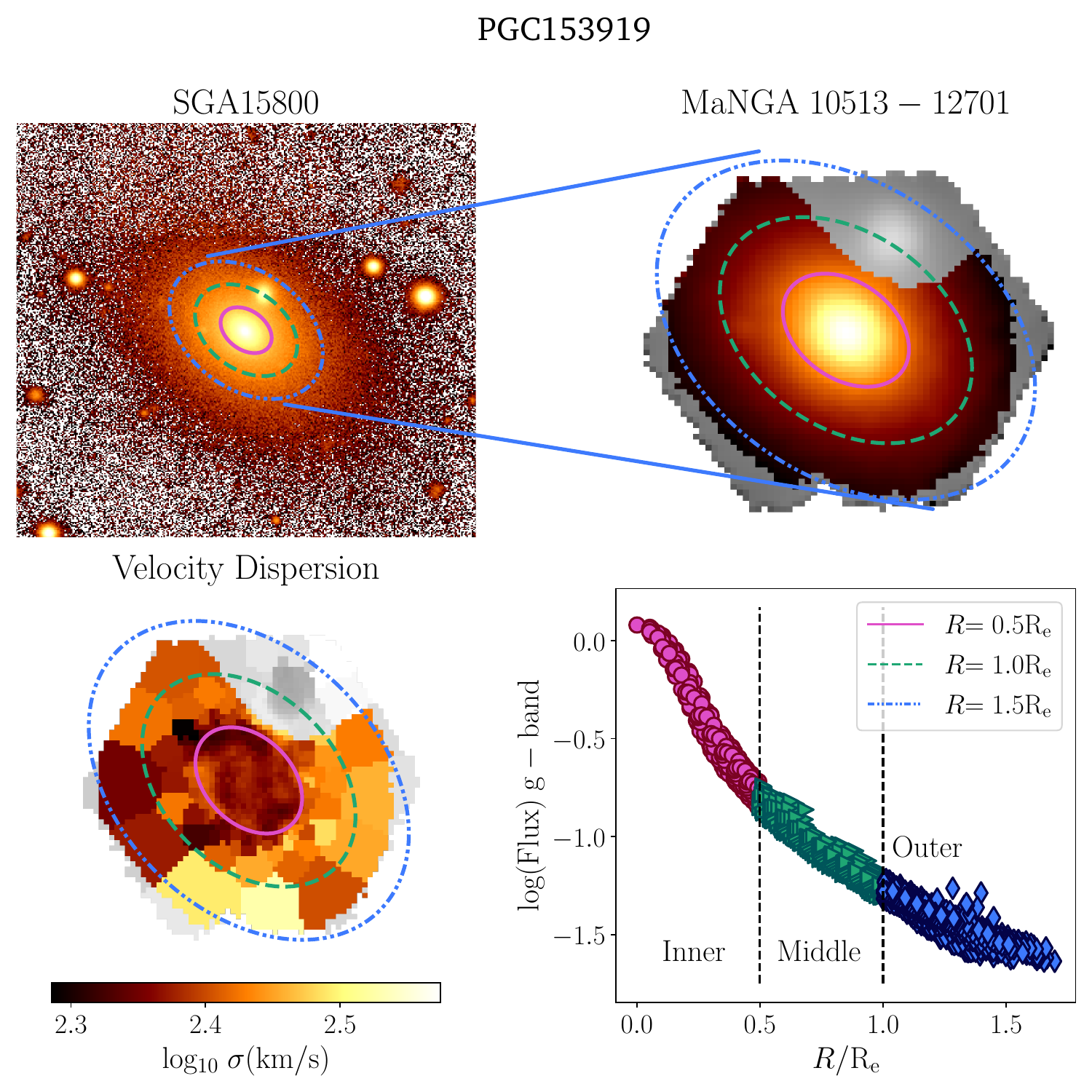}
        \caption{
            An example galaxy (PGC153919) in the final sample. Top left: SGA g-band image. The title is the \texttt{SGA id} of this galaxy. Top right: MaNGA g-band flux map with its \texttt{plate-ifu} as the title. The gray region represents the original data, and the colored pixels are the ones used in the stacking after the \snratio{} cut and multi-source masking. Bottom left: the Voronoi-binned stellar velocity dispersion map available in MaNGA \texttt{DAP} maps. Similarly, the gray color map covers the entire data cube, while the colored region represents the selected region. The similar ellipses in these three plots shown in blue dash-dot-dot, green dashed and pink solid ellipses have semi-major-axis of $\mathrm{0.5\ R_e)}$, $\mathrm{1.0\ R_e)}$ and $\mathrm{1.5\ R_e}$ respectively.
            \\ Bottom right: the radial profile of the g-band flux in MaNGA data. We can see that the additional masking results in a smoother profile, implying that our masks have successfully excluded contamination from other sources. The pink circles,  green triangles, and blue diamonds stand for pixels with major axis range in $\mathrm{Inner(}R\mathrm{\le 0.5R_e)}$, $\mathrm{Middle(0.5R_e\le}R\mathrm{\le1.0R_e}$, and $\mathrm{Outer(}R\mathrm{\ge 1.0R_e)}$ respectively.
            }
        \label{fig:qa}
    \end{figure}

    Based on MaNGA DR17 and the SGA-2020 data, we select a sample of massive early-type galaxies (ETGs) with both IFU observations and deep 1-D surface brightness profiles. First, we cross-match the MaNGA DR17 final data release catalog, which contains 11,273 unique records of observations, with the SGA-2020 summary catalog, comprising 383,620 galaxies. Using a 3.0\asec{} matching radius, we find 8,239 matched items as our initial sample (see Figure \ref{fig:flowchart}). Then, we perform quick data quality checks on both datasets and discard the matched records that satisfy the following three criteria: 1. Redundant observations in MaNGA: 107 galaxies have repeated observations. We only keep the data cube with the highest average \snratio{} and exclude 118 repeated data cubes for each unique galaxy in this group; 2. 337 Problematic data cubes or 1-D profiles: 196 galaxies have no SGA files, 131 galaxies have no profiles, 3 galaxies have no MaNGA data cubes, and 7 galaxies have contaminated LegacySurvey data; 3. 19 galaxies not included in NSA catalog are also excluded as we need to compare with the NSA stellar mass in later steps. These three criteria discard 474 galaxies, leaving us with 7765.

    In the second step, we perform morphological selection to focus on the ETGs. We exclude galaxies that are either identified as Late-Type Galaxies (LTG) or have minor-to-major axis ratios smaller than 0.3 ($\rm b/a \lt 0.3$). For the LTG rejection, we utilize the final release of the deep-learning-based morphological catalog from \citet{SanchezMNRAS2022}. In their work, LTGs are defined as those with $\rm T-Type\gt 0$, which contains both barred galaxies and spiral galaxies. And the ETGs are made up of S0 galaxies ($\rm -2 \lt T-Type\lt 0$) and Ellipticals ($\rm -4 \lt T-Type\lt -2$). We discard 5180 galaxies with $\rm T-Type\gt 0$. We also applied an edge-on rejection where we discard 3 ETGs with $\rm b/a \lt 0.3$ given that it is difficult to separate LTGs and ETGs when viewing from edge-on angles. 
    The morphological criterion eliminates an additional 5,183 galaxies from the sample. So far, among the rejected galaxies, only 91 have $M_{\rm star}>10^{11.2}\ M_{\odot}$ from the NSA catalog.
    
    Before proceeding to the next step, we must obtain reliable estimates of the remaining 2,583 galaxies' $R_{\rm e}$ and stellar mass density profiles to calculate the stellar masses within different apertures. For $R_{\rm e}$, we integrate the $r$-band surface brightness profiles following the geometric parameters provided by SGA-2020 to construct the Curve-of-Growth (CoG)\footnote{While SGA-2020 initially provided the CoG measurements in $grz$-band, they were affected by a \href{https://www.legacysurvey.org/sga/sga2020/\#known-issues}{known coding issue}. We therefore decided to make another attempt.} Using the CoG, we estimate $R_{\rm e}$ as the radius where the fractional light enclosed within reaches 50\% of the maximum flux value in the CoG through interpolation. While the NSA catalog also estimates the $R_{\rm e}$ of each MaNGA galaxy, it is based on a 2-D \ser{} model fitting of the much shallower SDSS data, which could lead to inaccurate size estimation \citep[e.g.,][]{NewmanAPJ2012}.
    
    To estimate the stellar mass, we first convert the LegacySurvey magnitudes into SDSS filters using the \texttt{filterTranBands} function in \texttt{ProSpect} \citep{RobothamMNRAS2020} that performs filter conversion at a given redshift. Next, we perform the Galactic extinction correction using the recalibrated {\tt SFD98} extinction map (\citealt{Schlegel1998}) and the new extinction coefficients for the SDSS filters by \citet{SchlaflyApJ2011}. We then $K$-correct the magnitudes using the \texttt{calc\_kcor} calculator \citep{ChilingarianMNRAS2012, ChilingarianMNRAS2010} based on single stellar population models. At $z<0.15$, the $K$-correction is small for the old stellar population and has a minimal effect on the stellar mass. The luminosities are then transformed to units of solar luminosity using the solar absolute magnitudes table by \citet{WillmerApJS2018}, and the luminosity distance are obtained using the cosmology stated in Section~\ref{sec:intro} ($\Lambda$CDM model with parameters ~$H_0=70$~km~s$^{-1}$ Mpc$^{-1}$, ~$\Omega_b=0.046$ and $\Omega_m=0.3$) and the redshift values come from. And finally, we use an average $g-r$ color of the galaxy and the mass-to-light-color-relation (MLCR) by \citet{RoedigerMNRAS2015} based on the BC03 stellar library \citep{BruzualMNRAS2003} assuming a Kroupa IMF ($\mathrm{log\ M_*/L_r = 1.629 * (g-r) - 0.792}$) to derive an average $r$-band $M_{\star}/L$ and estimate stellar masses. We define the average color as the mean color between $R={\rm FWHM}_{r}$ (FWHM$_r$ being the full-width-half-maximum of the PSF in $r$-band) and when the surface brightness profile reaches the threshold of 28 ${\rm mag}/{\rm arcsec}^2$. For massive ETGs, color gradients (or the $M_{\star}/L_r$ gradients) are typically shallow and show little dependence on luminosity or other key properties; therefore, applying an average $M_{\star}/L_r$ is unlikely to bias the stellar mass estimates or the comparison of stellar mass density profiles. We also deliberately choose to correct only for Galactic extinction, not the intrinsic dust attenuation within the galaxy, which can be derived through spectral fitting at $R<1.5\times R_{\rm e}$ using MaNGA data. We are uncertain whether the dust attenuation value from the inner region can be applied to the extended stellar halo without biasing the outskirt stellar mass. Moreover, massive ETGs commonly have very low average internal attenuation, and the derived value is sensitive to the stellar population model, so we do not wish to introduce new systematics into the analysis. With the average $M_{\star}/L_r$ in hand, we convert the Galactic extinction-corrected \& $K$-corrected $r$-band surface brightness profiles into stellar mass density profiles ($\mu_{\star}$) and use the CoG to derive the ``total'' stellar mass along with a series of aperture and outskirt stellar masses. Note that the ``total'' stellar mass only reflects the imaging depth provided by the LegacySurvey. This work focuses on the stellar mass within 10 kpc ($M_*(R\lt10\ {\rm kpc})$) and outside of 20 kpc ($M_*(R\gt20\ {\rm kpc})$). 
    
    Similar to the $R_{\rm e}$ case, while NSA provides the $M_{\star}$ estimates based on the 2-D \ser{} model ($M_*(\rm NSA)$), this over-simplified model, combined with the shallower SDSS image, could make the stellar mass less robust. We, therefore, rely on our CoG-based stellar masses ($M_*(\rm SGA)$) for further analysis. Meanwhile, $M_*(\rm NSA)$ and $M_*(\rm SGA)$ agree well with each other. To focus on the massive ETGs, we exclude 1,515 galaxies with $M_{*, {\rm SGA}} < 10^{11.2}\ M_{\odot}$. To be extra cautious, we also discard 23 outliers of the $M_*(\rm NSA)$ and $M_*(\rm SGA)$ relation, defined using the lower and upper 2\% of the residual distribution (Residual$ =M_*(\mathrm{NSA})-M_*(\rm SGA)$). These cuts result in 1,044 massive ETGs with reliable mass and size measurements. Given MaNGA's target selection strategy, even at the high-$M_{\star}$ end, our ETG sample is still incomplete. However, MaNGA's selection is not biased toward ETGs with any specific property. Hence, the statistical conclusion of this work will not be impacted by the incompleteness.

    With the $M_*(\rm SGA)$-selected sample, we also want to ensure the galaxy's 1-D profile is not heavily smeared by the PSF and has enough useful MaNGA spaxels for further analysis. So, we first exclude 273 galaxies whose $0.5 \times$ \re{} is smaller than ${\rm FWHM}_{r, \rm MaNGA}$. Given their smaller angular size, we cannot reliably recover the stellar population properties within 0.5 \re{}. The galaxies discarded in this step skew towards lower \mstar{} and higher $z$. While visually inspecting the MaNGA data, we noticed that many data cubes contain multiple objects.
    We first use the \texttt{sep.Background} function in \texttt{sep}\ \citep{BertinAAPS1996, BarbaryTHEJOURNALOFOPENSOURCESOFTWARE2016} for additional background subtraction to increase contrast, then we follow the deblending routine using the function \texttt{photutils.segmentation} in the python package \texttt{photutils}\ \citep{Bradley2023}, where we first detect the sources and then create a 'Segmentation Image' which we can deblend into separate sources and mask out the unwanted ones.
    However, for a few special cases, such as the major merger systems or galaxies with substantial foreground contamination, a sufficient mask will cover most of our target (especially the center), rendering the ETG unsuitable for analysis. We additionally exclude from further analysis any pixels that meet one or more of the following criteria: $\mathrm{SNR_{g}}\lt 2$; lack valid kinematic information from the MaNGA \texttt{DAP}; or flagged as \texttt{"do not use"} in either the \texttt{DAPPIXMASK} or \texttt{DRP3PIXMASK}. Afterwards, we reject 44 MaNGA data cubes that contain fewer than 20 useful spaxels. In the last step, we perform a final visual inspection of the LegacySurvey images \& profiles and MaNGA data. We exclude PGC3097770 due to corrupted inner 1-D surface brightness profiles. Ultimately, we have 726 massive ETGs as our final sample. 
    Figure \ref{fig:qa} is an example galaxy of our final sample. We can see that, compared with the MaNGA data cube, the LegacySurvey image has broader coverage. Note that this example also shows the additional mask for the small object near the galaxy's center. The three ellipses in the images represent the three ellipses with semi-major axes of $\mathrm{0.5\ R_e}$, $\mathrm{1.0\ R_e}$, and $\mathrm{1.5\ R_e}$, respectively.
    All the data {\it MaNGA} data used in this work can be found in MAST: \dataset[10.17909/40vg-r661]{http://dx.doi.org/10.17909/40vg-r661}

% -------------------------------------------------------------------------------------------- %
% Methods
% -------------------------------------------------------------------------------------------- %

\section{Methods} 
    \label{sec:methods}

    In this work, we seek to combine the advantages of resolved spectroscopic information enabled by MaNGA IFU data with the more complete stellar mass density profiles provided by the deeper LegacySurvey data to shed new light on the connection between the stellar mass distributions and stellar population properties of low-redshift massive ETGs. In particular, we seek to understand the influence of the extended stellar halo at $R>20$ kpc on the average stellar population properties or star formation history within the inner $R<1.5\times R_{\rm e}$ probed by MaNGA. Limited by the sample size and the quality of the spectra within a small radial range for a single galaxy, we adopt the strategy to 1. group the massive ETGs into two samples based on their central or outskirt properties; 2. then, for each group, we coadd their MaNGA spectra within three different radial bins based on their \re{} to achieve a much higher spectral \snratio{} per \r{A} to facilitate the inference of stellar age, metallicity (\feh), and $\alpha$-element abundance ratio (\mgfe), similar to several previous works, \citep[e.g.,][]{FerrerasMNRAS2019, ParikhMNRAS2018, ParikhMNRAS2021, OyarzunAPJ2022, OyarzunAPJ2023}. Since we focus on a small population of the most massive galaxies, the dynamic ranges of potential variations of key stellar population parameters should be moderate. Hence, high-quality stacked spectra are crucial for delivering reliable conclusions. We briefly introduce the sample split method in Section~\ref{ssec:sample_split} and the spectral stacking method in Section~\ref{ssec:stacking}. 

% -------------------------------------------------------------------------------------------- %

\subsection{Sample Split}
    \label{ssec:sample_split} 
    
    To explore how some of the essential global properties, especially those regarding the stellar mass distributions of massive ETGs, impact the stellar population properties, the average star formation history, and also the mass assembly history within $\sim 1.5\times R_{\rm e}$, 
    we split the sample into two-dimensional parameter spaces to investigate the average spectral features of galaxies occupying different regions of the same space.
    Note that previous works have explored the variation of the average stellar population properties (often about the central region), such as the stellar age, metallicity, and [Mg/Fe] over the parameter space defined by the Fundamental Plane \citep[e.g.,][]{GravesApJ2009, ZahidApJ2017}, the $M_{\star}$-$R_{\rm e}$ relation \citep[e.g.,][]{McDermidMNRAS2015, ScottMNRAS2017, BaroneMNRAS2022}, or the dynamical mass (e.g., $M_{\rm JAM}$)-$R_{\rm e}$ relation \citep[e.g.,][]{CappellariMNRAS2013, McDermidMNRAS2015, LuMNRAS2023}. These essential scaling relations encode information about the stellar mass distributions, but 1. they are often based on earlier shallower images that miss the outer stellar halos proved to be critical to understanding the formation of massive galaxies; 2. parameters such as $R_{\rm e}$ or $\mu_{\rm e}$ (surface brightness at or within $R_{\rm e}$) often depends on a particular (over)-simplified model and can lead to incomplete or even biased pictures. Therefore, in this work, we focus on the similarities or differences of the average stellar mass distributions within different radial ranges defined unambiguously using physical radial boundaries (i.e., using kpc as unit, instead of $R_{\rm e}$ for two sub-samples of massive galaxies that shared a similar distribution of the control variable. As we demonstrate later, this could enable us to directly comment on the connection between the assembly and star formation of massive ETGs within the two-phase formation scenario (e.g., \citealt{OserAPJ2010, OserAPJ2012, vanDokkumAPJ2010}) in mind. 
    
    Here, we split the MaNGA massive ETG sample into different sub-samples based on the following two criteria: 1. The ``$\sigma_{\star}$-split'': at fixed outskirt stellar mass (stellar mass at $R>20$ kpc, \mout{}), split the sample based on their stellar velocity dispersion within $0.5 \times R_{\rm e}$; 2. The ``extendedness-split'': at the fixed ``inner'' stellar mass (stellar mass at $R<10$ kpc), split the sample based on their ``extendedness'', represented using the outskirt stellar mass (\mout). Because we primarily focus on the average difference across galaxy populations, we exclude extreme outliers before conducting sample splitting. We achieve this by utilizing the Isolation Forest Algorithm from the \texttt{scikit-learn} package. We fit the entire sample and obtain the average anomaly score from the \texttt{decision function}. For this work, we chose a less aggressive approach by excluding the first 3\% of galaxies from the sample. This ensures that the most egregious galaxies do not contaminate the sample split while avoiding removing too many from the entire sample. We then use a $\log$-linear relation to describe the correlation, with the scatter along the relation related to the statistical uncertainties of the physical properties. Following a similar strategy with \citet{OyarzunAPJ2023}, we then estimate the 35th and 65th percentile boundaries using the distribution perpendicular to the best-fit relation. Based on the distribution of the scatter around the relation, we define the two sub-samples as those with scatter in the 1st-35th percentile and the 65th-99th percentile, respectively.

    % -------------------------------------------------------------------------------------------- %

\subsubsection{The ``\texorpdfstring{$\sigma_{\star}$}{sigma*}-Split''}
    \label{ssec:sigma_split}

\begin{figure*}[!th]
    \centering
    \includegraphics[width=0.9\linewidth]{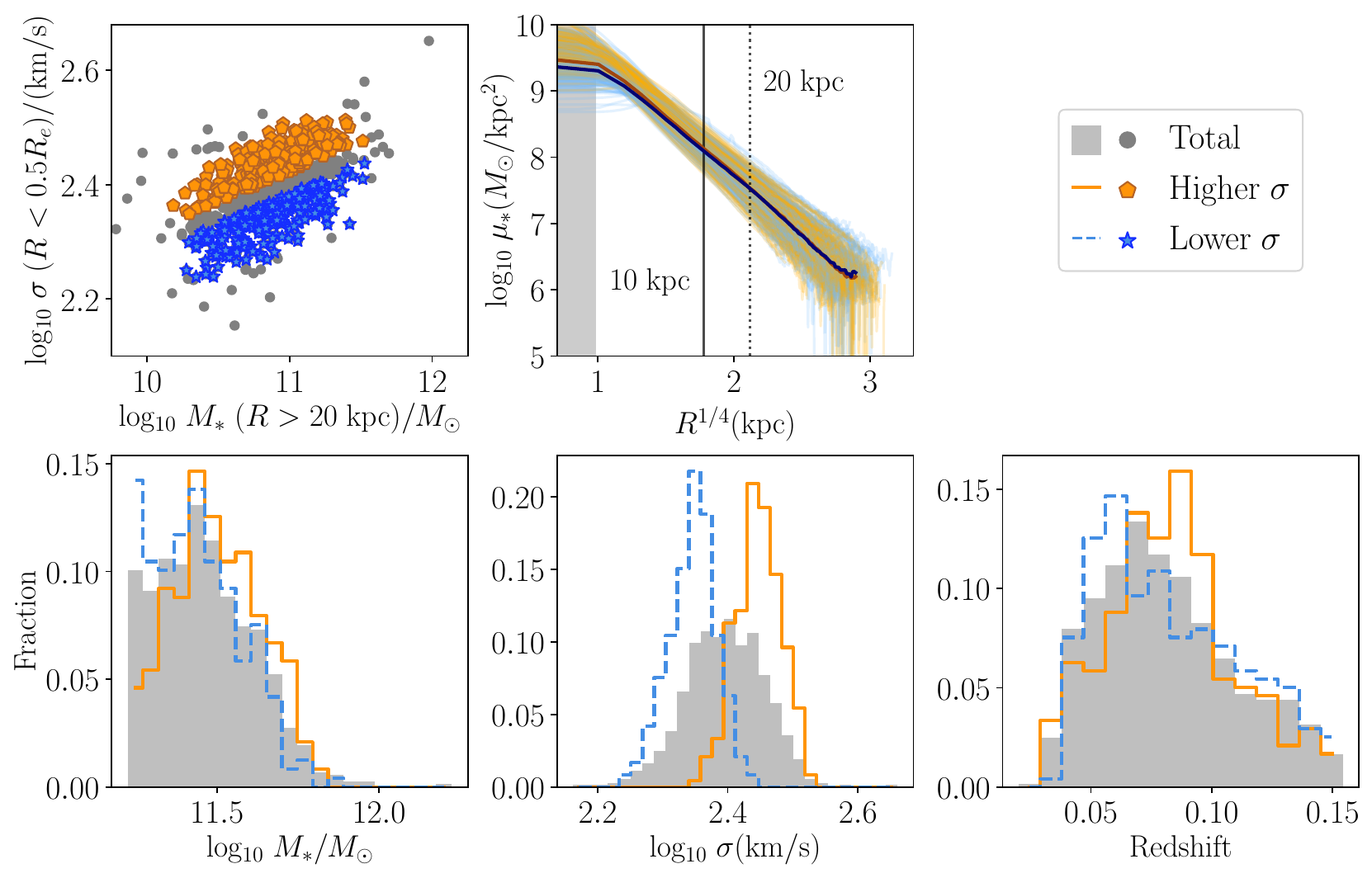}
    \caption{
        The two sub-samples are split by position on the \logmout{} vs. \sigmacen{} plane (top left). For galaxies with similar outskirt stellar mass, orange pentagons represent the galaxies with larger inner velocity dispersion(`\hsig' ), and blue stars represent galaxies with smaller inner velocity dispersion (`\lsig'). The background gray circles stand for the total sample. The top right plot displays the radial profiles of stellar mass density ($\rm M_\odot{/kpc^2}$), with the thick lines representing the median profiles.\\
        Using the same color-coding convention, the histograms show the distributions of their physical parameters, including \mstar{} (bottom left), \sigmacen{} (bottom middle), and redshift (bottom right). Orange solid lines and blue dashed lines correspond to `\hsig' and `\lsig' samples, respectively, and the gray-filled bar is the total sample. The \texttt{Jupyter} notebook for reproducing this figure can be found here: \href{https://github.com/xyzhangwork/mdensity_v_stellarpop/blob/main/plot_scripts/plot_split.ipynb}{\faGithub}. This repository is also available on \href{https://doi.org/10.5281/zenodo.17979404}{Zenodo}.
        }
    \label{fig:split_sigma}
    \end{figure*}

    First, as shown in Figure~\ref{fig:split_sigma}, we split the sample based on the relation between the stellar mass in the outskirts of these massive ETGs (\mout{}) and their velocity dispersions within $0.5\times R_{\rm e}$. Here, we use the velocity dispersion within $0.5\times R_{\rm e}$ (\sigmacen$\mathrm{=\sigma_{*, R<0.5R_e}}$), calculated based on the pixel-wise $\sigma_{\star}$ maps from the MaNGA {\tt DAP} products `\texttt{MAPS-VOR10-MILESHC-MASTARSSP}', where the spaxels are Voronoi-binned to improve \snratio. And the outer stellar mass is obtained using the same MLCR as we used for total stellar mass. The details are described in Section~\ref{ssec:selection}. 
    
    Stellar velocity dispersion is commonly believed to reflect the depth of the gravitational well. It is considered one of the most fundamental physical parameters for massive ETGs, especially for driving the scaling relations with stellar population properties \citep[e.g.,][]{BernardiAJ2003, GreeneAPJ2015, Martin-NavarroMNRAS2018}. Therefore, by splitting the sample based on differences in velocity dispersion at a fixed outskirt stellar mass, we aim to assess the variation in the average stellar population driven by central velocity dispersion among ETGs with similar accretion histories. As mentioned earlier, while ETGs grew primarily through mergers, they also experienced intense starbursts in the early stages of their evolution. Hence, the stellar population in a massive ETG's inner region reflects the unique star formation history of its \insitu{} component and a complicated assembly history in its \exsitu{} component. 
    
    Here, we divide our sample into two groups, the \hsig{} and \lsig{} sub-samples, based on their relative positions to the best-fit \mout{}-$\sigma_{\star}$ relation, which is well-fit by a $\log$-linear relation as well. The \lsig{} sub-sample corresponds to the galaxies below the relation with scatters in the 1st-35th percentile, while the \hsig{} sub-sample is the one above the relation, with scatters in the 65th-99th percentile. Each sub-sample contains 239 galaxies. 
    The \hsig{} sub-sample displays a higher central stellar mass density within the inner $\sim 5$ kpc. While the $R \lesssim 1$ kpc region is affected by the smearing of PSF, the difference is robust and consistent with the expectation of a higher $\sigma_{\star}$ value. We perform a two-sample Kolmogorov–Smirnov test (K-S test) on the \mout{} distributions of the \hsig{} and \lsig{} sub-samples, and obtain a p-value of $\rm 0.37\pm 0.23$ (To obtain the uncertainty of the p-value, we resample the \mout{} distributions by assuming a uniform error 0.1 dex.), well above the 0.05 threshold, suggesting the two sub-samples have the same \mout{} distributions. This difference also leads to slightly higher median total $M_{\star}$ \& lower $R_{\rm e}$ value in the \hsig{} sample than the \lsig{} one ($10^{11.5} M_{\odot}$ v.s. $10^{11.4} M_{\odot}$; 9.6 kpc v.s. 10.5 kpc). 

    In terms of the $\sigma_{\star}$, as expected, the \hsig{} sub-sample has a significantly higher average value than the \lsig{} one (274 km/s v.s. 218 km/s). It is worth noting that, while we refer to them as the \hsig{} and \lsig{} samples, this is not equivalent to the simple $\sigma_{\star}$-bins methods used in many previous works \citep[e.g.,][]{GreeneAPJ2015, Martin-NavarroMNRAS2018, BeverageApJ2023, ChengMNRAS2024}. The matched \emph{outer} stellar masses and profiles are the key here, as they demonstrate these two groups of massive galaxies should have, on average, very similar \exsitu{} mass growth while also showing systematic differences in their center that are more dominated by the \insitu{} star formation. 

    % -------------------------------------------------------------------------------------------- %

\subsubsection{The Extendedness-Split}
    \label{ssec:extend_split}
    
\begin{figure*}[!t]
    \centering
    \includegraphics[width=0.9\linewidth]{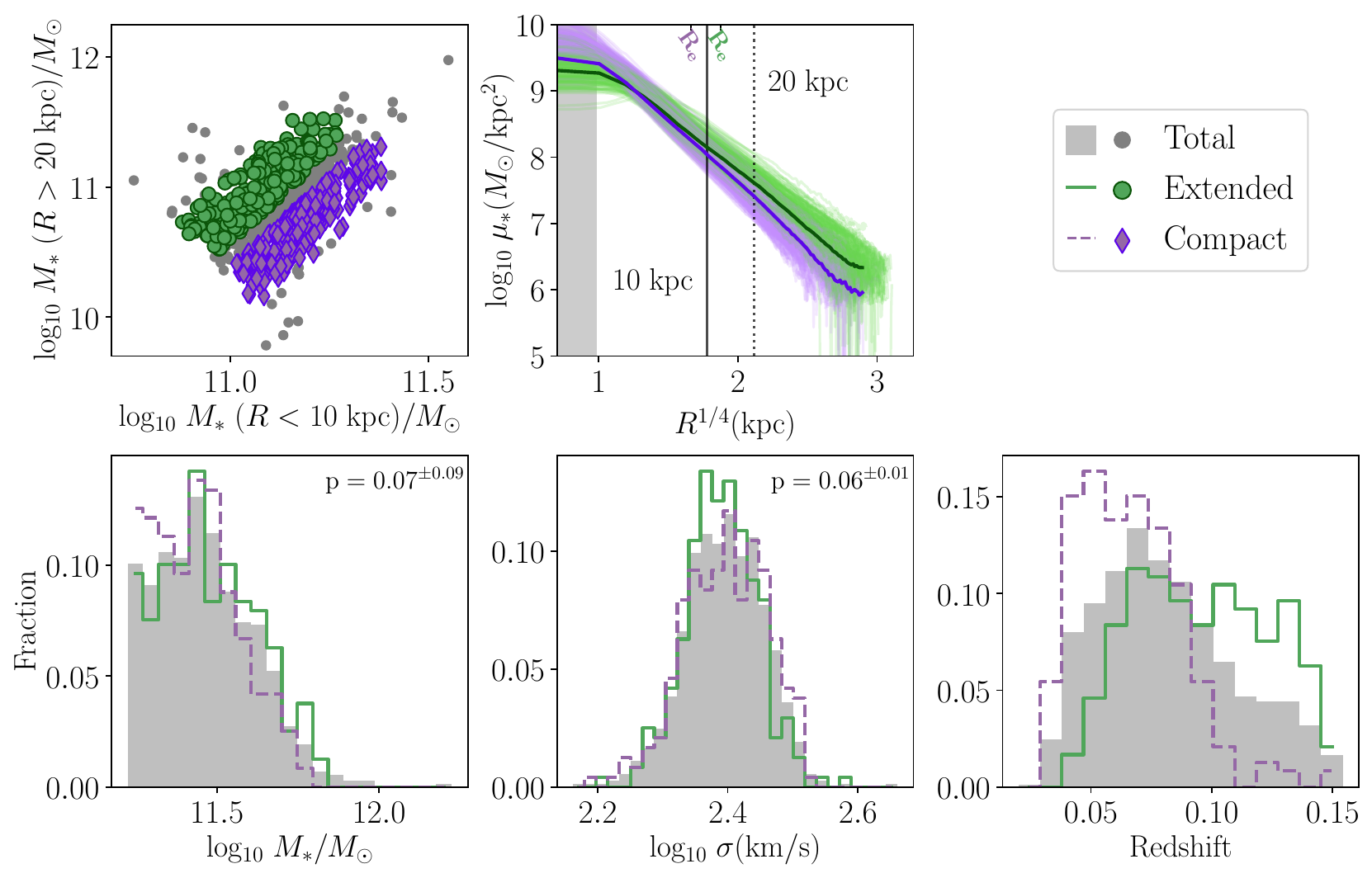}
    \caption{
        The top-left plot shows the sample-split scheme, where galaxies are categorized by position on the \logmin{} vs. \logmout{} plane. The two samples have similar inner stellar masses (\min); the green circles represent galaxies with more extended outskirts (`Extended'), and the purple diamonds represent more compact galaxies (`Compact'). As before, the total sample is in gray circles. The top right plot displays the radial profiles of stellar mass density ($\rm M_\odot{/kpc^2}$), with the thick lines representing the median profiles. The green solid lines and purple dashed lines correspond to the Extended and Compact samples, respectively. The gray-filled bar represents the total sample.\\
        Similar to Figure \ref{fig:split_sigma}, the histograms show the \mstar{} (bottom left), \sigmacen{} (bottom middle)， and redshift (bottom right). We also show the p-values (with errors from resampling the original sample in their respective uncertainties. For \mout{}, we assume a 0.1 dex error) of the Kolmogorov–Smirnov tests (K-S test) on the \mstar{} (bottom left) and \sigmacen{} (bottom middle) distributions between the two sub-samples on the first two histograms.  The \texttt{Jupyter} notebook for reproducing this figure can be found here: \href{https://github.com/xyzhangwork/mdensity_v_stellarpop/blob/main/plot_scripts/plot_split.ipynb}{\faGithub}. This repository is also available on \href{https://doi.org/10.5281/zenodo.17979404}{Zenodo}.}
        
    \label{fig:split_outskirt}
    \end{figure*} 
    
    Next, we want to compare the average stellar population properties in the stellar halos of massive galaxies with different assembly histories at the same stellar mass. It becomes increasingly clear that the massive ETGs grew mostly by accumulating stellar mass into their extended stellar halos through (primarily minor) mergers and accretions. Therefore, grouping galaxies by the prominence of their outer stellar halos can help us compare them. 
    
    As shown in Figure \ref{fig:split_outskirt}, we split the sample based on the relation between the ``inner'' stellar mass ($<10$ kpc) and the stellar mass beyond 20 kpc. Because our sample selection only focuses on total stellar mass, the distribution of the galaxies on the inner \& outer mass plane would naturally form a lower boundary with a negative slope (lower inner mass 'compensated' by a higher outer mass, see top left plot on Figure~\ref{fig:split_outskirt}). This means a simple least-$\chi^2$ linear regression fit results in a biased sample split. Therefore, we adopt the Principal Component Analysis (PCA) method instead. After rescaling the data to a zero mean and unit variance, we perform a 2-component PCA, with the direction of the 1st eigenvector serving as the scaling relation for the sample split.
    
    We then define the galaxies between the 1st and 35th percentiles as the ``Compact'' sub-sample and those in the 65th and 99th percentiles as the ``Extended'' sub-sample. Each sub-sample contains 239 galaxies. This approach reduces the impact of statistical uncertainties on any potential difference in the average stellar population properties of the two sub-samples and excludes outliers. Note that the ``Compact'' term describes the sample in a relative term, as these ``Compact'' galaxies are still among the largest and most massive ETGs, not the compact elliptical \citep[cE, e.g.,][]{FaberApJ1973} or the ``relic'' galaxies \citep[e.g.,][]{TrujilloApJ2009, TrujilloApJ2014, Ferre-MateuMNRAS2017}. 

    In Figure \ref{fig:split_outskirt}, we highlight the average stellar mass density profiles of these two sub-samples on top of the individual profiles of the whole sample. As expected, the ``Extended'' sub-sample displays lower surface mass density within the inner 10 kpc while showing a more prominent stellar halo outside 20 kpc than the ``Compact'' sub-sample. This contrast perfectly highlights the motivation behind this split: We aim to explore the connection between the average stellar population properties in the inner regions of massive ETGs and their stellar mass density profiles, which reflect the assembly history of not only the galaxies but also possibly their dark matter halos. As shown in Figure \ref{fig:split_outskirt}, these two sub-samples' total stellar mass distributions match well with each other \emph{by design}. Interestingly, these two sub-samples also have very similar distributions of their stellar velocity dispersions despite showing systematically different central stellar mass density profiles. We perform two-sample K-S tests on the \sigmacen{} and \mstar{} distributions of the two sub-samples and find the p-value of the \mstar{} is $\rm 0.07\pm 0.09$ and the p-value of \sigmacen{} is $\rm 0.06\pm 0.01$, and the values are also shown in Figure~\ref{fig:split_outskirt}. This suggests the two sub-samples have statistically indistinguishable \sigmacen{} and \mstar{} distributions. Controlling the sample split at the same average stellar mass and velocity dispersion is a great advantage. As both stellar mass and velocity dispersion correlate with important stellar population properties in ETGs, this allows us to interpret any potential differences without worrying that they are indirectly caused by underlying scaling relations for either. 
    
    It is worth noting that 20 kpc is a relatively small value to define the ``outskirt'' for the massive ETGs in our sample, as even deeper imaging data (e.g., from the Hyper-Surprime Cam or HSC) can already trace their distributions to $>100$ kpc for individual galaxies. And, in \citet{HuangMNRAS2022}, the authors demonstrate that it is the stellar mass in the extremely outer region (e.g., the mass between 50 and 100 kpc) that traces the underlying halo mass more accurately through galaxy-galaxy weak lensing measurements. Hydrosimulations also recently demonstrated this conclusion \citep[e.g.,][]{Xu2024}. While the LegacySurvey images are available for most MaNGA galaxies, they are still shallower than the HSC ones (e.g., \citealt{LiMNRAS2022}). Therefore, 20 kpc is a more practical choice and already extends beyond the radial coverage of MaNGA or SDSS data. 

    % -------------------------------------------------------------------------------------------- %

% -------------------------------------------------------------------------------------------- %
\subsection{Stacking Analysis of MaNGA Spectra}
    \label{ssec:stacking}     

    While the MaNGA survey provides spatially resolved spectra of individual galaxies, spectral differences caused by variations in star formation or mass assembly history \emph{at fixed stellar mass or velocity dispersion} can be subtle, requiring high \snratio{} stacked spectra to derive reliable results. This strategy has been adopted by many past works using MaNGA data \citep[e.g.,][]{ParikhMNRAS2018, BelfioreMNRAS2017, OyarzunAPJ2022}. Here, we describe our spectral stacking procedure. 

    First, for each MaNGA data cube, following the elliptical isophotal shape defined by the SGA-2020 catalog, we divide the spaxels into three radial bins: $R\lt 0.5 \times R_{\rm e}$ (Inner), $0.5 \times R_{\rm e}\lt R \lt 1.0 \times R_{\rm e}$ (Middle), and $R\gt 1.0 \times R_{\rm e}$ (Outer). As mentioned, we require each spaxel to have $\mathrm{SNR_{g}}\ge 2$. The masking and radial binning procedure is shown in where the gray area in the MaNGA data cube represents the masked spaxels.

    For spaxels within each radial bin, we shift them to the same radial velocity based on the Voronoi-binned stellar line-of-sight velocity measurements in MaNGA. The kinematic information extracted from each spaxel is often less precise due to the limited \snratio{}, especially in the outer region of the cube. However, MaNGA provides kinematic maps of all galaxies based on Voronoi bins, where a group of spaxels within a regularly shaped area are binned to reach the same \snratio{} threshold to ensure reliable radial velocity and velocity dispersion measurements (see the bottom-left panel of Figure \ref{fig:qa}). Based on these Voronoi-binned radial velocity maps, we shift the spaxels in each radial bin to the same rest-frame reference and integrate them into a higher-\snratio{} spectrum for further analysis. And, using the bootstrap method within each bin, we estimate the statistical uncertainties of the integrated spectra. This procedure yields three stacked spectra for each galaxy, each accompanied by statistical error bars.

    As telluric residuals remain prevalent in the reduced MaNGA spectra, it is necessary to identify and mask them correctly. To achieve this, we Gaussian-smooth all integrated spectra to a standard velocity resolution of $500\ \rm km/s$, which is much higher than the stellar velocity dispersions of the massive ETGs in our sample. Subtracting the highly smoothed spectrum from the original spectrum enhances the contrast of the sharp residual features. First, on the residual spectra, we mask any 1-$\sigma$ outlier data point within $3$\r{A} of the strong telluric lines identified in \citet{NollAAP2014}. We also mask out any 10-$\sigma$ outlier from the residual spectra, regardless of whether it is close to a known telluric feature. 
    
    During this procedure, we noticed that the $\lambda \ \gt 7000$\r{A} region begins to be heavily affected by the telluric residuals. In addition, in the old stellar population, the $\lambda \ \gt 6300$\r{A} wavelength range is often dominated by strong molecular absorption bands, such as TiO, VO, and CaH features, making spectral fitting or physical interpretation challenging. As this work primarily focuses on the stellar metallicity and [$\alpha$/Fe] properties that can be well-constrained at $\lambda \ \lt 6300$\r{A}, we limit our spectral index or full-spectrum fitting analysis within the 4000\r{A} $\lt \ \lambda \ \lt 6300$\r{A} wavelength range. Meanwhile, we also perform full-spectrum fitting with a broader wavelength coverage to include IMF-sensitive features and reach the same qualitative conclusions (see Section \ref{ssec:imf}).

    We then fit the spectra with a 10-degree polynomial as the pseudo-continua and remove them from the original spectra by division. We then use the \texttt{smoothspec} function from the package \texttt{Prospector} \citep{JohnsonApJS2021} to smooth all the spectra from the original resolution to 300 km/s at 5000\AA.
    This is crucial for the next step, when we stack spectra from different galaxies, as the differences in their velocity dispersions would be eliminated by the smoothing to a uniform resolution. 

\begin{figure}[!h]
    \centering
    \includegraphics[width=1\linewidth]{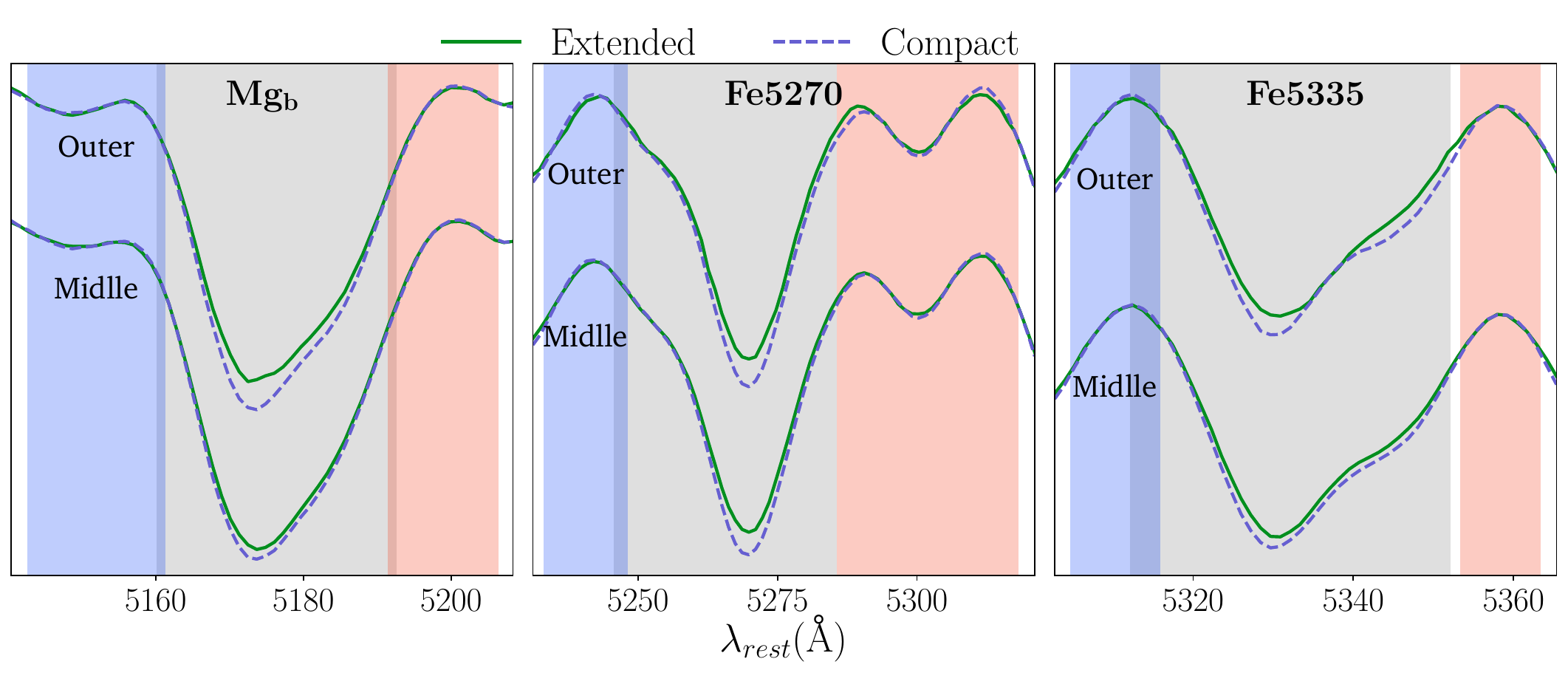}
    \caption{
        Normalized spectra in the three absorption index regions of the extended (solid green line) and compact (dashed purple line) sub-samples. From left to right: $\mathrm{Mg_b,\ Fe5270,\ Fe5335}$}
    \label{fig:zoomin}
\end{figure}

    To further improve the \snratio{} of the cleaned and integrated spectra of each galaxy, we shift all the smoothed spectra to the rest frame and then median-stack the spectra of each radial bin of the four sub-samples defined by the two sample-split methods. We choose the median as the stacking method because it has been shown in various studies to have better resistance to emission lines, telluric features, and bad pixels. In the end, we have six spectra for each sample-split method: the two sub-samples $\mathrm{\times}$ three radial bins. The final stacked spectra reach a statistical \snratio{}$> 100$ even in the outer radial bins. Figure \ref{fig:zoomin} shows the normalized stacked spectra in three absorption index regions. The middle and outer spectra are vertically shifted for visual distinction. We can directly visualize the subtle but clear systematic differences around these features. 

% -------------------------------------------------------------------------------------------- %

\section{Results} 
    \label{sec:results}
    
    Our primary goal in this work is to understand how stellar mass distributions relate to the stellar population and star-formation (assembly) histories of massive ETGs. In this section, we study stacked spectra by measuring Lick/IDS absorption-line indices and performing full-spectrum fitting. Although absorption indices (Section \ref{sec:index}) can provide a model-independent evaluation of the spectral differences among different sub-samples, they have been known to suffer from age-metallicity degeneracy\citep[e.g.,][]{WortheyAPJS1994, WortheyASPC1999} and have complicated relations with the underlying stellar population properties \citep[e.g.,][]{ThomasMNRAS2003, KornA&A2005, ConroyAPJ2012a}. And to interpret the differences in the spectra requires using stellar population models, which account for the variations in parameters such as stellar age. This is achieved in Section \ref{sec:fsf} where we employ the full-spectrum fitting code to investigate the spectra.
    These two methods compensate for each other and, together, can provide a more reliable description of the trends of physical parameters within these massive ETGs.

    Considering the strategy for spatial coverage of the MaNGA data cubes, we construct radial bins in units of \re{}. However, since we split the sample based on properties related to the stellar mass distributions, which will inevitably introduce differences in \re{} within sub-samples, we present the radial gradients in physical units (kpc) in this section to acknowledge the size difference. For each radial bin defined by \re{}, we calculate the median galactocentric distance of the bin centers for each sample, in kpc units, and use these values to represent the radial trends in stellar population properties presented here.
    
 \subsection{Spectral Indices}
    \label{sec:index}

    We use \texttt{pyphot} by \citet{Fouesneau2022} (following the definition of \citet{WortheyAPJS1994, WortheyAPJS1997}) to measure the Lick/IDS indices. This work mainly focuses on $\mathrm{Mg_b}$, which is most sensitive to the Magnesium abundance, and two iron-sensitive indices: Fe5270 and Fe5335. Apart from these features, we also construct composite indices $$\mathrm{[MgFe]' =\sqrt{Mg_b\cdot(0.72\cdot Fe5270+0.28\cdot Fe5335)}}$$ and $$\mathrm{Mg_b/\langle Fe\rangle=\frac{Mg_b}{0.5\cdot(Fe5270+Fe5335)}}$$ following the formulation of \citet{Gonzalez1993}. The former serves as a proxy for total metallicity and is less sensitive to \alphafe. The latter is a good empirical indicator of \alphafe{} \citep{ThomasMNRAS2003}. The errors associated with each index are estimated using the bootstrap method. This technique involves 1000 iterations of randomly sampling the same number of galaxies (with repetitions) from the original sample and stacking the available spectra, and the error is quantified as the standard deviation across these iterations.

\begin{figure}[!h]
    \centering
    \includegraphics[width=1.0\linewidth]{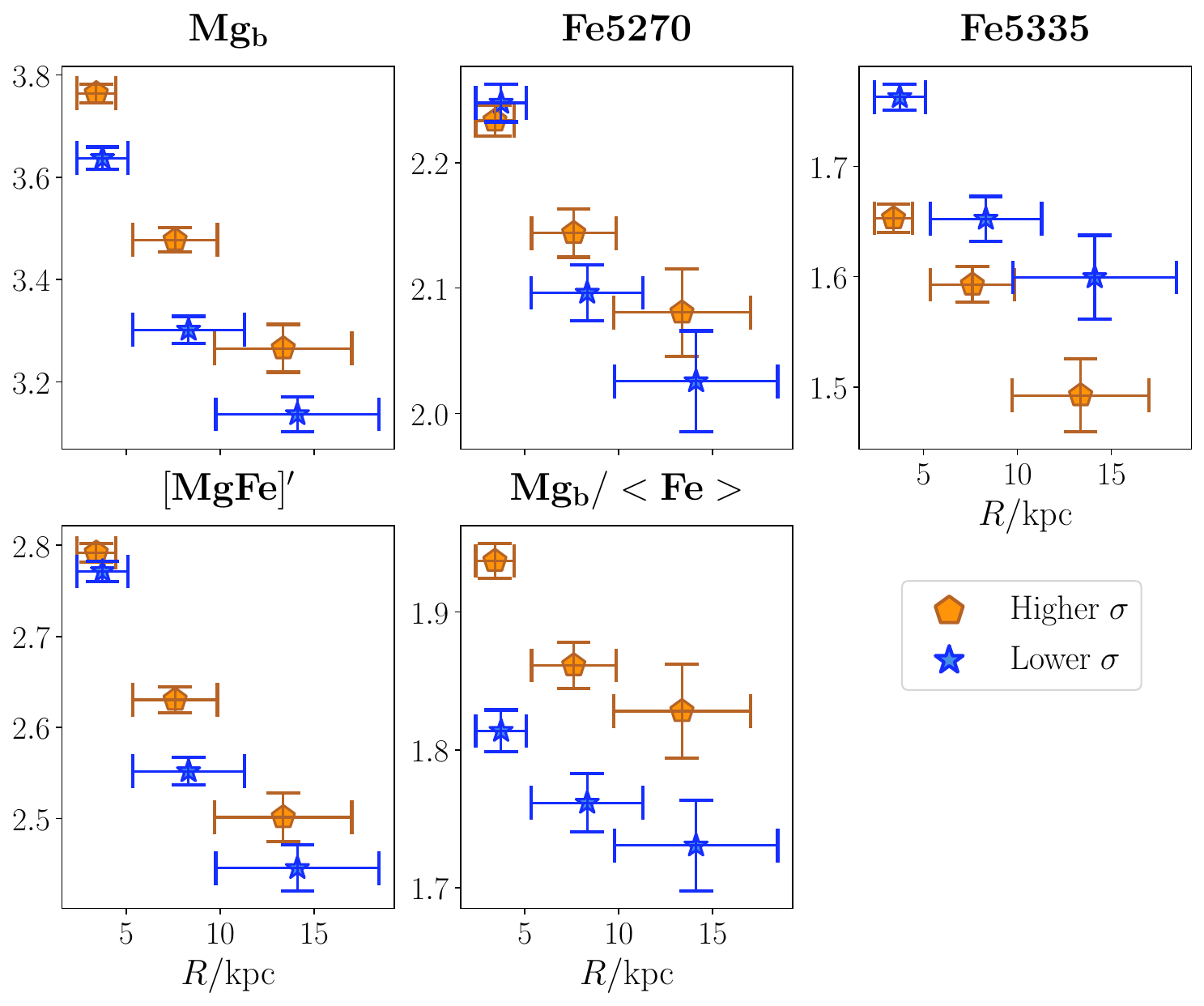}
    \caption{
        The radial gradients of spectral indices of the `\hsig'(orange pentagons) and `\lsig'(blue stars) sub-samples. The top row shows the three absorption indices, and the bottom row contains the composite index, which serves as a proxy for total metallicity (\mgfex) and \alphafe{} (\mgbfex). The x-axis represents the average galactocentric distance in kpc. The \texttt{Jupyter} notebook for reproducing this figure can be found here: \href{https://github.com/xyzhangwork/mdensity_v_stellarpop/blob/main/plot_scripts/plot_index.ipynb}{\faGithub}. This repository is also available on \href{https://doi.org/10.5281/zenodo.17979404}{Zenodo}.
        }
    \label{fig:index_sigma}
\end{figure}

    Firstly, Figure \ref{fig:index_sigma} shows the radial profiles of five indices for the \hsig{} and \lsig{} sub-samples. We recall from the \ref{ssec:sample_split} section that the \hsig{} sample is defined as having higher \sigmacen{} but the same \mout as the \lsig{} sub-sample, which will result in a slightly smaller \re{} since they have more stellar mass in the center. While both sub-samples exhibit negative slopes for all five indices, there are some notable differences. In particular, the \hsig{} sub-sample consistently exhibits higher \mgb{} and lower Fe5335 across all radial bins. 

    As for the two composite indices, both sub-samples show negative radial gradients, while the \hsig{} galaxies show a clear trend of higher \mgbfex{} than the \lsig{} ones. As for the \mgfex{} index, although the \hsig{}-sample shows a consistently higher value than the \lsig{}-sample in all three radial bins, the differences are much subtler than the \mgbfex{} index. We note that two iron-related indices exhibit different behaviors between the \hsig{} and \lsig{} sub-samples: the two sub-samples have similar Fe5270 values, whereas the \lsig{} galaxies have higher Fe5335 values in all three radial bins. This is not entirely surprising as these indices have been known to rely on many other properties such as \alphafe, stellar age, and total metallicity \citep[e.g.,][]{ThomasMNRAS2003, KornA&A2005}, and Fe5335 is slightly more sensitive to \alphafe{} than Fe5270 \citep{ThomasMNRAS2003}. We note that the \hsig{} sub-sample shows higher \mgbfex{} at all radii, driven by consistent trends in both the \mgb{} and Fe5335 indices. 

    Following the empirical interpretations of these two indices, we find that, for massive galaxies with the identical outer stellar mass distributions, the ones with higher \sigmacen{} tend to be more \alph{}-rich than the lower-\sigmacen{} ones in the entire radial range explored here. At the same time, their difference in total metallicity is much subtler. Meanwhile, the consistent difference within $R < 1.5\times R_{\rm e}$ is interesting. Assuming a higher \alphafe{} indicates a more efficient early star formation history, this result could suggest that the radial trend is driven by the \insitu{} component that dominates the stellar mass budget in the inner region or that the average star formation history in the \exsitu{} component (e.g., at $R > R_{\rm e}$) was also influence by the physical conditions that shaped the \sigmacen{} values.

\begin{figure}[!h]
    \centering
    \includegraphics[width=1.0\linewidth]{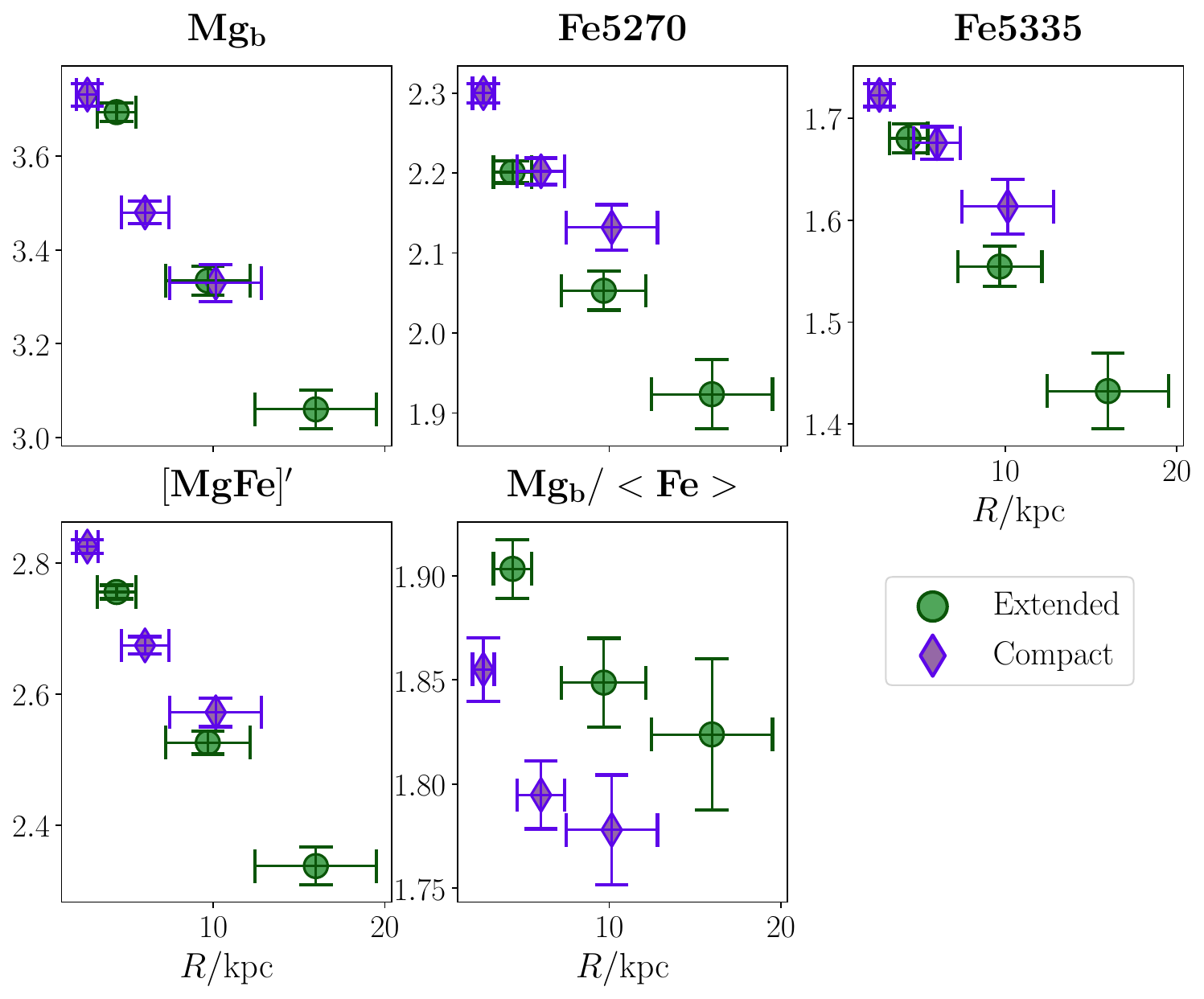}
    \caption{
        The radial gradients of the 'Compact'(green circles) and 'Extended'(purple diamonds). The top row shows the three absorption indices and the bottom row contains the two composite indices, which serve as proxies for total metallicity (\mgfex, bottom left) and \alphafe{} (\mgbfex, bottom right). The \texttt{Jupyter} notebook for reproducing this figure can be found here: \href{https://github.com/xyzhangwork/mdensity_v_stellarpop/blob/main/plot_scripts/plot_index.ipynb}{\faGithub}. This repository is also available on \href{https://doi.org/10.5281/zenodo.17979404}{Zenodo}.
        }
    \label{fig:index_mass}
\end{figure}

\begin{figure*}[!th]
\centering
\gridline{\fig{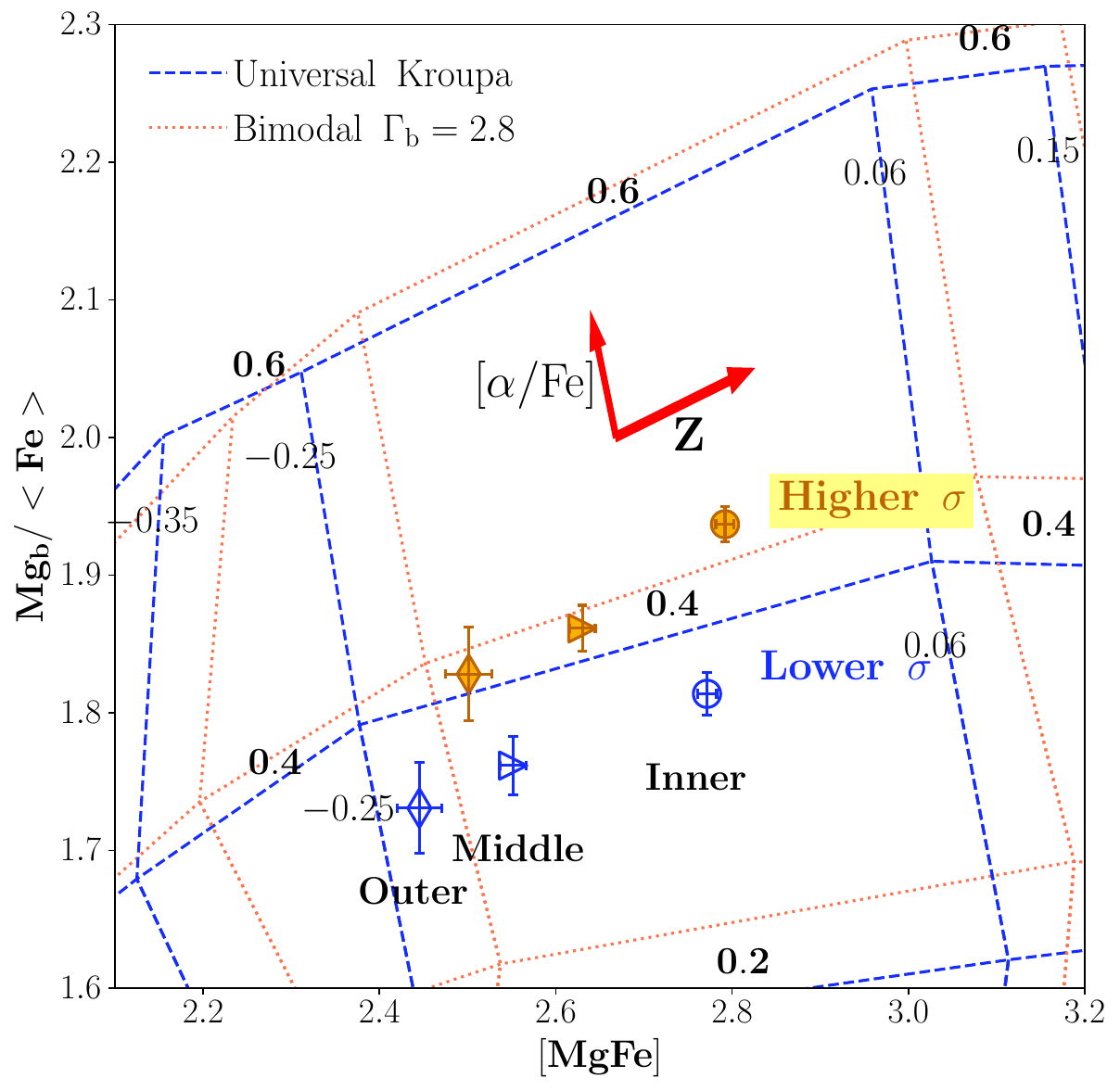}{0.45\textwidth}{(a)}
\fig{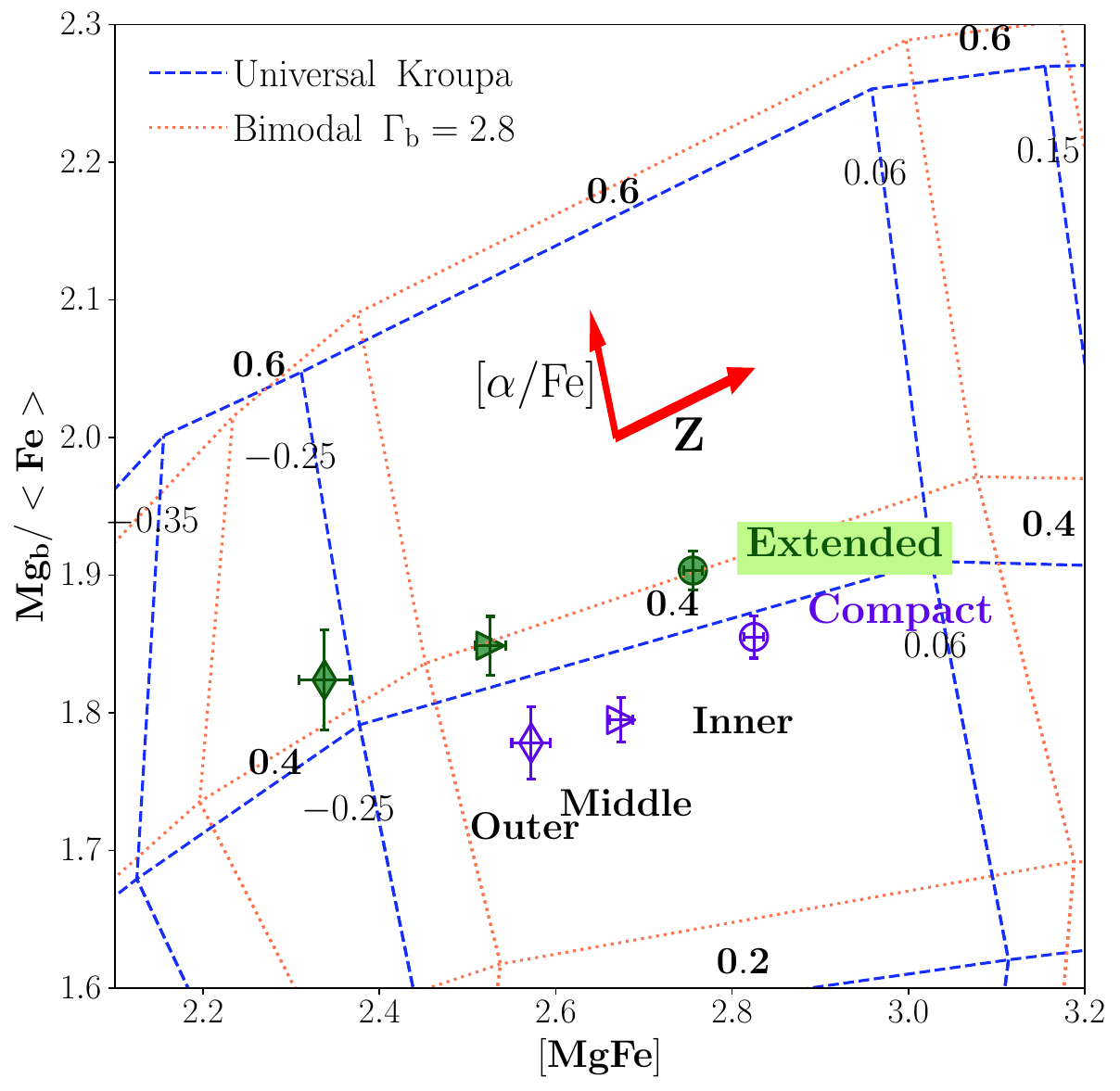}{0.45\textwidth}{(b)}}
    \caption{
         The \mgfex{} vs. \mgbfex{} plane where the grids are from the \texttt{sMILES} SSPs, which have been convolved to the same resolution. The thicker red arrow points to the direction of increasing metallicity, and the thinner red arrow points to the direction of increasing \alphafe. We show the relations in SSPs with two different IMFs: Universal Kroupa \citep{KroupaMNRAS2001} (blue dashed lines) and Bimodal IMF \citep{VazdekisAPJS1996} with $\Gamma=2.8$. The stellar age of both grids is set at 10 Gyr to reflect the old stellar population of our galaxies.\\
         The six data points are the values of two samples and three radial bins. (a): \hsig{} (filled orange) and \lsig{} (empty blue). (b) Compact (filled green) and Extended (empty purple).\\
         The \texttt{Jupyter} notebook for reproducing this figure can be found here: \href{https://github.com/xyzhangwork/mdensity_v_stellarpop/blob/main/plot_scripts/simle_plot.ipynb}{\faGithub}. This repository is also available on \href{https://doi.org/10.5281/zenodo.17979404}{Zenodo}.
        }
    \label{fig:sspgrid}
\end{figure*}

    Next, we turn to the extendedness-split samples to examine the differences between the extended and compact sub-samples. The first thing to notice in Figure \ref{fig:index_mass} is that the more extended sub-sample has a much larger radial coverage than the compact ones in units of kpc. Therefore, between these two samples, a direct comparison can only extend to $\sim 10$ kpc, which is still within the ``inner'' region defined in Figure \ref{fig:split_outskirt}. However, we can already spot systematic differences within the shared radial coverage between these two samples: while they share a similar \mgb{} radial trend within 10 kpc, the more compact sample demonstrates a consistently elevated iron-related index value, particularly in the outermost bin. For the total metallicity indicator \mgfex{}, the two samples fall on the same negative \mgfex{} radial trend to their respective outermost radius. The compact sample reveals a moderately higher \mgfex{} value in the $>R_{\rm e}$ region than the more extended sample, leaving the question of whether this deviation will continue to the $>10$ kpc region an intriguing one for the future. As for the \alphafe{} indicator \mgbfex{}, the more extended galaxies exhibit higher \mgbfex{} out to $\sim$10 kpc. Assuming that these two samples share a similar total metallicity within 10 kpc, the higher \mgbfex{} index throughout the galaxy may suggest that the overall star formation efficiency for the stellar population in the more extended galaxies is systematically higher than the more compact ones. It is worth noting that the two samples here share similar \sigmacen{} distributions; their \mgbfex{} values differ in the innermost radial bin. This is not entirely surprising, as there is typically an intrinsic scatter around the scaling relation. However, differences in the average values between sub-samples could imply that the intrinsic scatter is not truly random with respect to galaxy extension.

    Finally, we put our index measurements on a model grid using the simple stellar population (SSP) predictions based on the empirical MILES  \citep{VazdekisMNRAS2010} and semi-empirical sMILES stellar libraries \citep{KnowlesMNRAS2019, KnowlesMNRAS2023}. The semi-empirical libraries extend the coverage of the parameter spaces, especially for the \alphafe{}. In short, the sMILES library performed differential corrections on individual MILES stars whose \alphafe{} values have been measured by \citep{MiloneMNRAS2011}. The \alphafe{} in the sMILES library covers the range from −0.2 to +0.6 dex in steps of 0.2 dex, particularly useful for massive ETGs with, on average, super-Solar \alph{} abundance. Figure \ref{fig:sspgrid} shows the SSP indices, which are measured after being smoothed to the same resolution (\sig$\sim 300\ \rm km/s$ at $\lambda\mathrm{=5000}$\r{A}) as our spectra. The red arrows indicate the general direction of increasing metallicity (thick) and increasing \alphafe{} (thin). Previous studies have shown some evidence of IMF variation in massive ETGs, with the centers being more bottom-heavy than the outskirts. Therefore, we present the model grids with two different IMFs, a bottom-heavy IMF: bimodal IMF $\gamma=2.8$ (red dotted line) and the Milky Way-like IMF Universal Kroupa \citep{KroupaMNRAS2001} (blue dashed line). Because the galaxies in our sample have, in general, old stellar populations, we set the stellar age of both grids to 10 Gyr.

    Figure \ref{fig:sspgrid} makes it clear that the two composite indices cannot trace the variation of \alphafe{} and metallicity perfectly. According to these SSP models, the massive galaxies in our samples all have super-Solar \alphafe{} at around 0.4 dex and a sub-to-near Solar total metallicity. All samples show clear negative metallicity gradients and flat radial profiles of \alphafe{}. While it is difficult to more qualitatively interpolate the grids to disentangle the exact \alphafe{} and metallicity values of our samples, the empirical trends we inferred from the comparisons of index gradients still stand: 

    \begin{itemize}
    
        \item For massive galaxies with higher central stellar velocity dispersion values, they exhibit slightly higher total metallicity and significantly higher \alphafe{} values in all three radial bins. While these two samples demonstrate almost identical stellar mass distributions at $R>20$ kpc, the differences in their average stellar population properties persist to the outermost bin.

        \item At fixed \sigmacen{} and total stellar mass, the stellar mass distributions of massive galaxies show connections to the average stellar population properties too: the more extended sample shows systematically higher \alphafe{} values than the more compact ones in all the radial bins. However, the quantitative difference could be slightly smaller than the difference between the \hsig{} and \lsig{} samples. Even after accounting for the significant difference in their average $R_{\rm e}$, the result remains true within the shared radial coverage at $R< 10$ kpc. When using $R_{\rm e}$ as the unit, the more compact sample shows higher total metallicity than the more extended ones in the corresponding radial bins. However, at the fixed radius using kpc as the unit, the compact sample only shows a higher stellar metallicity at $\sim 10$ kpc.
        
    \end{itemize}

    In this section, we present the absorption indices of the stacked spectra, which already reveal tantalizing empirical trends. We emphasize that, even without comparison to the SSP model grids, the `model-independent' index values faithfully reflect the systematic differences in the average spectra among massive galaxies with different properties. Meanwhile, we still need a more quantitative approach to compare the element abundance between these sub-samples. We now explore this direction using the full-spectrum fitting method. 
    
\subsection{Full Spectrum Fitting}
    \label{sec:fsf}

\begin{figure*}[!th]
\centering
    \includegraphics[width=1\linewidth]{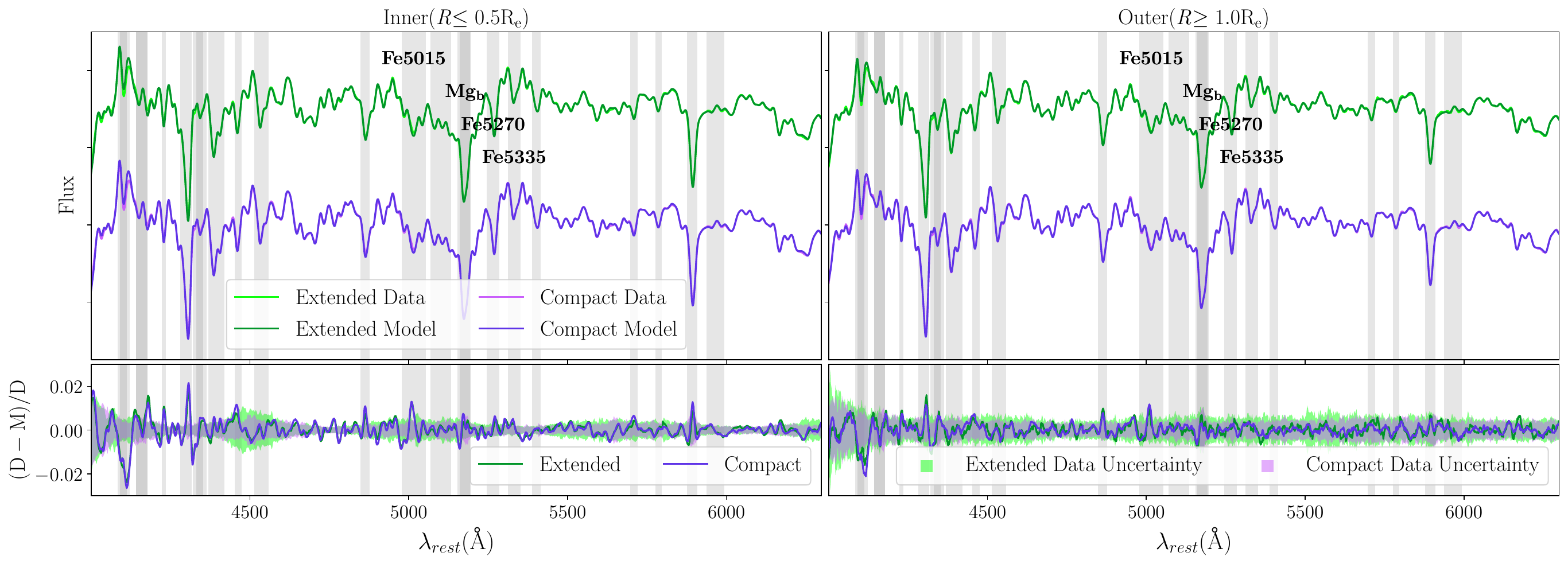}
    \caption{
        Top row: continuum-extracted input data(lighter-colored lines) and the model spectra (darker-colored lines), Bottom row: relative residual=(Data-Model)/Data, and the relative data uncertainty ($3\sigma$/Data) are shown in lines and shaded areas, respectively. The inner and outer spectra are shown in the left and right subplots, respectively. The extended sub-sample is in green, and the compact sub-sample is in purple. The \texttt{Jupyter} notebook for reproducing this figure can be found here: \href{https://github.com/xyzhangwork/mdensity_v_stellarpop/blob/main/plot_scripts/plot_alf_spectra.ipynb}{\faGithub}. This repository is also available on \href{https://doi.org/10.5281/zenodo.17979404}{Zenodo}.
        }
    \label{fig:alf_spec}
\end{figure*}

\begin{figure*}[!th]
\centering
    \gridline{\fig{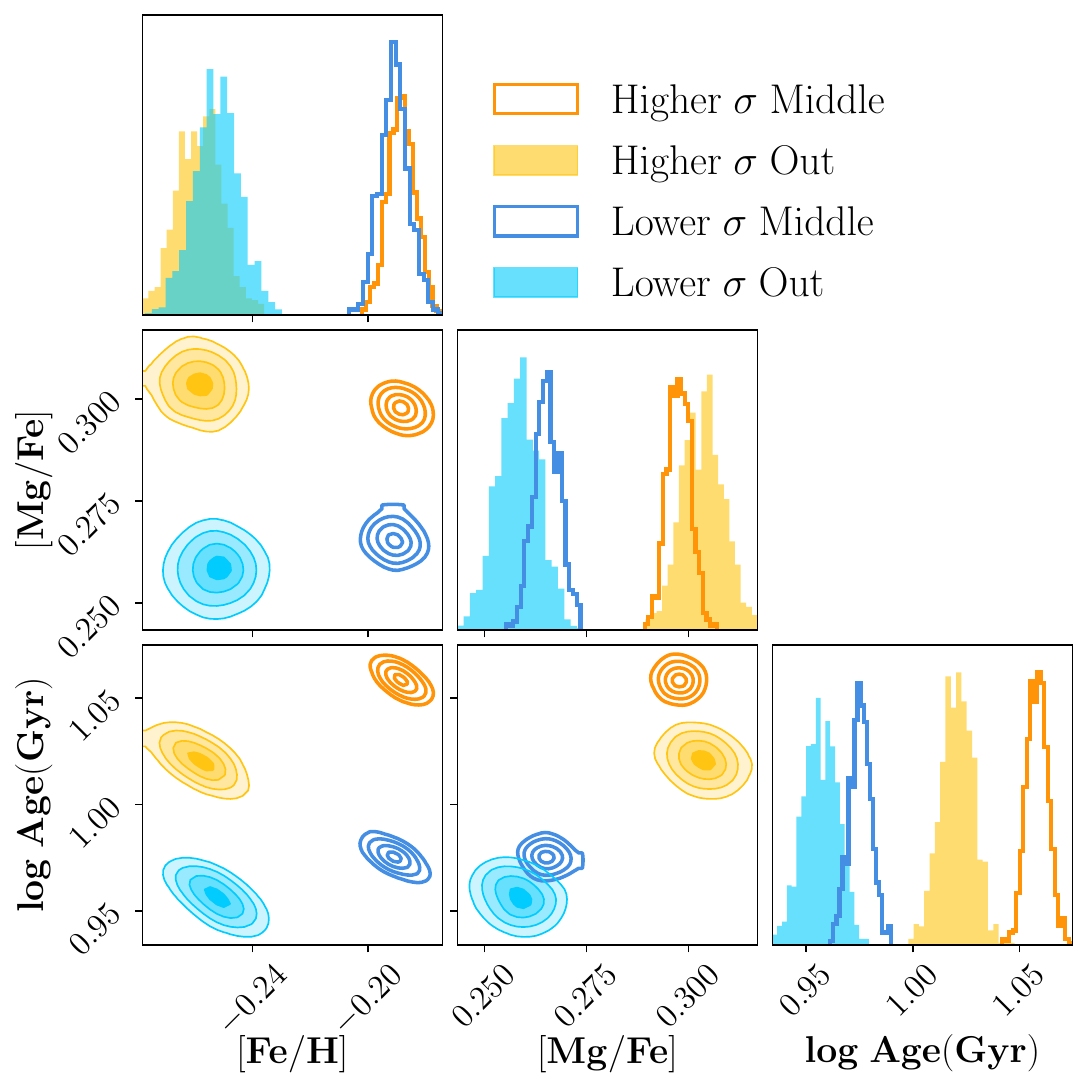}{0.5\textwidth}{(a)}
    \fig{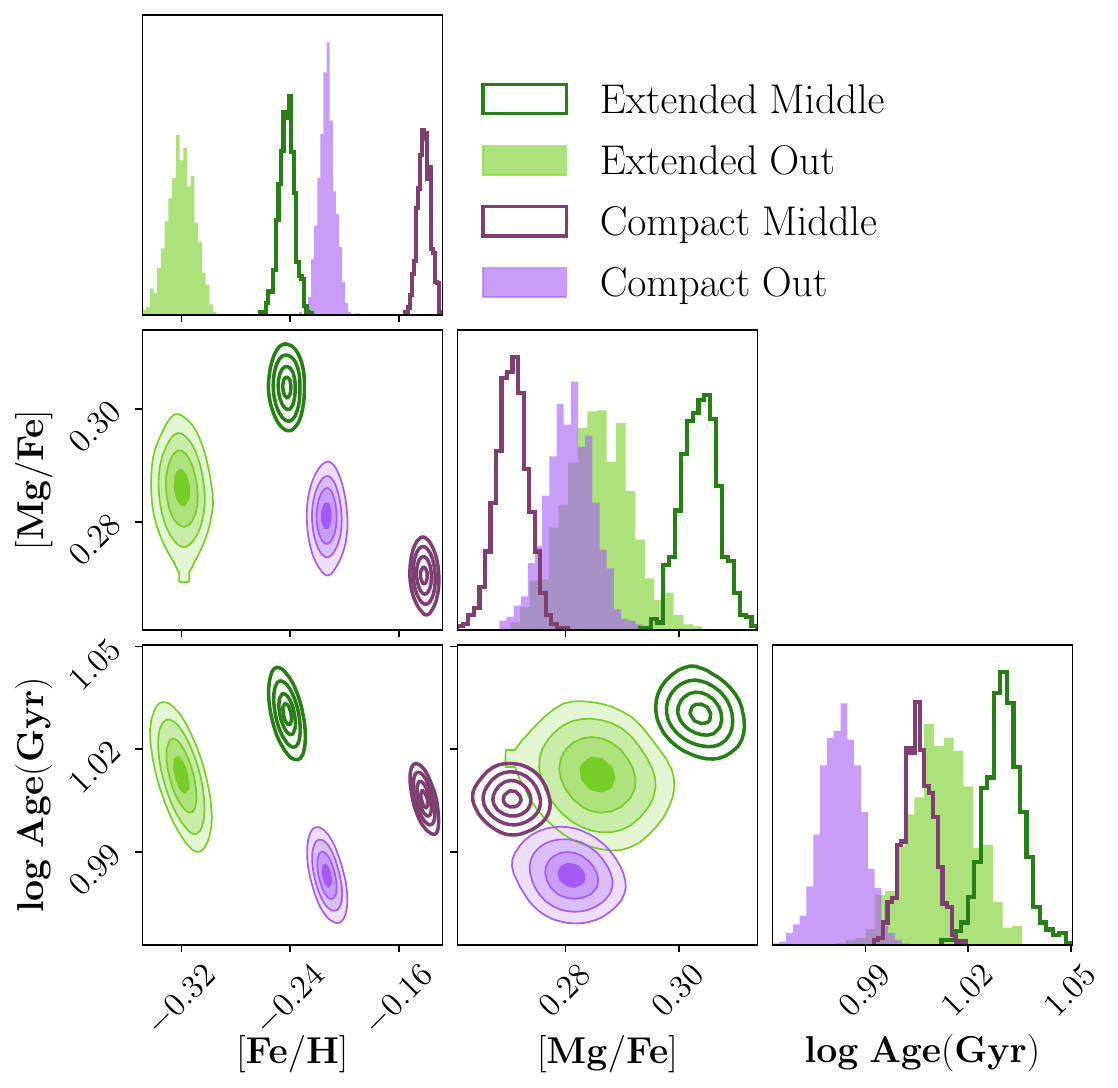}{0.5\textwidth}{(b)}}
        \caption{
        Corner plots of \feh, \mgfe, and stellar age(\logage) from the \alf{} fitting for the two sample-split methods. We only show the middle (empty) and outer (filled) bins in the plots. The left figure shows the \hsig{} (orange) vs. \lsig{} (blue) sub-samples. The right figure shows the Extended (green) vs. Compact (purple) sub-samples. The \texttt{Jupyter} notebook for reproducing this figure can be found here: \href{https://github.com/xyzhangwork/mdensity_v_stellarpop/blob/main/plot_scripts/plot_alf_corner.ipynb}{\faGithub}. This repository is also available on \href{https://doi.org/10.5281/zenodo.17979404}{Zenodo}.
        }
        \label{fig:corner}
\end{figure*}

    In this section, we employ the full-spectrum fitting method to address some limitations of relying on empirical trends based on absorption indices and to facilitate more quantitative comparisons of stellar population properties across our samples. Here, we utilize the full-spectrum absorption line fitting code \alf{} \citep{ConroyAPJ2012, ConroyAPJ2014, ConroyAPJ2018}, which can simultaneously estimate individual element abundances using synthetic response functions as functions of stellar age and metallicity \citep{ConroyAPJ2012} - a key advantage for our goals. Following the literature, we apply \alf{} to fit our stacked spectra using the MIST \citep{ChoiAPJ2016} isochrones with MILES and Extended-IRTF empirical stellar libraries \citep{VillaumeAPJS2017}. \alf{} implements the Markov-Chain Monte Carlo (MCMC) algorithm from the \texttt{emcee} package\citep{ForemanMackeyPASP2013} to infer the posterior distributions of the explored stellar population parameters. 

\begin{figure*}[!th]
    \centering
    \includegraphics[width=0.8\linewidth]{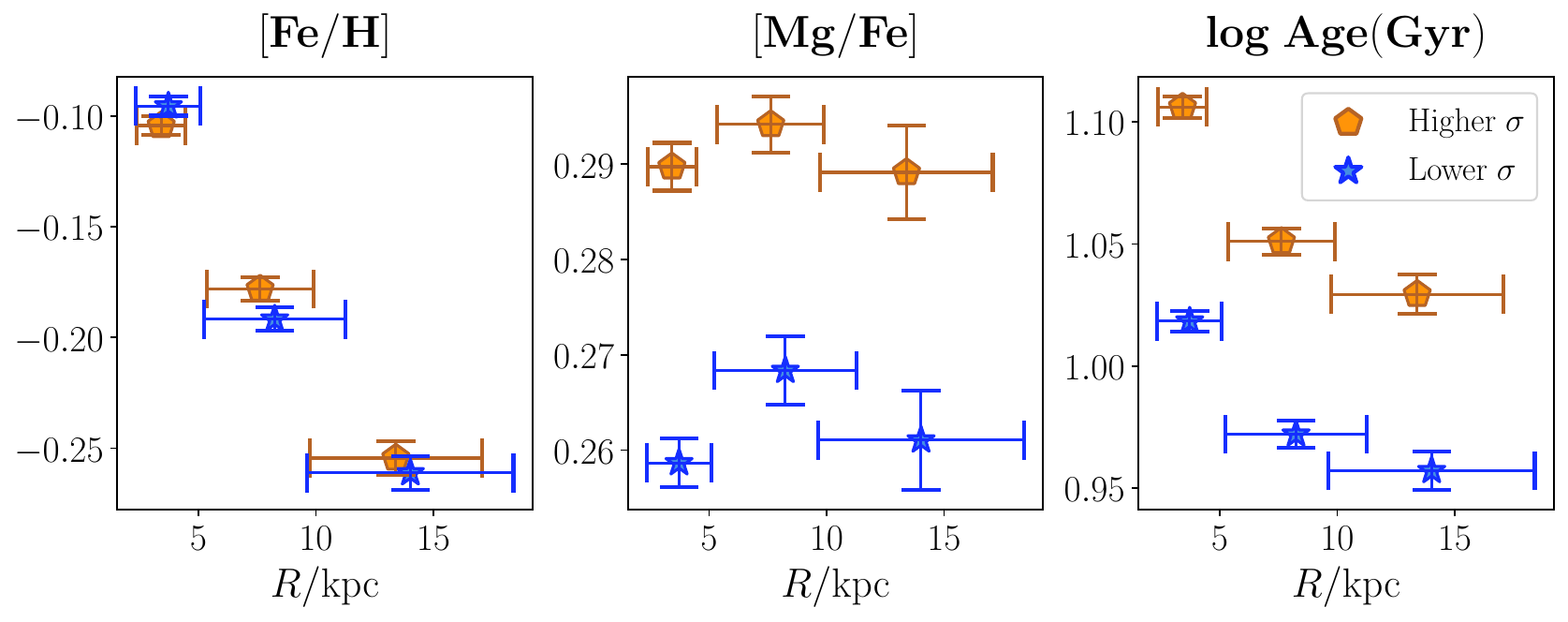}
    \caption{
        The radial profiles of the \hsig{}(orange pentagons) and \lsig{} (blue stars) sub-samples. From left to right: \feh, \mgfe, and \logage. The y-axis error is the standard deviation of the posterior, and the distance error is the standard deviation of the radial distance in each bin. The \texttt{Jupyter} notebook for reproducing this figure can be found here: \href{https://github.com/xyzhangwork/mdensity_v_stellarpop/blob/main/plot_scripts/plot_alf_gradient.ipynb}{\faGithub}. This repository is also available on \href{https://doi.org/10.5281/zenodo.17979404}{Zenodo}.
        }
    \label{fig:alf_gradients_sigma}
\end{figure*}

\begin{figure*}[!th]
    \centering
    \includegraphics[width=0.8\linewidth]{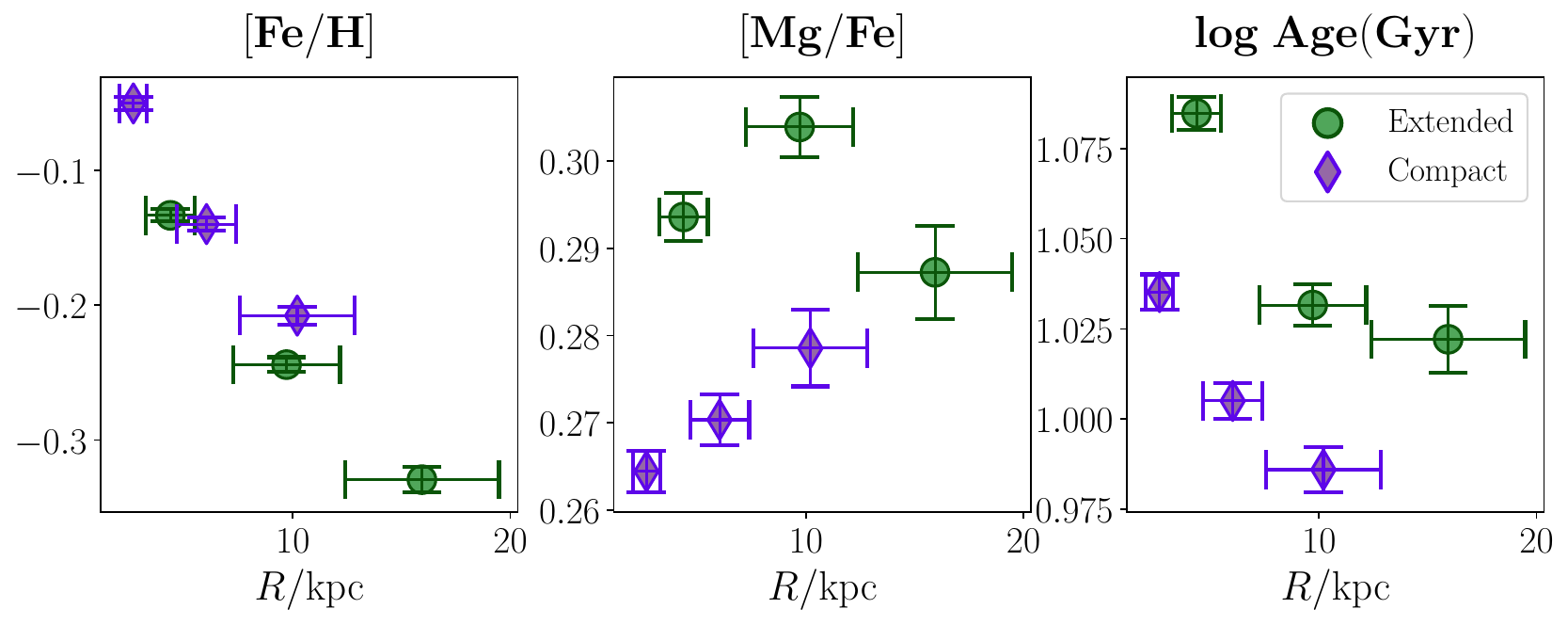}
    \caption{
        The radial profiles of the Extended (green circles) and Compact (purple diamonds) sub-samples. From left to right: \feh, \mgfe, and \logage. The y-axis error is the standard deviation of the posterior, and the distance error is the standard deviation of the radial distance in each bin. The \texttt{Jupyter} notebook for reproducing this figure can be found here: \href{https://github.com/xyzhangwork/mdensity_v_stellarpop/blob/main/plot_scripts/plot_alf_gradient.ipynb}{\faGithub}. This repository is also available on \href{https://doi.org/10.5281/zenodo.17979404}{Zenodo}.
        }
    \label{fig:alf_gradients_mass}
\end{figure*}

\begin{table*}[!th]
    \centering
    \begin{tabular}{|l|l|l|l|}
    \hline
         & \delt \feh /\delt log R(kpc) & \delt \mgfe /\delt log R(kpc) & \delt \logage{} /\delt log R(kpc) \\
         \hline
        {\color[HTML]{CE6E01} \hsig{}} & $-0.25\pm 0.01$ dex & $-0.00\pm 0.01$ dex &  $-0.13 \pm 0.02$\\
        \hline
        {\color[HTML]{3166FF} \lsig{}} & $-0.26 \pm 0.02$ dex & $-0.00 \pm 0.01$ dex & $-0.11 \pm 0.02$\\
        \hline
        {\color[HTML]{009901} Extended} & $-0.35 \pm 0.02$ dex & $-0.01\pm 0.01$ dex & $-0.11 \pm 0.02$\\
        \hline
        {\color[HTML]{6434FC} Compact} & $-0.27 \pm 0.01$ dex  & $0.02 \pm 0.01$ dex & $-0.08 \pm 0.01$\\
        \hline
    \end{tabular}
    \caption{The radial gradients of \feh{}, \mgfe{} and \logage{} for all four sub-samples.}
    \label{table:1}
\end{table*}

    Given the wavelength coverage and the statistical \snratio{} of our stacked spectra, we adopt the \smode{} of \alf{} to fit 12 parameters simultaneously within the wavelength range of $\mathrm{4000\sim 6300}$\r{A}: the SSP stellar age, stellar radial velocity, stellar velocity dispersion, \feh, and the elemental abundances of C, N, O, Mg, Si, Ca, Ti, and Na. To ensure the coverage of the MCMC chains, we use 200 `walkers', a 20,000-step burn-in process, and a final 512 steps to infer the posteriors of all the parameters. For the fiducial fitting run, we assume a Milky Way-like Kroupa IMF. As more and more observational evidence suggests that the center of massive ETGs could have a more bottom-heavy IMF than the Kroupa model, we present the results of a varying IMF \alf{} model in Section \ref{ssec:imf}. While we observe evidence of a more bottom-heavy IMF in the center and its systematic impact on stellar population properties such as \mgfe{} and \logage{} (towards lower \mgfe{} and older stellar age compared to a fixed Kroupa IMF), the effect on different sub-samples is similar, and the primary conclusions concerning the two sample-split methods still hold.

    For stellar metallicity, we use \feh{} as the proxy for reasons already articulated in \citet{GreeneAPJ2015}: the iron abundance directly comes from the various iron-related indices in the optical range, while the total metallicity requires that we know all the elemental abundances including oxygen, which is difficult to estimate in the optical range due to the lack of relevant absorption features. For the same reason, we choose \mgfe{} instead of \alphafe. It also enables us to compare our results directly with those of many previous studies and the results based on absorption indices in Section \ref{sec:index}.

    In Figure \ref{fig:alf_spec}, we display the continuum-subtracted, median-stacked spectra of the extendedness-split samples as an example. We only present the inner (left) and outer (right) radial bins to emphasize the differences in the spectra (top) and their uncertainties (shaded area in the bottom row), as described in Section \ref{ssec:stacking}. The bottom rows highlight the relative residuals of the \alf{} models (represented by the green and purple lines). The more compact galaxies show higher statistical uncertainties in the outskirts due to their steeper light profiles \citep[e.g.,][]{ParikhMNRAS2018}. These residuals show us that, even with the \smode{} mode, \alf{} can recover the input spectra with relative error less than $2\%$. We also note that, despite differences in input uncertainties, the two samples exhibit similar residuals across the full radial range. There is, however, a region of relatively high residual where the index Fe5015 resides, which could be caused by weak emission lines such as $\rm [OIII]4959$\r{A},\ 5007\r{A}. To evaluate its impact on the results, we also test fitting the spectra with the Fe5015 region masked out and verify the robustness of the main results. 
    
    We visualize the \alf{} results from the fiducial models in Figure \ref{fig:corner} using corner plots, which show the \feh{}, \mgfe{}, and stellar age for the middle (empty histograms) and outer (filled histograms) radial bins. On the left, the corner plot for the \sigmacen{}-split samples shows a clear negative gradient for \feh{} and \logage{}, with a flat \mgfe{} profile within these two radial bins. And, on the right side, for the extendedness-split samples, while the negative \feh{} and \logage{} gradients are still visible, the radial trends for \mgfe{} become more complicated. These corner plots visualize the statistical capability enabled by the stacked spectra and the \alf{} models, revealing subtle yet systematic differences in the physical properties of old stellar populations. 
    
    Next, we analyze the radial trends of these average stellar population properties in physical units of kpc, similar to those shown in Figure \ref{fig:index_sigma} and Figure \ref{fig:index_mass} in Section \ref{sec:index}. Figure \ref{fig:alf_gradients_sigma} and Figure \ref{fig:alf_gradients_mass} show the radial gradients of \feh{}, \mgfe{}, and \logage{} of the \sigmacen{}-split and extendedness-split samples. We also calculate a simple $\log$-linear radial slope value, defined as $\Delta\ X /R = (X_{\rm outer}-X_{\rm inner})/(\mathrm{log}\ R_{\rm outer}(\mathrm{kpc})-\mathrm{log}\ R_{\rm inner}(\mathrm{kpc}))$, for each profile, and summarize these values in Table \ref{table:1} to provide a more quantitative comparison.

    For the \sigmacen{}-split samples, the \hsig{} and \lsig{} galaxies follow an almost identical negative \feh{} gradients in all three radial bins, which is very different from the Fe5335 index profiles and similar to the result of Fe5270 and the composite \mgfex{} index. Both samples also show a negative stellar age gradient and a flat \mgfe{} profile, consistent with several previous works \citep[e.g.,][]{ParikhMNRAS2021, OyarzunAPJ2023}. However, these two sub-samples demonstrate significant differences in their \mgfe{} and \logage{} values: the \hsig{} sub-samples exhibit higher \mgfe{} and older stellar populations consistently within $R< 15$ kpc. It is worth emphasizing that, while the qualitative differences between the \hsig{} and \lsig{} samples seen here are in line with the expectations based on the known scaling relations between \sigmacen{} and stellar population properties, they are not equivalent. Here, we match these two sub-samples using the outskirts stellar mass, and they have identical average stellar mass distributions at $R > 20$ kpc, which could indicate a similar merger or accretion history. At the same time, the \hsig{} sub-sample does show a steeper light profile and higher stellar mass density in the central region. Combine these characterizations with the stellar population trends, this observation suggests a potential link between the more efficient and intense starburst at an earlier time and the buildup of the high-density, high-\sigmacen{} stellar component, presumably related to the \insitu{} component, in the central region of massive ETGs. Note that, while the \hsig{} sample is slightly more massive than the \lsig{} samples in total stellar mass, a significant fraction of that difference could be accounted for by the inner region. Although it is challenging to infer the \insitu{} or \exsitu{} fraction according to the definition in simulation, our conclusion could suggest that massive galaxies with different early \insitu{} formation could share a similar merger history later, as indicated by the identical outer stellar distributions of these two samples. However, both \lsig{} and \hsig{} sub-samples exhibit a flat \mgfe{} profile, and their difference shows no sign of narrowing as we move to a larger radius (1.5\re{}). This means that either the outer regions covered by this work still have a significant contribution from \insitu{} components, or this is actually a sign of coordinated quenching between the accreted satellites and the central galaxy, leading to a similarly high \mgfe{} in both \insitu{} and \exsitu{} components.

    As for the extendedness-split samples, after addressing the differences in physical sizes between the compact and extended samples, we see on the left panel of Figure \ref{fig:alf_gradients_mass} that the more compact massive galaxies show higher \feh{} within the inner 10 kpc than the more extended ones, consistent with the results from the iron-sensitive indices in Section \ref{sec:index}. It is interesting to see that, while a \sigmacen{}-split does not reveal a clear difference in \feh{} among massive galaxies, we achieve this using a more ``global'' property -- the stellar mass distribution, \emph{at the fixed \sigmacen{} and total stellar mass}. Both samples show a negative \feh{} gradient with comparable slopes ($0.28\pm 0.01$ dex for the compact sample and $0.31\pm 0.02$ dex for the extended sample). As for the stellar age (right panel of Figure \ref{fig:alf_gradients_mass}), the more extended galaxies are slightly older than the more compact samples in the shared radial coverage. Note that the age difference is slightly narrower than that between the \hsig{} and \lsig{} samples, and should be treated with caution given the ancient nature of these samples. More importantly, the more extended galaxies also exhibit systematically higher \mgfe{} values within the inner 10 kpc than their more compact counterparts. We also notice that both sub-samples have a weak positive gradient from the center to 10 kpc, in contrast to the negative gradient of \mgbfex{} in Section \ref{sec:index}. However, the \mgfe{} of the extended sub-sample drops at \~ 15 kpc, albeit with a higher uncertainty. Even with the cautious stacking method employed in this work, the statistical uncertainties of the median spectra increase significantly with increasing radius. We recommend exercising greater caution when interpreting the trends in the outer radial bins of both sub-samples. For instance, taking the \ofe{} abundance ratio inferred from \alf{} at face value, the more extended sample shows an even higher \ofe{} value in the outer bin instead (see Appendix \ref{appendix:abundance}). We still need to wait for IFU data with higher \snratio{} and larger radial coverage to confirm the hints seen here. 
    
    Meanwhile, regardless of these systematics, our results demonstrate that, \emph{at the same \sigmacen{} and total stellar mass}, there is a connection between the stellar mass distributions in massive galaxies, hence their assembly history, and the average stellar population properties, which reflect their integrated star formation histories. When the total amount of stars formed or accreted into a massive galaxy is controlled, the galaxies that started to form stars at an earlier epoch or had a more intense \& efficient starburst period (or were quenched more rapidly) tend to accumulate more stars later in their stellar halos. In contrast, the ones that could enjoy a more prolonged period of star formation at their centers (hence the higher \feh{}, lower \mgfe{}, and slightly younger age) failed to do so. We acknowledge that the absolute difference in the \mgfe{} values is small and subject to model uncertainties. However, as our conclusion is robust against the choice of using a varying IMF (see Appendix \ref{ssec:imf}), and the \mgfe{} difference in the central region is close to the expected difference between the \hsig{} and \lsig{} samples ($\sim0.03$-0.04 dex), we believe the qualitative conclusion draw from this work is safe.

% -------------------------------------------------------------------------------------------- %
% Discussions
% -------------------------------------------------------------------------------------------- %

\section{Discussions} 
    \label{sec:discussions}
    
% -------------------------------------------------------------------------------------------- %
\subsection{Scaling Relations based on Central Velocity Dispersion}
    \label{ssec:sigma_cen}

    The central stellar velocity dispersion (\sigmacen) of a galaxy is a critical metric, reflecting the depth of the gravitational well and profoundly influencing critical baryonic processes (e.g., gas accretion, star formation, metal recycling, and different types of feedback). It is often considered the "governing factor" for stellar population properties, showing tighter correlations than luminosity or stellar mass \citep{GreeneAPJ2015, BaroneAPJ2018}. Therefore, we want to sanity check our results against the \sigmacen{}-based scaling relations in the literature. Therefore, we want to sanity check our results against the \sigmacen{}-based scaling relations. Note our sample covers a narrow, overlapping \sigmacen{} range ($\sim 200\ \rm km/s$ to $300\ \rm km/s$) with moderate average differences ($\sim 220\ \rm km/s$ v.s. $280\ \rm km/s$).

    Firstly, massive ETGs with higher \sigmacen{} show slightly lower Fe5335 in the center while the difference in Fe5270 is much smaller (Figure \ref{fig:index_sigma}). However, the difference in Fe5335 between \hsig{} and \lsig{} is $\sim 0.1$ dex, consistent with the scatter in the scaling relation by \citet{KuntschnerMNRAS2001}.

    And when we turn to the \feh{} estimated from \alf{}, we see little difference between the \hsig{} and \lsig{} sub-samples, consistent with previous work that either found none or only a weak correlation between velocity dispersion and \feh{} \citep[e.g.,][]{PriceMNRAS2011, JohanssonMNRAS2012, ConroyAPJ2014, WalcherA&A2015}, or a steeper slope at low-$\sigma$ end and flattens towards higher $\sigma$ \citep[e.g.,][]{GallazziMNRAS2006, ParikhMNRAS2021}
    We also find that the impact of \sigmacen{} on \feh{} declines with radius (Figure \ref{fig:alf_gradients_sigma}). This is in agreement with several previous works \citep[e.g.,][]{GreeneAPJ2015, ParikhMNRAS2019, FerrerasMNRAS2019}.
    
    For stellar ages, we find that the \hsig{} sub-sample has older, higher \mgfe{} stellar populations in all three radial bins, consistent with the hierarchical picture in which systems with higher \sigmacen{} undergo a more intense, shorter star-formation process that started earlier. This is in agreement with several previous works \citep[e.g.,][]{TragerAJ2000, ThomasAPJ2005, JohanssonAPJ2012, ConroyAPJ2014, GreeneAPJ2015, FeldmeierKrauseAPJ2021,ParikhMNRAS2021}. 

    Interestingly, we note that although the \hsig{} and \lsig{} sub-samples have similar \mout, the difference in \mgfe{} persists to the outer radial bin and shows no signs of narrowing. This could be because the MaNGA  `outer' radial bin (to 10-15 kpc) is typically smaller than our definition of the `outer' stellar mass ($>20$ kpc), so the \exsitu{} population might not dominate. Meanwhile, as shown by \citet{GuApJ2018}, the environments of massive galaxies or their host dark matter halos could help `coordinate' their assembly. For instance, a more massive dark matter halo can not only affect the star-formation history of its central galaxy (or the brightest cluster galaxy, BCG), but also shape the galaxy's morphology. Still, it can also help quench the satellites, or even the surrounding low-mass galaxies, more abruptly, leading to higher \alphafe{} values in these systems than predicted by the \sigmacen{}-\alphafe{} scaling relation. And when these less massive galaxies eventually fell into the BCG, they contribute to an \alphafe{}-enhanced stellar halo. Although most of the massive ETGs in MaNGA are not BCGs, this scenario could apply to all dark matter halos above a certain threshold, resulting in a flat \alphafe{} or \mgfe{} profiles from the center to $>1.5\times R_{\rm e}$. 
    
    In \citet{HuangMNRAS2022}, the authors demonstrate that the outskirt $M_{\star}$ (stellar mass between 50 and 100 kpc) correlates with the underlying dark matter halo mass. Although this is beyond the reach of the LegacySurvey data, the matched average mass density profiles out to $>50$ kpc should provide a good indication of a similar average halo mass. At the same time, the central $\sigma_{\star}$ is considered an excellent proxy of the velocity dispersion of the dark matter halo, especially for the massive ETGs, making it also a potential halo mass proxy (e.g., \citealt{Zahid2018, Sohn2020, Utsumi2020}). Under this assumption, the \hsig{} sample should have a higher average halo mass than the \lsig{} sample. To verify this, we used the catalog \citet{Tinker2021}, which provides a self-calibrated group for SDSS galaxies and predicts a halo mass for each MaNGA massive galaxy. Based on this data, 89\% (94\%) of the \hsig{} (\lsig{}) samples are central galaxies of their groups. For these centrals, the \hsig{} and \lsig{} sub-samples share very similar halo mass distributions with median values of $10^{13.5}M_{\odot}$ vs. $10^{13.3}M_{\odot}$. While the \hsig{} sample, in theory, has a higher average halo mass, it is too early to tie this insignificant difference to the \alphafe{} profiles. Future works using a larger sample of massive ETGs and more reliable halo mass estimates (e.g., using weak gravitational lensing) could further shed light on this.

\subsection{Impact of IMF Choice}
    \label{ssec:imf}

\begin{figure*} 
    \centering
    \includegraphics[width=1.0\linewidth]{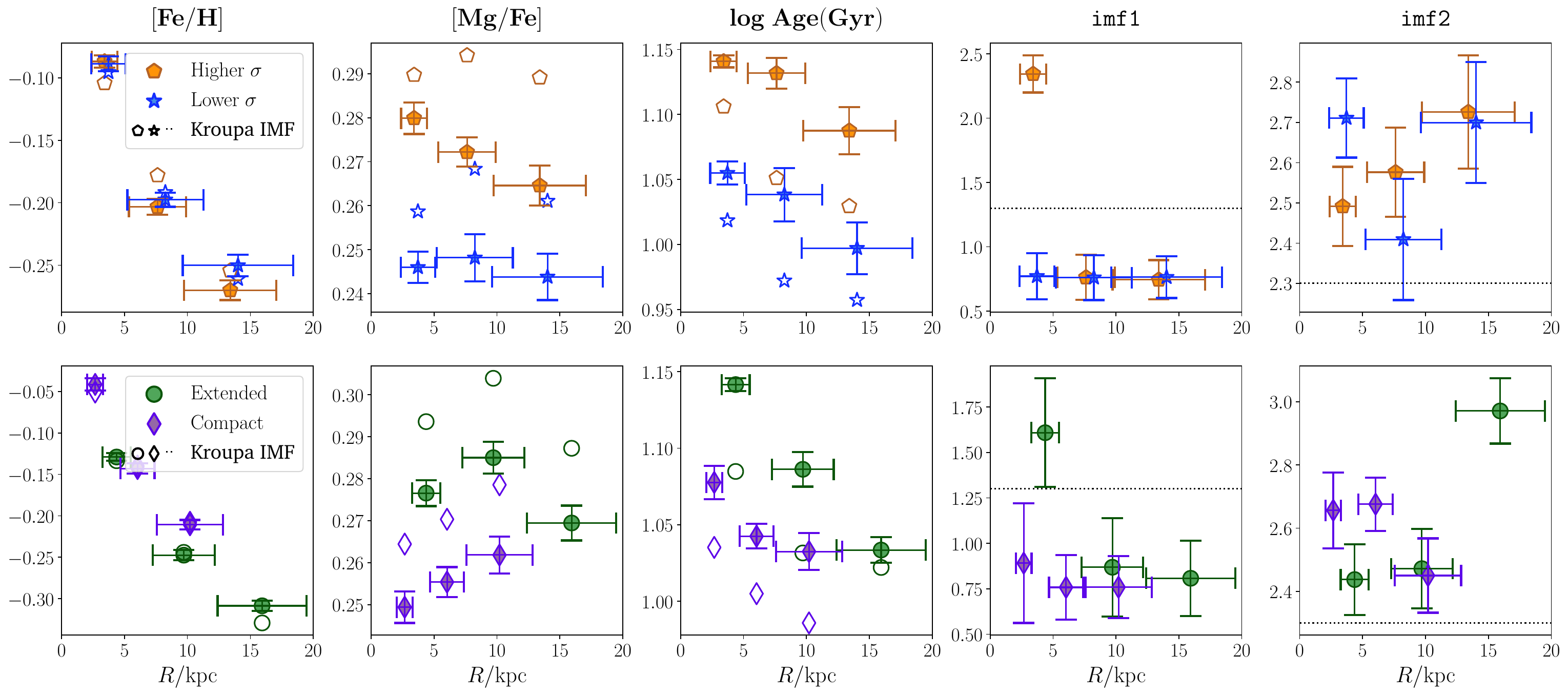}
    \caption{
        Radial gradients under \fmode{} fitting. From left to right: \feh, \mgh, \mgfe, \imfi{} and \imfii. Top row: the \sig-split samples. The bottom row: the \mstar{} profile-split samples. The empty markers in the first columns represent the \mgfe{} values from the previous \smode{} fitting. The dashed line in the last two columns shows the slopes of a Kroupa IMF. The \texttt{Jupyter} notebook for reproducing this figure can be found here: \href{https://github.com/xyzhangwork/mdensity_v_stellarpop/blob/main/plot_scripts/plot_alf_gradient_imf.ipynb}{\faGithub}. This repository is also available on \href{https://doi.org/10.5281/zenodo.17979404}{Zenodo}.}
    \label{fig:alf_gradients_imf}
\end{figure*}

    The variation of IMF in massive ETGs is a heavily debated topic in the field. Some \citep[e.g.,][]{LaBarberaMNRAS2016, vanDokkumAPJ2017, ConroyAPJ2017} argue for a more bottom-heavy IMF in the center of massive ETGs and a more Kroupa-like IMF in larger radii, while others \citep[e.g.,][]{AltonMNRAS2017} think a varying elemental abundance can explain the spectral difference observed. An oversimplified IMF assumption could mislead the interpretation of spectral features and even have unexpected effects on overall abundance estimates. Although we use the \smode{} (which assumes a Kroupa IMF) of \alf{} in this work to perform full-spectrum fitting to avoid over-fitting in a relatively short spectral range, we want to take extra precautions to make sure our results are not affected by uncertainties in IMF, especially when the massive galaxies in our sample could have a bottom-heavy IMF at its center\citep[e.g.,][]{CenarroMNRAS2003, vanDokkumNatur2010, SpinielloApJ2012, ConroyAPJ2012, GuAPJ2022} and also show an IMF gradient \citep[e.g.,][]{Martin-NavarroApJ2015, vanDokkumAPJ2017, ParikhMNRAS2018, SarziMNRAS2018, LaBarberaMNRAS2019}. 

    To test this, we perform the \alf{} \fmode{} fitting to stacked spectra with an additional segment within the spectral range between $\mathrm{8000\sim 8700}$\r{A} to include the IMF-sensitive features like the NaI and CaII triplets. The \fmode{} fits in a total of 46 parameters. We choose a double power-law IMF function with fixed cutoffs at 0.08 and 100 \msun{} and two varying slopes at the low mass end (\imfi{} for 0.08\msun $\lt$ M $\lt$ 0.5\msun, and \imfii{} for 0.5\msun $\lt$ M $\lt$ 1.0\msun). For $\gt$ 1.0\msun stars, the slope is fixed at $\gamma$=2.3 in the Salpeter IMF. For a Kroupa IMF, we would have \imfi = 1.3 and \imfii = 2.3, respectively. We choose a larger burn-in step of 25,000 and 100 chains, each with 1024 steps.

    In Figure \ref{fig:alf_gradients_imf}, we show the radial profiles of \feh, \mgfe, \logage{}, and the two IMF slope gradients from the \fmode{} fitting (The radial profiles of other elemental abundances can be found in Appendix \ref{appendix:abundance}.). The empty markers in the left three columns show the \smode{} fitting results for comparison. The upper and lower rows show the \sigmacen-split and extendedness-split results, respectively. We can see that allowing the IMF slope to vary has little impact on the estimated \feh{} values. In the meantime, it results in systematic changes in other elemental abundances and stellar age: the \fmode{} fitting gives a lower \mgfe{} and older \logage{} for all 12 spectra. Most notably, the outer \mgfe{} of the \hsig{} decreases drastically compared with the \smode{} estimation, resulting in the difference between \hsig{} and \lsig{} sub-samples narrowing with radius. Although the current data quality prevents us from providing a definitive estimate of the stellar population properties, this section confirms that the trend observed between the sub-samples still holds across varying IMFs.

    Regarding the IMF slopes, we observe that the \hsig{} sample exhibits a bottom-heavy IMF in the center and a steep IMF gradient within the central 5 kpc, as indicated by the \imfi{} slope of $<0.5$ \msun. For the extendedness-split samples, both the compact and extended samples show evidence of a bottom-heavy IMF in the center. This is consistent with several previous studies on spatially resolved IMF in massive galaxies using IFU data \citep[e.g.,][]{Martin-NavarroMNRAS2015, ConroyAPJ2017, GuAPJ2022}. Outside the central radial bins, all samples show \imfi{} values lower than those for a Kroupa IMF. As our wavelength coverage and data quality are not ideal for accurately inferring IMF slopes, we do not wish to confirm a bottom-heavy IMF or estimate \imfi{} values in the outer radial bins. But the \fmode{} fitting can provide reassurance of the robustness of the \mgfe{} difference against any IMF slope variations.

\subsection{Diagnosing the Two-Phase Assembly of Massive Galaxies}
    \label{ssec:assembly}

\begin{figure*}[!ht]
\centering
    \includegraphics[width=0.7\linewidth]{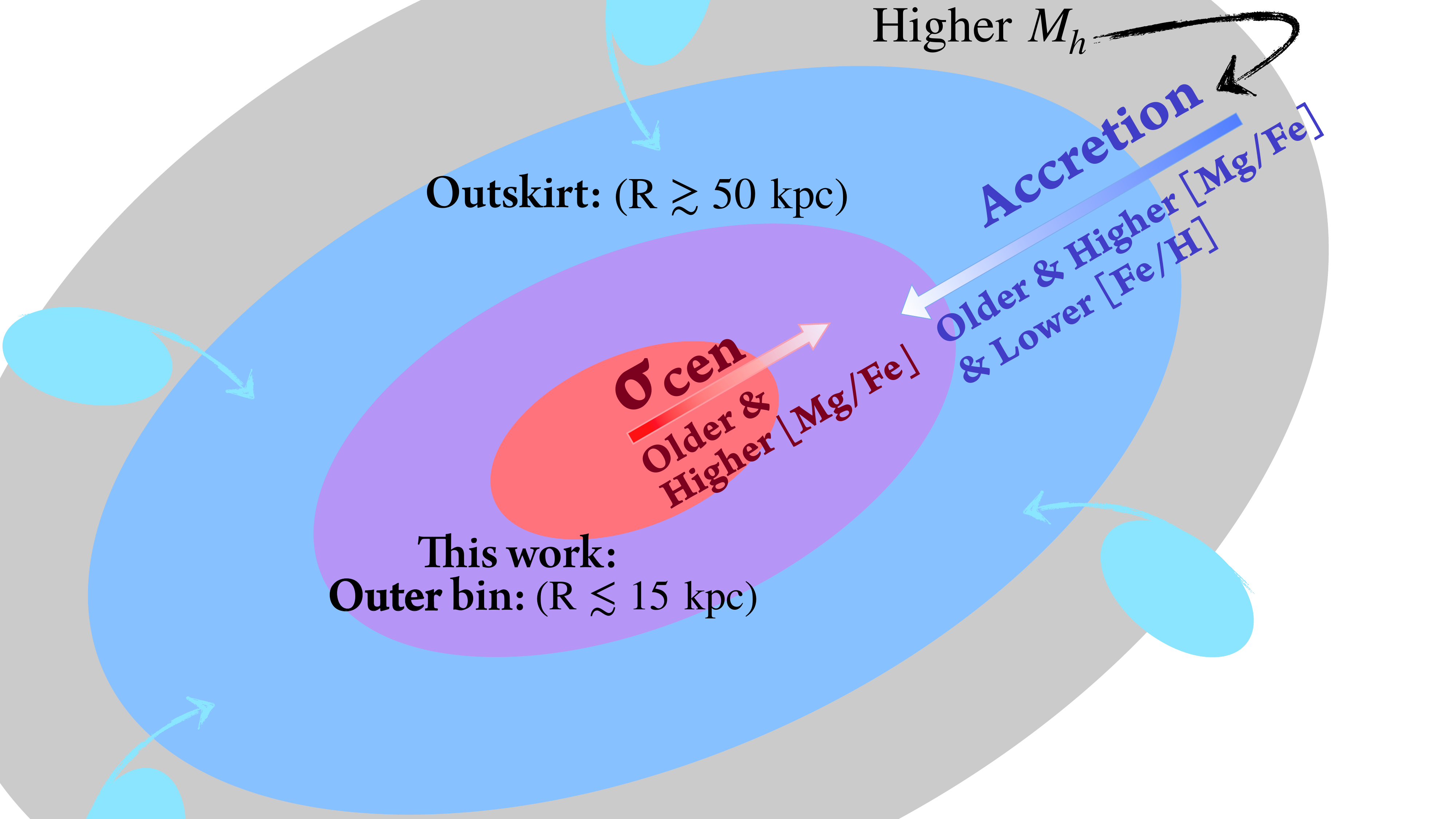}
    \caption{
        An illustration showing how the internal (\sigmacen{}) and external (Accretion) properties and their impacts on the stellar populations of massive ETGs: Higher \sigmacen{} and Accretion from dry minor mergers (\exsitu{} component) both can elevate the \alphafe{} of the average stellar population, as well as make the average stellar age slightly older. In the meantime, the \exsitu{} stellar population also has lower \feh{}, causing an iron-deficiency in the outskirts. And at the outer radial bin in this work ($\sim 1.5$ \re{}), both factor plays a non-negligible role, which can be seen from the spectral features.
        }
    \label{fig:illustration}
\end{figure*}

    According to the two-phase formation scenario (e.g., \citealt{OserAPJ2010, OserAPJ2012}), after $z\sim 2$, massive ETGs should have quenched the in-situ star formation and mostly grown by accreting nearby satellite galaxies (e.g., \citealt{vanDokkumAPJ2010}). These (mostly ``dry'') minor mergers result in a gradually accumulated accreted component (\exsitu) in the outskirts of the host galaxy \citep[e.g.,][]{NaabAPJL2009, JohanssonAPJ2012, HirschmannMNRAS2015, Rodriguez-GomezMNRAS2016, QuMNRAS2017, RemusApJ2022}. While this population picture has been widely accepted as the fiducial model for massive galaxy formation and has been qualitatively supported by many observations, its physical framework is based on the \insitu{} and \exsitu{} components, which are conveniently defined in hydro-simulations, but remain challenging to decompose in observations. Therefore, the quantitative details of this scenario still await examination with novel diagnostics. 

    In this work, the combination of deep imaging and IFU observation reveals several intriguing details about the average assembly and star formation histories of nearby massive ETGs out to $<1.5\times R_{\rm e}$. Our main results show that, when controlled for different variables, the systematic variations in the radial profiles of stellar population properties highlight the impact of different physical drivers at different epochs. Both the average stellar population properties and their spatial distributions could not be described by a simple scaling relation or predicted based on a `fundamental physical property' with a purely random, statistical scatter (see also \citealt{ScholzDiaz2022MNRAS}). While this complicates the situation significantly, it also presents unique opportunities to better diagnose the two-phase formation scenario.

    Compared to the trends in the sigma-split sub-samples discussed in ~\ref{ssec:sigma_cen}, where we find no \feh{} difference and a higher \mgfe{} in \hsig{} galaxies, we see a clear \alph-enhancement and \feh{} deficiency in extended galaxies even with the same \sigmacen{} and \mstar{}. One of the plausible explanations is an \exsitu{} related scenario: the more extended stellar halo is the result of a rich accretion history enabled by a higher halo mass, which can also quench the satellites more efficiently and lead to the higher \alphafe{} and \feh{} deficiency in the satellite population and then in the stellar halo after merging.

    Focusing on dwarf galaxies, \citet{LiuAPJL2016} found that the increasing scatter of the \sigmacen{}-\alphafe{} relation depends on the environment at the low \sigmacen{} end: dwarf galaxies in denser environments show elevated \alphafe{} values.  Similarly, other works using the SAMI IFU survey also find elevated \alphafe{} for dwarf galaxies in high-density environments \citep{WatsonMNRAS2022, RomeroGomezMNRAS2023}.
    Using the MANGA IFU data, \citet{OyarzunAPJ2022} found that satellites in more massive dark matter halos have higher \mgfe{} in general and exhibit more \alph-enhanced outskirts than the centers, further supporting the environmentally driven, outside-in-quenching scenario. While this scenario was supported by other works as well \citep[e.g.,][]{Annibali2011AA, Scott2017MNRAS}, others have pointed to a different direction \citep[e.g.,][]{LaBarbera2014MNRAS}. So far, the path to a solid conclusion appears to be complicated by various definitions of `environment', by stellar population analysis methods, and by potential aperture effects across different datasets.

%However, regardless of these complications, these results propose an interesting scenario related to our conclusion: if a more massive dark matter halo can indeed promote a more rapid quenching process for its satellite population, then it would create a more $\alpha{}$-rich dwarf galaxy population than their counterparts from a less massive halo, and also result in a stellar halo with higher \alphafe{} around the central galaxy. This scenario is also consistent with the `coordinated assembly' picture proposed by \citet{GuApJ2018}, where the building blocks of BCGs are more dominated by low mass populations that have, in general, higher \alphafe{} and older stellar age. While our work does not focus on massive galaxy clusters, the higher \mgfe{} value and older stellar age in the outer bin of the more extended massive ETGs at fixed \sigmacen{} seems to be qualitatively consistent with the scenario in a more general sample. 
    
    Regardless of these limitations, these results propose an interesting scenario where a more massive dark matter halo promotes a more rapid satellite quenching process, resulting in a more $\alpha{}$-rich dwarf galaxy population that, upon accretion, forms a stellar halo with higher \alphafe{} around the central galaxy. This is also consistent with the `coordinated assembly' picture proposed by \citet{GuApJ2018}, which states that the building blocks of BCGs are more dominated by low mass populations with higher \alphafe{} and older stellar age. While our work does not focus on massive galaxy clusters, the higher \mgfe{} value and older stellar age in the outer bin of the more extended massive ETGs at fixed \sigmacen{} seem to be qualitatively consistent with the scenario in a more general sample. 

    At the same time, we acknowledge the limitation of the conclusion we could draw from an average spectrum within the $1.5\times R_{\rm e}$ of a massive ETG. We believe that further observational pursuits in the following directions, with the help of hydro-simulations or semi-analytic models, could help us put the proposed scenario, along with the entire `two-phase' formation scenario of massive galaxies, on a more solid and quantitative basis. Here are just a few of the directions we can take in the future:

\begin{itemize}

    \item \textbf{Mapping the connection between halo mass, stellar mass distributions, and stellar population properties}: 
    Recent work shows that the outer stellar halo of massive ETGs could be a powerful halo-mass proxy \citep[e.g.,][]{HuangMNRAS2022}.  A model developed by \citet{Huang2020ASAP}, which constrains the halo mass variation across a 2-D parameter space defined by an inner and outer stellar mass\footnote{The original work uses an inner stellar mass and the total stellar mass. However, using an outskirt stellar mass was proven to have more advantage in constraining the halo mass.} also predicted that massive ETGs with more extended stellar halos reside in more massive dark matter halos. To confirm whether the \alphafe{} increases along with (or deviates from) the direction for higher halo mass on this 2-D plane (see Figure~\ref{fig:illustration}), beyond the simple sample-split method used here, will help us refine this critical scenario to understand the formation of massive galaxies. In \citet{ScholzDiaz2022MNRAS}, the authors also found that, when focusing on ETGs with high \sigmacen{}, halo mass plays a ''secondary yet noticeable role''. In the near future, we believe that 1. the \feh{} and \alphafe{} of the stellar halo could be a more powerful tool to disentangle the different physical processes that drive the \insitu{} and \exsitu{} growth; 2. jointly constraining the galaxy-halo connection model with weak lensing data will be a significant improvement compared to the stellar-halo mass relation (SHMR) today.
    
    \item \textbf{The intrinsic scatters of stellar age, \feh{}, \mgfe{} in the stellar halos of massive ETGs}: as we enter the extended stellar halo dominated by the \exsitu{} component, both merger history's stochasticity and its dependence on halo properties could leave their fingerprints on the stellar population. Compared to the tight \sigmacen{}-\alphafe{} relation in the center, diagnosing the two-phase scenario requires checking whether the \alphafe{} in the stellar halo still correlates well with \sigmacen{} or instead shows increasing intrinsic scatter. If the scatter increases, its dependence on halo mass or environment could be a crucial constraint for the above scenario. 

    \item \textbf{Comparison between the stellar halo and satellite galaxies today}: 
    Several above-mentioned works explored the \sigmacen{}-\alphafe{} relations of low-mass galaxies in today's massive halos. However, many of these recently accreted dwarf galaxies may have different star-formation and quenching histories compared to the satellites that have been accreted into the stellar halos. Comparing the current dwarf population with the `fossil' one could provide new insights into the details of the two-phase formation scenario. We acknowledge the potentially important role of past major (dry) mergers on the stellar mass distribution and stellar population properties of the stellar halo, and that this could complicate the comparison, which may require assistance from simulations or models for physical interpretation.
        
    \end{itemize}

    As we have begun to work on these directions (Xiao-Ya Zhang, in preparation), we want to emphasize the challenging nature of the required observations for the above goals: while deep imaging data can regularly reveal the extended stellar halos around low-redshift massive ETGs toward $R>100$ kpc, spectroscopic observation at the matched surface brightness limit is still extremely rare. We argue that this could be a crucial motivation for developing future IFU instruments or a strategy for low-surface-brightness (LSB) spectroscopic observations.

% -------------------------------------------------------------------------------------------- %
% Summary and Conclusions
% -------------------------------------------------------------------------------------------- %

\section{Summary and Conclusions} 
    \label{sec:summary}
    
    In this work, we explore the potential connections between the stellar mass distribution of low-redshift massive ETGs, the mass assembly history, and the star formation timescales using the MaNGA and LegacySurvey dataset. We focus on the stellar population properties of these galaxies, especially the \alphafe{} abundance ratio. Through measuring the absorption line indices and full-spectrum fitting, we quantitatively compared the average stellar population in three radial bins (from the center to $1.5\ R_{\rm e}$) using the stacked spectra of massive galaxies split based on 1. their central velocity dispersion values (\sigmacen) at fixed stellar mass beyond 20 kpc radius; 2. their stellar mass distributions characterized by the outskirt stellar mass at $>20$ kpc at fixed stellar mass within 10 kpc radius. \sigmacen{} is widely considered as one of the most fundamental properties to define the scaling relations with stellar population properties. And work suggests the stellar mass distributions in the extended stellar halo of massive galaxies could reflect their dark matter halo mass and assembly history. Both the empirical absorption indices and the full-spectrum fitting approaches reach the same key conclusions:

    \begin{itemize}
    
        \item Massive ETGs with the identical stellar mass distributions outside of 20 kpc (see Figure \ref{fig:split_sigma}) show a significant difference in their \alphafe{} abundance ratio when split based on \sigmacen{}: higher \sigmacen{} tend to be more \alph{}-rich and older than those with lower \sigmacen within $\sim 1.5\times R_{\rm e}$. As we expect the inner region probed by MaNGA contains the majority of the \insitu{} stars, this suggests that the \sigmacen{} or the central gravitational potential plays a crucial role in determining the \insitu{} star-formation history, where a deeper potential (higher \sigmacen) leads to an earlier, more efficient, and shorter star formation. Assuming that the outer stellar halo mass can act like a dark matter halo mass proxy and reflects the accumulation of \exsitu{} stars, this may suggest that massive galaxies with a similar accretion history can still show a significant difference in the early history of their \insitu{} star formation.
    
        \item At the similar \sigmacen{} and total stellar mass, massive ETGs with more extended stellar halo mass profiles (higher stellar mass at $R>20$ kpc) show systematically lower \feh{}, higher \mgfe{} abundance ratio, and older stellar ages than their more compact counterparts within the same physical radius ($<10$ kpc). Under the framework of the ``two-phase'' formation of massive galaxies, although it is still premature to infer the stellar population properties and mass distributions of the \insitu{} and \exsitu{} components, our results reveal an interesting level of ``coordination'' between the early \insitu{} star formation and the second-phase \exsitu{} assembly of the stellar halo. At fixed stellar mass and \sigmacen{}, the galaxies that experienced an earlier and more intense starburst, and/or the ones that were quenched more abruptly, also tend to be the ones that can accumulate a more extended stellar halo through subsequent mergers and accretions. 
     
    \end{itemize}

        We have carefully examined and ruled out any potential bias from \mstar{} difference and unrealistic IMF assumptions. Our primary conclusions demonstrate that the star-formation and assembly histories of massive galaxies are too complex and intricate to be summarized by a few scaling relations anchored solely in stellar mass or velocity dispersion. 
        
        We have shown in this work that, even with the modest spatial coverage of the MaNGA data, there is clear evidence that the different assembly histories leave their traces in the stellar population. As \citet{HuangMNRAS2022} shows in their study, the extended stellar halo of massive galaxies beyond the coverage of MaNGA (e.g., from 50 to 100 kpc) could be excellent proxy of the galaxy's dark matter halo mass and indicator of its assembly history, as the stellar halo mainly consists of stars accreted from recent mergers and is less affected by the in-situ star formation that shapes massive galaxies' inner region. Therefore, our work demonstrates the scientific potential in extending the stellar population analysis \& comparison to the further outskirt regions of massive galaxies near and far, which is still quite challenging in observation.
    
        In the near future, we look forward to taking advantage of the next-generation IFU surveys, such as the Hector project \citep{Bryant2016}, to expand our analysis to a larger sample (a parent sample of 15,000 galaxies) with slightly better spatial coverage ($2\times R_{\rm e}$). At the same time, more powerful IFU on large telescopes, such as the Multi-Unit Spectrographic Explorer \citep[MUSE, ][]{Bacon2010}, has started to accumulate high-quality, high spatial resolution observations of nearby or low-$z$ massive galaxies that cover their stellar halos beyond $2\times R_{\rm e}$. In the following work, we will attempt to jointly constrain the functional form of the IMF along with other stellar population properties. And, with the new generation of powerful instruments such as the James Webb Space Telescope \citep[JWST, ][]{GardnerPASP2023}, our chances of finding the progenitors of the \alph{}-enriched galaxies before their massive neighbor accretes them have never been greater. We hope these promising new projects will further advance our understanding of assembly history and, potentially, the galaxy-halo connection in massive galaxies.

        The scripts to execute the analysis in this paper and the Jupyter Notebooks to generate the figures in this paper are hosted at the \href{https://github.com/xyzhangwork/mdensity_v_stellarpop}{GitHub repository} and also is preserved on Zenodo at \href{https://doi.org/10.5281/zenodo.17979404}{10.5281/zenodo.17979404}.
    
% -------------------------------------------------------------------------------------------- %

% -------------------------------------------------------------------------------------------- %
% Acknowledgements
% -------------------------------------------------------------------------------------------- %

\section{Acknowledgment}

    We thank Charlie Conroy for sharing stellar population modeling files for using {\tt ALF}.
    
    % TAHPC
    The authors acknowledge the Tsinghua Astrophysics High-Performance Computing platform at Tsinghua University for providing computational and data storage resources that have contributed to the research results reported within this paper. 

    % SH's funding
    SH acknowledges support from the Ministry of Science and Technology of China, Grant No.~2023YFA1605601, the National Natural Science Foundation of China, Grant No.~12273015, No.~12433003, and the China Crewed Space Program through its Space Application System. 

    % SDSS & MaNGA part
    The Alfred P. Sloan Foundation, the U.S. Department of Energy Office of Science, and the Participating Institutions have provided funding for the Sloan Digital Sky Survey IV. SDSS acknowledges support and resources from the Center for High-Performance Computing at the University of Utah. The SDSS website is \url{www.sdss.org}.

    SDSS is managed by the Astrophysical Research Consortium for the Participating Institutions of the SDSS Collaboration, including Caltech, The Carnegie Institution for Science, Chilean National Time Allocation Committee (CNTAC) ratified researchers, The Flatiron Institute, the Gotham Participation Group, Harvard University, Heidelberg University, The Johns Hopkins University, L’Ecole polytechnique f\'{e}d\'{e}rale de Lausanne (EPFL), Leibniz-Institut f\"{u}r Astrophysik Potsdam (AIP), Max-Planck-Institut f\"{u}r Astronomie (MPIA Heidelberg), Max-Planck-Institut f\"{u}r Extraterrestrische Physik (MPE), Nanjing University, National Astronomical Observatories of China (NAOC), New Mexico State University, The Ohio State University, Pennsylvania State University, Smithsonian Astrophysical Observatory, Space Telescope Science Institute (STScI), the Stellar Astrophysics Participation Group, Universidad Nacional Aut\'{o}noma de M\'{e}xico, University of Arizona, University of Colorado Boulder, University of Illinois at Urbana-Champaign, University of Toronto, University of Utah, University of Virginia, Yale University, and Yunnan University.
  
    % LegacySurvey part
    The Legacy Surveys consist of three individual and complementary projects: the Dark Energy Camera Legacy Survey (DECaLS; Proposal ID \#2014B-0404; PIs: David Schlegel and Arjun Dey), the Beijing-Arizona Sky Survey (BASS; NOAO Prop. ID \#2015A-0801; PIs: Zhou Xu and Xiaohui Fan), and the Mayall z-band Legacy Survey (MzLS; Prop. ID \#2016A-0453; PI: Arjun Dey). DECaLS, BASS, and MzLS together include data obtained, respectively, at the Blanco telescope, Cerro Tololo Inter-American Observatory, NSF’s NOIRLab; the Bok telescope, Steward Observatory, University of Arizona; and the Mayall telescope, Kitt Peak National Observatory, NOIRLab. Pipeline processing and data analysis were supported by NOIRLab and the Lawrence Berkeley National Laboratory (LBNL). The Legacy Surveys project is honored to be permitted to conduct astronomical research on Iolkam Du’ag (Kitt Peak), a mountain with particular significance to the Tohono O’odham Nation. Please see the complete acknowledgments for the LegacySurvey \href{https://www.legacysurvey.org/acknowledgment/}{here}.

    % Siena Atlas
    The Siena Galaxy Atlas was made possible by funding support from the U.S. Department of Energy, Office of Science, Office of High Energy Physics under Award Number DE-SC0020086, and the National Science Foundation under grant AST-1616414.
    
    This research made use of the ``K-corrections calculator'' service available at http://kcor.sai.msu.ru/

% Keep track of all the software used in this work
\software{
    \href{https://www.astropy.org/}{\texttt{Astropy}} \citep{astropy:2013, astropy:2018, astropy:2022},
    \href{https://matplotlib.org}{\texttt{Matplotlib}} \citep{HunterCOMPUTINGINSCIENCEANDENGINEERING2007},
    \href{http://www.numpy.org}{\texttt{NumPy}} \citep{HarrisNAT2020},
    \href{https://www.scipy.org}{\texttt{SciPy}} \citep{VirtanenNATUREMETHODS2020} 
}

% -------------------------------------------------------------------------------------------- %
% Appendix
% -------------------------------------------------------------------------------------------- %
\appendix
\section{Mass and Size Comparison}
    \label{appendix:comparison}
    
\begin{figure*}[!th]
\centering
    \includegraphics[width=1\linewidth]{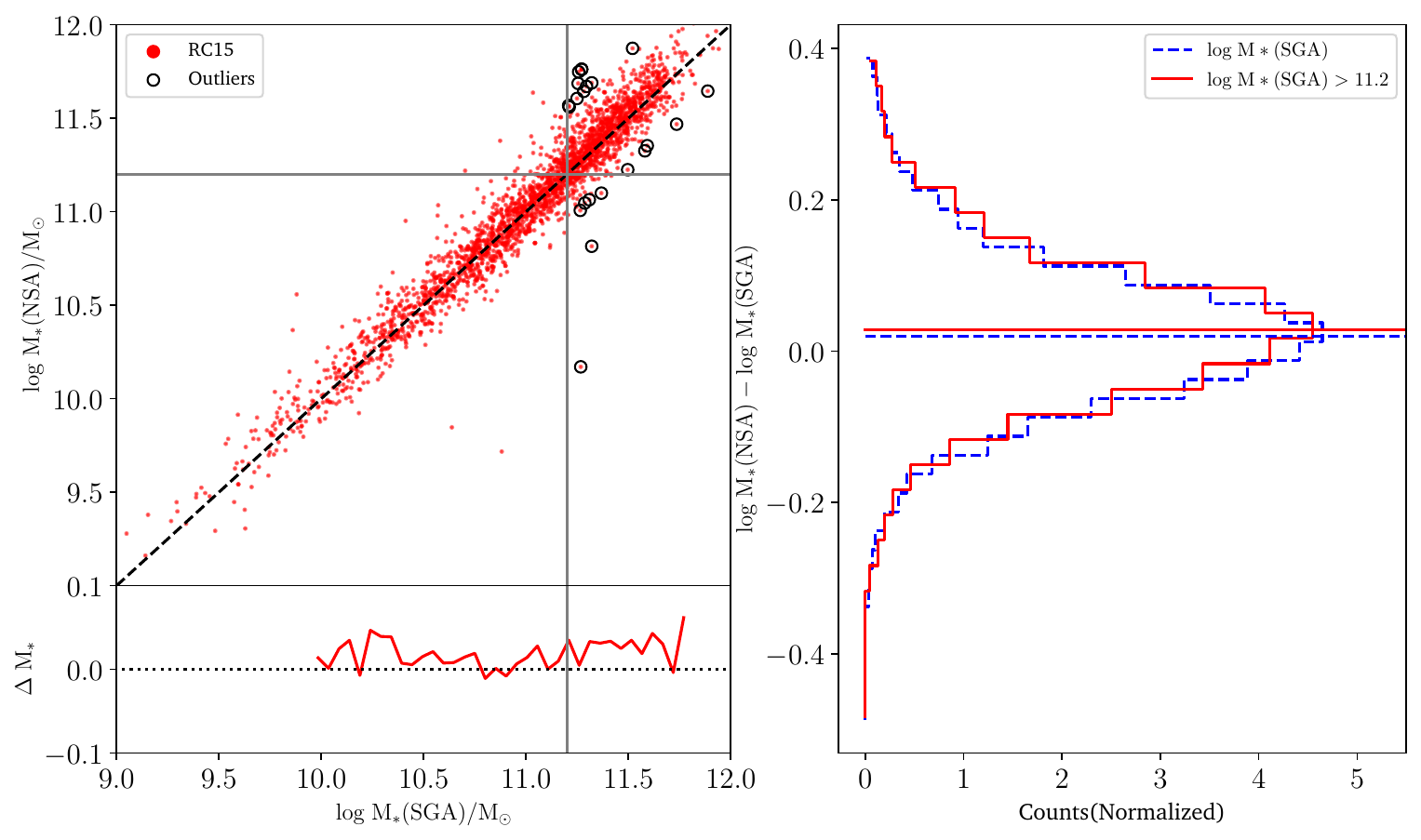}
    \caption{
        Stellar Mass comparison between the K-correction fit for \ser{} fluxes ($\Omega_m=0.3$, $\Omega_\Lambda=0.7$, $h=0.7$) from the NSA catalog (y-axis), and the Stellar Mass estimated from the SGA profiles and MLCR from \citet{RoedigerMNRAS2015} (x-axis) with the BC03 \citep{BruzualMNRAS2003}. SSP. Left sub-splot shows the $\rm M_*(SGA)$ (x-axis) v. $\rm M_*(NSA)$ (y-axis) scatter plot (top) with the $M_*=10^{11.2}M_\odot$ lines in gray and 1:1 relation in black dashed line. The left bottom plot shows the average difference between the two stellar masses ($\rm \Delta M_*=M_*(NSA)-M_*(SGA)$. The right sub-plot shows the histogram and the average of the stellar mass differences, with the total sample in blue dashed lines and the massive sub-sample ($\rm M_*(SGA)>10^{11.2}M_\odot$ in red solid lines. 
        }
    \label{fig:mstar}
\end{figure*}

    In this section, we present a comparison of stellar mass and effective radius. Figure~\ref{fig:mstar} shows the stellar mass estimated from the SGA profiles ($\rm M_*(SGA)$, x-axis) and the stellar mass from the K-correction fit for the \ser{} fluxes from the NSA catalog. We have marked $M_*=10^{11.2}M_\odot$ for both axes (solid gray lines) as well as the 1:1 relation (black dashed line). we can see from the comparison that $\rm M_*(SGA)$ is in general $\lesssim$ 0.05 dex smaller than the NSA stellar mass. We also present the distribution of the stellar mass difference on the right sub-plot, with the total sample in blue dashed lines and the massive sub-sample ($\rm M_*(SGA)>10^{11.2}M_\odot$) in red solid lines. The average difference is 0.020 dex for the total sample, and 0.028 dex for the massive sample. 

\begin{figure*}[!th]
\centering
    \gridline{\fig{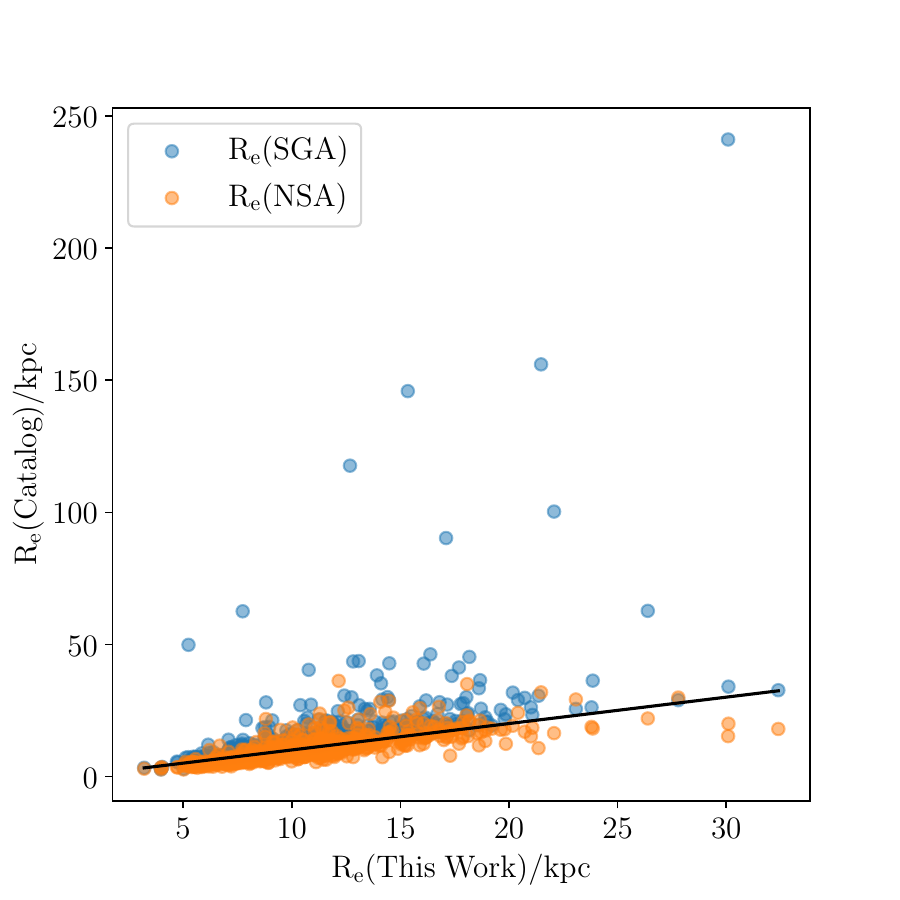}{0.5\textwidth}{(a)}
    \fig{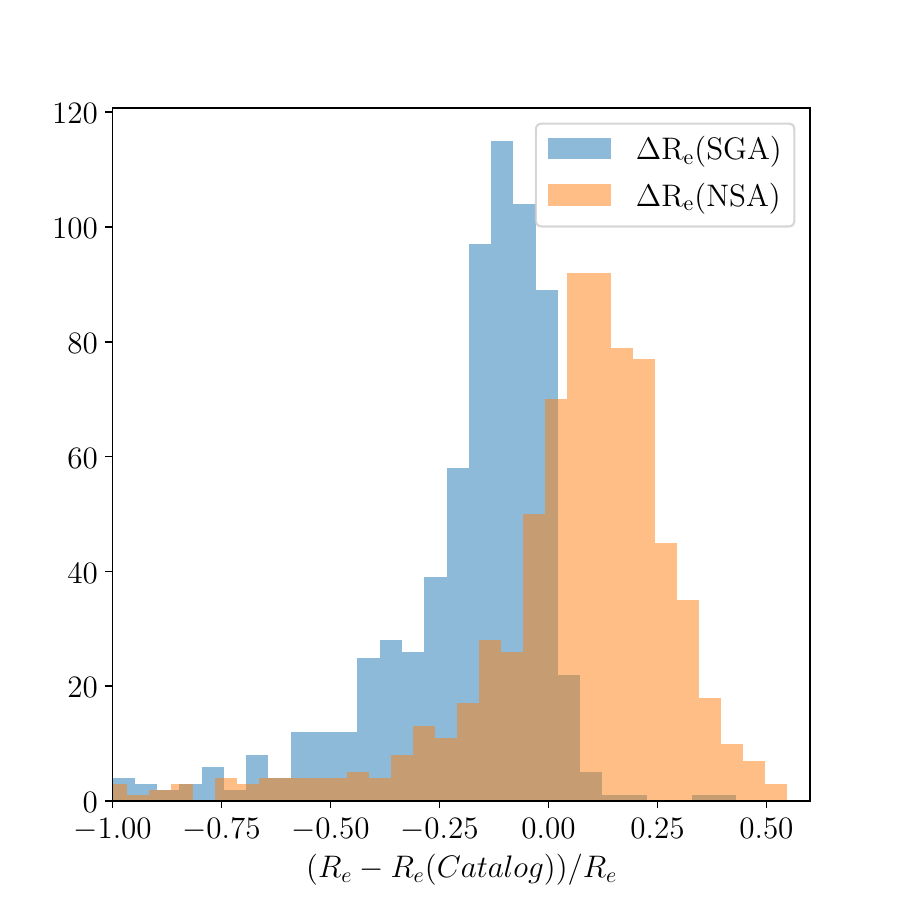}{0.5\textwidth}{(b)}}
        \caption{ Left: Effective radius comparison between the \re{} estimated for this work (x axis) and the \re{} from other catalogs (y axis), including the SGA fit (blue) and NSA catalog, which is estimated from \ser{} fit on r-band flux (orange). The 1:1 relation is shown in the black solid line. 
        \\ Right: the distribution of the relative difference of \re{} compared with $\rm R_e (SGA)$ (blue) and $\rm R_e (NSA)$ (orange), defined as $\rm \Delta R_e (Catalog)=(R_e(This Work)-R_e(Catalog))/R_e(This Work).$} 
        \label{fig:compare_re}
\end{figure*}

    Figure~\ref{fig:compare_re} compares our estimated effective radius (\re{}) with the values from SGA (blue) and NSA (orange). The left figure shows the \re{} values from different catalogs, and the black solid line is the 1:1 relation. We can see that although our \re{} mostly agrees well with both $\rm R_e(SGA)$ and $\rm R_e(NSA)$, there are a few where the SGA value is much larger than our estimation. This is likely caused by our decision to cut the profile at 28 $mag/arcsec^2$. We can also see the larger galaxies tend to have $\rm R_e(This\ Work)>R_e(SGA)$, which is due to the steep light profiles of massive ETGs, which means the shallower SDSS photometry would underestimate \re{}. The right figure shows the difference in \re{}, where $\rm (R_e(This\ Work)- R_e(SGA))/R_e(This\ Work)$ is in blue and $\rm (R_e(This\ Work)- R_e(NSA))/R_e(This\ Work)$ is in orange. From this plot, we can see that the relative differences between methods are mostly less than 25\%. 

\section{Total Stellar Mass Bins}
\label{appendix: totm}

\begin{figure*}[!th]
    \centering
    \includegraphics[width=0.7\linewidth]{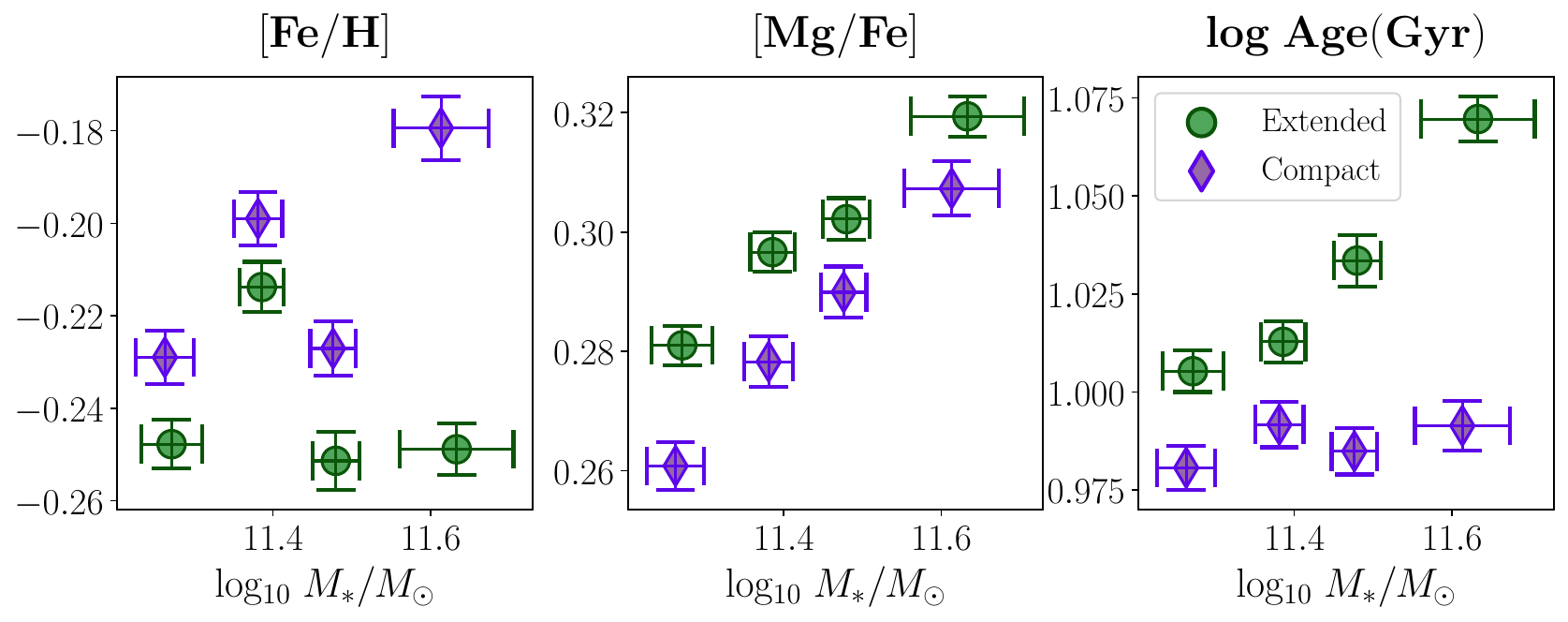}
    \caption{
        The stellar population properties (from left to right: \feh{},\mgfe{}, and \logage{}) of the Extended and Compact sub-samples are split into four narrow \mstar{} bins. The error bars on the x-axis reflect the spread of \mstar{} in each bin, and the y-axis error bar is the width of the posterior from the \alf{} fitting.
        }
    \label{fig:mbins}
\end{figure*}

    In this section, we further divide the extended and compact galaxies into narrower stellar-mass bins and compare the stacked spectra within each bin. We choose 4 stellar mass bins: $11.20<log\ M_*<11.33$, $11.33<log\ M_*<11.43$, $43<log\ M_*<53$, $11.53<log\ M_*<13$, and each bin has $\sim$ 60 galaxies. Because the two sub-samples have different physical sizes, we choose the middle radial bin of the extended sub-sample and the outer radial region of the compact sub-sample; these two radial bins are both around 10 kpc (see Section~\ref{sec:results}), which is the largest radius at which we can conduct a direct comparison between the two sub-samples. We then perform \alf{} fitting on all the spectra, and Figure~\ref{fig:mbins} presents the stellar population properties at around 10 kpc with the average stellar mass in each bin. The error bars on the x-axis stand for the spread of total stellar mass in the respective bin, and the y-axis error bars stand for the width of the posterior distribution. We can see the same trends observed in the entire sub-samples shown in Section~\ref{sec:fsf}: Extended galaxies have lower \feh{}, higher \mgfe{}, and older stellar age in all four stellar mass bins.
    
\section{Other elemental abundances}
    \label{appendix:abundance}
    
\begin{figure}[!h]
    \centering
    \includegraphics[width=1.0\linewidth]{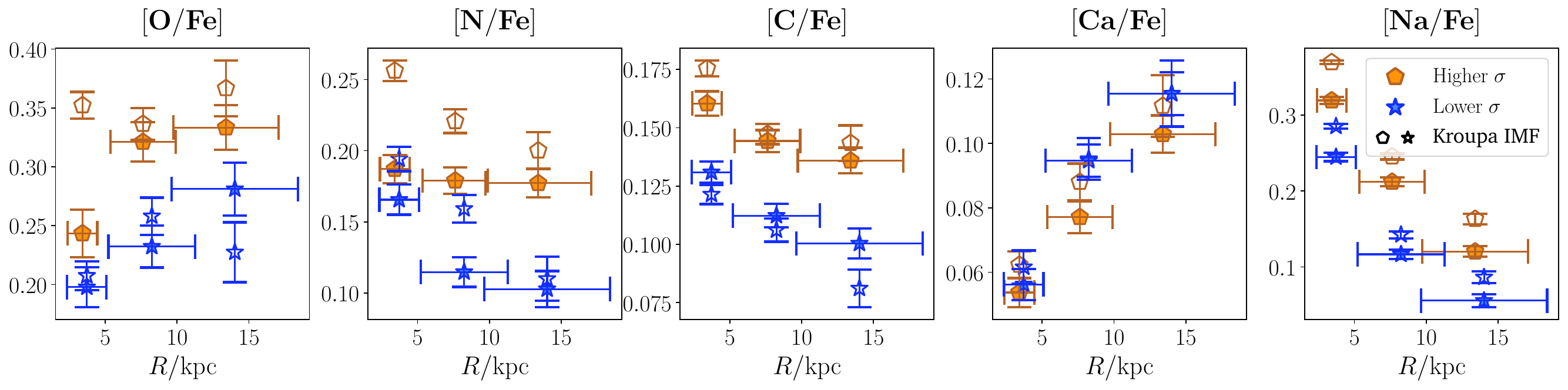}
    \caption{
        Additional elemental abundances from \alf{} \smode{} (empty markers) and \fmode{} (filled markers) fitting of the \hsig{} and \lsig{} sub-samples. We present here the abundances of oxygen, nitrogen, carbon, calcium, and sodium from both \smode{} (empty markers) and \fmode{} (filled markers) fits. The \texttt{Jupyter} notebook for reproducing this figure can be found here: \href{https://github.com/xyzhangwork/mdensity_v_stellarpop/blob/main/plot_scripts/plot_alf_gradient.ipynb}{\faGithub}. This repository is also available on \href{https://doi.org/10.5281/zenodo.17979404}{Zenodo}.
        }
    \label{fig:sigma_ele}
\end{figure}

\begin{figure}[!h]
    \centering
    \includegraphics[width=1.0\linewidth]{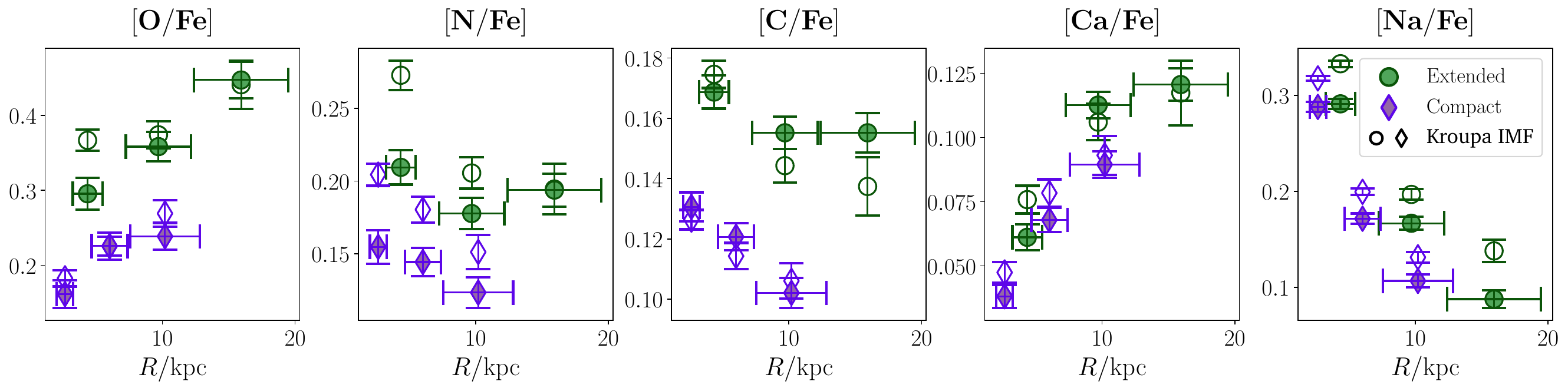}
    \caption{
        Additional elemental abundances from \alf{} \smode{} (empty markers) and \fmode{} (filled markers) fitting of the \hsig{} and \lsig{} sub-samples. We present here the abundances of oxygen, nitrogen, carbon, calcium, and sodium from both \smode{} (empty markers) and \fmode{} (filled markers) fits. The \texttt{Jupyter} notebook for reproducing this figure can be found here: \href{https://github.com/xyzhangwork/mdensity_v_stellarpop/blob/main/plot_scripts/plot_alf_gradient.ipynb}{\faGithub}. This repository is also available on \href{https://doi.org/10.5281/zenodo.17979404}{Zenodo}.
        }
    \label{fig:mass2d_ele}
\end{figure}

    In this section, we present some of the other elemental abundances in the \alf{} fitting with both \smode{} (empty markers) and \fmode{} (filled markers). Figure \ref{fig:sigma_ele} and figure \ref{fig:mass2d_ele} show the elemental abundances of O (\alph), N, C, Ca, and Na. Although oxygen is the most abundant \alph{} element in the universe, and there have been studies showing that magnesium and oxygen do not evolve in lock-step and should not be used interchangeably, we use magnesium as our main \alphafe{} indicator because it is easier to constrain given our spectral range and the lack of atomic oxygen lines in the optical range. Here we show that the estimated oxygen abundance qualitatively agrees with our main conclusion: the \hsig{} sub-sample shows higher \alphafe{} than the compact galaxies at 10 kpc, and extended galaxies have higher \alphafe{} than the compact galaxies at 10 kpc. We also see an elevated C and N abundance in both \hsig{} and extended sub-samples, respectively. This could indicate an excess of intermediate-mass stars.
    Neither sample-split method shows any difference in Ca abundance between sub-samples.
    We also note that the IMF choice does not affect the comparison between sub-samples in any of the elements presented here. 

% End of the Draft
% -------------------------------------------------------------------------------------------- %

\bibliography{manuscript}

\begin{thebibliography}{}
\expandafter\ifx\csname natexlab\endcsname\relax\def\natexlab#1{#1}\fi
\providecommand{\url}[1]{\href{#1}{#1}}
\providecommand{\dodoi}[1]{doi:~\href{http://doi.org/#1}{\nolinkurl{#1}}}
\providecommand{\doeprint}[1]{\href{http://ascl.net/#1}{\nolinkurl{http://ascl.net/#1}}}
\providecommand{\doarXiv}[1]{\href{https://arxiv.org/abs/#1}{\nolinkurl{https://arxiv.org/abs/#1}}}

\bibitem[{{Aguado} {et~al.}(2019){Aguado}, {Ahumada}, {Almeida}, {Anderson},
  {Andrews}, {Anguiano}, {Aquino Ort{\'\i}z}, {Arag{\'o}n-Salamanca},
  {Argudo-Fern{\'a}ndez}, {Aubert}, {Avila-Reese}, {Badenes}, {Barboza
  Rembold}, {Barger}, {Barrera-Ballesteros}, {Bates}, {Bautista}, {Beaton},
  {Beers}, {Belfiore}, {Bernardi}, {Bershady}, {Beutler}, {Bird}, {Bizyaev},
  {Blanc}, {Blanton}, {Blomqvist}, {Bolton}, {Boquien}, {Borissova}, {Bovy},
  {Brandt}, {Brinkmann}, {Brownstein}, {Bundy}, {Burgasser}, {Byler}, {Cano
  Diaz}, {Cappellari}, {Carrera}, {Cervantes Sodi}, {Chen}, {Cherinka}, {Choi},
  {Chung}, {Coffey}, {Comerford}, {Comparat}, {Covey}, {da Silva Ilha}, {da
  Costa}, {Dai}, {Damke}, {Darling}, {Davies}, {Dawson}, {de Sainte Agathe},
  {Deconto Machado}, {Del Moro}, {De Lee}, {Diamond-Stanic}, {Dom{\'\i}nguez
  S{\'a}nchez}, {Donor}, {Drory}, {du Mas des Bourboux}, {Duckworth}, {Dwelly},
  {Ebelke}, {Emsellem}, {Escoffier}, {Fern{\'a}ndez-Trincado}, {Feuillet},
  {Fischer}, {Fleming}, {Fraser-McKelvie}, {Freischlad}, {Frinchaboy}, {Fu},
  {Galbany}, {Garcia-Dias}, {Garc{\'\i}a-Hern{\'a}ndez}, {Garma Oehmichen},
  {Geimba Maia}, {Gil-Mar{\'\i}n}, {Grabowski}, {Gu}, {Guo}, {Ha},
  {Harrington}, {Hasselquist}, {Hayes}, {Hearty}, {Hernandez Toledo}, {Hicks},
  {Hogg}, {Holley-Bockelmann}, {Holtzman}, {Hsieh}, {Hunt}, {Hwang},
  {Ibarra-Medel}, {Jimenez Angel}, {Johnson}, {Jones}, {J{\"o}nsson},
  {Kinemuchi}, {Kollmeier}, {Krawczyk}, {Kreckel}, {Kruk}, {Lacerna}, {Lan},
  {Lane}, {Law}, {Lee}, {Li}, {Lian}, {Lin}, {Lin}, {Lintott}, {Long},
  {Longa-Pe{\~n}a}, {Mackereth}, {de la Macorra}, {Majewski}, {Malanushenko},
  {Manchado}, {Maraston}, {Mariappan}, {Marinelli}, {Marques-Chaves},
  {Masseron}, {Masters}, {McDermid}, {Medina Pe{\~n}a}, {Meneses-Goytia},
  {Merloni}, {Merrifield}, {Meszaros}, {Minniti}, {Minsley}, {Muna}, {Myers},
  {Nair}, {Correa do Nascimento}, {Newman}, {Nitschelm}, {Olmstead}, {Oravetz},
  {Oravetz}, {Ortega Minakata}, {Pace}, {Padilla}, {Palicio}, {Pan}, {Pan},
  {Parikh}, {Parker}, {Peirani}, {Penny}, {Percival}, {Perez-Fournon},
  {Peterken}, {Pinsonneault}, {Prakash}, {Raddick}, {Raichoor}, {Riffel},
  {Riffel}, {Rix}, {Robin}, {Roman-Lopes}, {Rose}, {Ross}, {Rossi}, {Rowlands},
  {Rubin}, {S{\'a}nchez}, {S{\'a}nchez-Gallego}, {Sayres}, {Schaefer},
  {Schiavon}, {Schimoia}, {Schlafly}, {Schlegel}, {Schneider}, {Schultheis},
  {Seo}, {Shamsi}, {Shao}, {Shen}, {Shetty}, {Simonian}, {Smethurst}, {Sobeck},
  {Souter}, {Spindler}, {Stark}, {Stassun}, {Steinmetz}, {Storchi-Bergmann},
  {Stringfellow}, {Su{\'a}rez}, {Sun}, {Taghizadeh-Popp}, {Talbot}, {Tayar},
  {Thakar}, {Thomas}, {Tissera}, {Tojeiro}, {Troup}, {Unda-Sanzana},
  {Valenzuela}, {Vargas-Maga{\~n}a}, {V{\'a}zquez-Mata}, {Wake}, {Weaver},
  {Weijmans}, {Westfall}, {Wild}, {Wilson}, {Woods}, {Yan}, {Yang}, {Zamora},
  {Zasowski}, {Zhang}, {Zheng}, {Zheng}, {Zhu}, {Zinn}, \&
  {Zou}}]{AguadoAPJS2019}
{Aguado}, D.~S., {Ahumada}, R., {Almeida}, A., {et~al.} 2019, \apjs, 240, 23,
  \dodoi{10.3847/1538-4365/aaf651}

\bibitem[{{Aihara} {et~al.}(2018){Aihara}, {Armstrong}, {Bickerton}, {Bosch},
  {Coupon}, {Furusawa}, {Hayashi}, {Ikeda}, {Kamata}, {Karoji}, {Kawanomoto},
  {Koike}, {Komiyama}, {Lang}, {Lupton}, {Mineo}, {Miyatake}, {Miyazaki},
  {Morokuma}, {Obuchi}, {Oishi}, {Okura}, {Price}, {Takata}, {Tanaka},
  {Tanaka}, {Tanaka}, {Uchida}, {Uraguchi}, {Utsumi}, {Wang}, {Yamada},
  {Yamanoi}, {Yasuda}, {Arimoto}, {Chiba}, {Finet}, {Fujimori}, {Fujimoto},
  {Furusawa}, {Goto}, {Goulding}, {Gunn}, {Harikane}, {Hattori}, {Hayashi},
  {He{\l}miniak}, {Higuchi}, {Hikage}, {Ho}, {Hsieh}, {Huang}, {Huang},
  {Imanishi}, {Iwata}, {Jaelani}, {Jian}, {Kashikawa}, {Katayama}, {Kojima},
  {Konno}, {Koshida}, {Kusakabe}, {Leauthaud}, {Lee}, {Lin}, {Lin},
  {Mandelbaum}, {Matsuoka}, {Medezinski}, {Miyama}, {Momose}, {More}, {More},
  {Mukae}, {Murata}, {Murayama}, {Nagao}, {Nakata}, {Niida}, {Niikura},
  {Nishizawa}, {Oguri}, {Okabe}, {Ono}, {Onodera}, {Onoue}, {Ouchi}, {Pyo},
  {Shibuya}, {Shimasaku}, {Simet}, {Speagle}, {Spergel}, {Strauss}, {Sugahara},
  {Sugiyama}, {Suto}, {Suzuki}, {Tait}, {Takada}, {Terai}, {Toba}, {Turner},
  {Uchiyama}, {Umetsu}, {Urata}, {Usuda}, {Yeh}, \& {Yuma}}]{AiharaPASJ2018}
{Aihara}, H., {Armstrong}, R., {Bickerton}, S., {et~al.} 2018, \pasj, 70, S8,
  \dodoi{10.1093/pasj/psx081}

\bibitem[{{Alton} {et~al.}(2017){Alton}, {Smith}, \& {Lucey}}]{AltonMNRAS2017}
{Alton}, P.~D., {Smith}, R.~J., \& {Lucey}, J.~R. 2017, \mnras, 468, 1594,
  \dodoi{10.1093/mnras/stx464}

\bibitem[{{Annibali} {et~al.}(2011){Annibali}, {Gr{\"u}tzbauch}, {Rampazzo},
  {Bressan}, \& {Zeilinger}}]{Annibali2011AA}
{Annibali}, F., {Gr{\"u}tzbauch}, R., {Rampazzo}, R., {Bressan}, A., \&
  {Zeilinger}, W.~W. 2011, \aap, 528, A19, \dodoi{10.1051/0004-6361/201015635}

\bibitem[{{Astropy Collaboration} {et~al.}(2013){Astropy Collaboration},
  {Robitaille}, {Tollerud}, {Greenfield}, {Droettboom}, {Bray}, {Aldcroft},
  {Davis}, {Ginsburg}, {Price-Whelan}, {Kerzendorf}, {Conley}, {Crighton},
  {Barbary}, {Muna}, {Ferguson}, {Grollier}, {Parikh}, {Nair}, {Unther},
  {Deil}, {Woillez}, {Conseil}, {Kramer}, {Turner}, {Singer}, {Fox}, {Weaver},
  {Zabalza}, {Edwards}, {Azalee Bostroem}, {Burke}, {Casey}, {Crawford},
  {Dencheva}, {Ely}, {Jenness}, {Labrie}, {Lim}, {Pierfederici}, {Pontzen},
  {Ptak}, {Refsdal}, {Servillat}, \& {Streicher}}]{astropy:2013}
{Astropy Collaboration}, {Robitaille}, T.~P., {Tollerud}, E.~J., {et~al.} 2013,
  \aap, 558, A33, \dodoi{10.1051/0004-6361/201322068}

\bibitem[{{Astropy Collaboration} {et~al.}(2018){Astropy Collaboration},
  {Price-Whelan}, {Sip{\H{o}}cz}, {G{\"u}nther}, {Lim}, {Crawford}, {Conseil},
  {Shupe}, {Craig}, {Dencheva}, {Ginsburg}, {Vand erPlas}, {Bradley},
  {P{\'e}rez-Su{\'a}rez}, {de Val-Borro}, {Aldcroft}, {Cruz}, {Robitaille},
  {Tollerud}, {Ardelean}, {Babej}, {Bach}, {Bachetti}, {Bakanov}, {Bamford},
  {Barentsen}, {Barmby}, {Baumbach}, {Berry}, {Biscani}, {Boquien}, {Bostroem},
  {Bouma}, {Brammer}, {Bray}, {Breytenbach}, {Buddelmeijer}, {Burke},
  {Calderone}, {Cano Rodr{\'\i}guez}, {Cara}, {Cardoso}, {Cheedella}, {Copin},
  {Corrales}, {Crichton}, {D'Avella}, {Deil}, {Depagne}, {Dietrich}, {Donath},
  {Droettboom}, {Earl}, {Erben}, {Fabbro}, {Ferreira}, {Finethy}, {Fox},
  {Garrison}, {Gibbons}, {Goldstein}, {Gommers}, {Greco}, {Greenfield},
  {Groener}, {Grollier}, {Hagen}, {Hirst}, {Homeier}, {Horton}, {Hosseinzadeh},
  {Hu}, {Hunkeler}, {Ivezi{\'c}}, {Jain}, {Jenness}, {Kanarek}, {Kendrew},
  {Kern}, {Kerzendorf}, {Khvalko}, {King}, {Kirkby}, {Kulkarni}, {Kumar},
  {Lee}, {Lenz}, {Littlefair}, {Ma}, {Macleod}, {Mastropietro}, {McCully},
  {Montagnac}, {Morris}, {Mueller}, {Mumford}, {Muna}, {Murphy}, {Nelson},
  {Nguyen}, {Ninan}, {N{\"o}the}, {Ogaz}, {Oh}, {Parejko}, {Parley}, {Pascual},
  {Patil}, {Patil}, {Plunkett}, {Prochaska}, {Rastogi}, {Reddy Janga},
  {Sabater}, {Sakurikar}, {Seifert}, {Sherbert}, {Sherwood-Taylor}, {Shih},
  {Sick}, {Silbiger}, {Singanamalla}, {Singer}, {Sladen}, {Sooley},
  {Sornarajah}, {Streicher}, {Teuben}, {Thomas}, {Tremblay}, {Turner},
  {Terr{\'o}n}, {van Kerkwijk}, {de la Vega}, {Watkins}, {Weaver}, {Whitmore},
  {Woillez}, {Zabalza}, \& {Astropy Contributors}}]{astropy:2018}
{Astropy Collaboration}, {Price-Whelan}, A.~M., {Sip{\H{o}}cz}, B.~M., {et~al.}
  2018, \aj, 156, 123, \dodoi{10.3847/1538-3881/aabc4f}

\bibitem[{{Astropy Collaboration} {et~al.}(2022){Astropy Collaboration},
  {Price-Whelan}, {Lim}, {Earl}, {Starkman}, {Bradley}, {Shupe}, {Patil},
  {Corrales}, {Brasseur}, {N{"o}the}, {Donath}, {Tollerud}, {Morris},
  {Ginsburg}, {Vaher}, {Weaver}, {Tocknell}, {Jamieson}, {van Kerkwijk},
  {Robitaille}, {Merry}, {Bachetti}, {G{"u}nther}, {Aldcroft},
  {Alvarado-Montes}, {Archibald}, {B{'o}di}, {Bapat}, {Barentsen}, {Baz{'a}n},
  {Biswas}, {Boquien}, {Burke}, {Cara}, {Cara}, {Conroy}, {Conseil}, {Craig},
  {Cross}, {Cruz}, {D'Eugenio}, {Dencheva}, {Devillepoix}, {Dietrich},
  {Eigenbrot}, {Erben}, {Ferreira}, {Foreman-Mackey}, {Fox}, {Freij}, {Garg},
  {Geda}, {Glattly}, {Gondhalekar}, {Gordon}, {Grant}, {Greenfield}, {Groener},
  {Guest}, {Gurovich}, {Handberg}, {Hart}, {Hatfield-Dodds}, {Homeier},
  {Hosseinzadeh}, {Jenness}, {Jones}, {Joseph}, {Kalmbach}, {Karamehmetoglu},
  {Ka{l}uszy{'n}ski}, {Kelley}, {Kern}, {Kerzendorf}, {Koch}, {Kulumani},
  {Lee}, {Ly}, {Ma}, {MacBride}, {Maljaars}, {Muna}, {Murphy}, {Norman},
  {O'Steen}, {Oman}, {Pacifici}, {Pascual}, {Pascual-Granado}, {Patil},
  {Perren}, {Pickering}, {Rastogi}, {Roulston}, {Ryan}, {Rykoff}, {Sabater},
  {Sakurikar}, {Salgado}, {Sanghi}, {Saunders}, {Savchenko}, {Schwardt},
  {Seifert-Eckert}, {Shih}, {Jain}, {Shukla}, {Sick}, {Simpson},
  {Singanamalla}, {Singer}, {Singhal}, {Sinha}, {Sip{H{o}}cz}, {Spitler},
  {Stansby}, {Streicher}, {{{S}}umak}, {Swinbank}, {Taranu}, {Tewary},
  {Tremblay}, {Val-Borro}, {Van Kooten}, {Vasovi{'c}}, {Verma}, {de Miranda
  Cardoso}, {Williams}, {Wilson}, {Winkel}, {Wood-Vasey}, {Xue}, {Yoachim},
  {Zhang}, {Zonca}, \& {Astropy Project Contributors}}]{astropy:2022}
{Astropy Collaboration}, {Price-Whelan}, A.~M., {Lim}, P.~L., {et~al.} 2022,
  \apj, 935, 167, \dodoi{10.3847/1538-4357/ac7c74}

\bibitem[{{Bacon} {et~al.}(2010){Bacon}, {Accardo}, {Adjali}, {Anwand},
  {Bauer}, {Biswas}, {Blaizot}, \& {Boudon}}]{Bacon2010}
{Bacon}, R., {Accardo}, M., {Adjali}, L., {et~al.} 2010, in Society of
  Photo-Optical Instrumentation Engineers (SPIE) Conference Series, Vol. 7735,
  Ground-based and Airborne Instrumentation for Astronomy III, ed. I.~S.
  {McLean}, S.~K. {Ramsay}, \& H.~{Takami}, 773508, \dodoi{10.1117/12.856027}

\bibitem[{{Barbary}(2016)}]{BarbaryTHEJOURNALOFOPENSOURCESOFTWARE2016}
{Barbary}, K. 2016, The Journal of Open Source Software, 1, 58,
  \dodoi{10.21105/joss.00058}

\bibitem[{{Barone} {et~al.}(2018){Barone}, {D'Eugenio}, {Colless}, {Scott},
  {van de Sande}, {Bland-Hawthorn}, {Brough}, {Bryant}, {Cortese}, {Croom},
  {Foster}, {Goodwin}, {Konstantopoulos}, {Lawrence}, {Lorente}, {Medling},
  {Owers}, \& {Richards}}]{BaroneAPJ2018}
{Barone}, T.~M., {D'Eugenio}, F., {Colless}, M., {et~al.} 2018, \apj, 856, 64,
  \dodoi{10.3847/1538-4357/aaaf6e}

\bibitem[{{Barone} {et~al.}(2022){Barone}, {D'Eugenio}, {Scott}, {Colless},
  {Vaughan}, {van der Wel}, {Fraser-McKelvie}, {de Graaff}, {van de Sande},
  {Wu}, {Bezanson}, {Brough}, {Bell}, {Croom}, {Cortese}, {Driver}, {Gallazzi},
  {Muzzin}, {Sobral}, {Bland-Hawthorn}, {Bryant}, {Goodwin}, {Lawrence},
  {Lorente}, \& {Owers}}]{BaroneMNRAS2022}
{Barone}, T.~M., {D'Eugenio}, F., {Scott}, N., {et~al.} 2022, \mnras, 512,
  3828, \dodoi{10.1093/mnras/stac705}

\bibitem[{{B{\'e}dorf} \& {Portegies Zwart}(2013)}]{BedorfMNRAS2013}
{B{\'e}dorf}, J., \& {Portegies Zwart}, S. 2013, \mnras, 431, 767,
  \dodoi{10.1093/mnras/stt208}

\bibitem[{{Belfiore} {et~al.}(2017){Belfiore}, {Maiolino}, {Tremonti},
  {S{\'a}nchez}, {Bundy}, {Bershady}, {Westfall}, {Lin}, {Drory}, {Boquien},
  {Thomas}, \& {Brinkmann}}]{BelfioreMNRAS2017}
{Belfiore}, F., {Maiolino}, R., {Tremonti}, C., {et~al.} 2017, \mnras, 469,
  151, \dodoi{10.1093/mnras/stx789}

\bibitem[{{Bernardi} {et~al.}(2017){Bernardi}, {Fischer}, {Sheth}, {Meert},
  {Huertas-Company}, {Shankar}, \& {Vikram}}]{BernardiMNRAS2017}
{Bernardi}, M., {Fischer}, J.~L., {Sheth}, R.~K., {et~al.} 2017, \mnras, 468,
  2569, \dodoi{10.1093/mnras/stx677}

\bibitem[{{Bernardi} {et~al.}(2013){Bernardi}, {Meert}, {Sheth}, {Vikram},
  {Huertas-Company}, {Mei}, \& {Shankar}}]{BernardiMNRAS2013}
{Bernardi}, M., {Meert}, A., {Sheth}, R.~K., {et~al.} 2013, \mnras, 436, 697,
  \dodoi{10.1093/mnras/stt1607}

\bibitem[{{Bernardi} {et~al.}(2003){Bernardi}, {Sheth}, {Annis}, {Burles},
  {Finkbeiner}, {Lupton}, {Schlegel}, {SubbaRao}, {Bahcall}, {Blakeslee},
  {Brinkmann}, {Castander}, {Connolly}, {Csabai}, {Doi}, {Fukugita}, {Frieman},
  {Heckman}, {Hennessy}, {Ivezi{\'c}}, {Knapp}, {Lamb}, {McKay}, {Munn},
  {Nichol}, {Okamura}, {Schneider}, {Thakar}, \& {York}}]{BernardiAJ2003}
{Bernardi}, M., {Sheth}, R.~K., {Annis}, J., {et~al.} 2003, \aj, 125, 1882,
  \dodoi{10.1086/367795}

\bibitem[{{Bertin} \& {Arnouts}(1996)}]{BertinAAPS1996}
{Bertin}, E., \& {Arnouts}, S. 1996, \aaps, 117, 393,
  \dodoi{10.1051/aas:1996164}

\bibitem[{{Bevacqua} {et~al.}(2022){Bevacqua}, {Cappellari}, \&
  {Pellegrini}}]{BevacquaMNRAS2022}
{Bevacqua}, D., {Cappellari}, M., \& {Pellegrini}, S. 2022, \mnras, 511, 139,
  \dodoi{10.1093/mnras/stab3732}

\bibitem[{{Beverage} {et~al.}(2023){Beverage}, {Kriek}, {Conroy}, {Sandford},
  {Bezanson}, {Franx}, {van der Wel}, \& {Weisz}}]{BeverageApJ2023}
{Beverage}, A.~G., {Kriek}, M., {Conroy}, C., {et~al.} 2023, \apj, 948, 140,
  \dodoi{10.3847/1538-4357/acc176}

\bibitem[{{Blanton} {et~al.}(2011){Blanton}, {Kazin}, {Muna}, {Weaver}, \&
  {Price-Whelan}}]{BlantonAJ2011}
{Blanton}, M.~R., {Kazin}, E., {Muna}, D., {Weaver}, B.~A., \& {Price-Whelan},
  A. 2011, \aj, 142, 31, \dodoi{10.1088/0004-6256/142/1/31}

\bibitem[{{Blanton} {et~al.}(2005){Blanton}, {Schlegel}, {Strauss},
  {Brinkmann}, {Finkbeiner}, {Fukugita}, {Gunn}, {Hogg}, {Ivezi{\'c}}, {Knapp},
  {Lupton}, {Munn}, {Schneider}, {Tegmark}, \& {Zehavi}}]{BlantonAJ2005}
{Blanton}, M.~R., {Schlegel}, D.~J., {Strauss}, M.~A., {et~al.} 2005, \aj, 129,
  2562, \dodoi{10.1086/429803}

\bibitem[{{Blanton} {et~al.}(2017){Blanton}, {Bershady}, {Abolfathi},
  {Albareti}, {Allende Prieto}, {Almeida}, {Alonso-Garc{\'\i}a}, {Anders},
  {Anderson}, {Andrews}, {Aquino-Ort{\'\i}z}, {Arag{\'o}n-Salamanca},
  {Argudo-Fern{\'a}ndez}, {Armengaud}, {Aubourg}, {Avila-Reese}, {Badenes},
  {Bailey}, {Barger}, {Barrera-Ballesteros}, {Bartosz}, {Bates}, {Baumgarten},
  {Bautista}, {Beaton}, {Beers}, {Belfiore}, {Bender}, {Berlind}, {Bernardi},
  {Beutler}, {Bird}, {Bizyaev}, {Blanc}, {Blomqvist}, {Bolton}, {Boquien},
  {Borissova}, {van den Bosch}, {Bovy}, {Brandt}, {Brinkmann}, {Brownstein},
  {Bundy}, {Burgasser}, {Burtin}, {Busca}, {Cappellari}, {Delgado Carigi},
  {Carlberg}, {Carnero Rosell}, {Carrera}, {Chanover}, {Cherinka}, {Cheung},
  {G{\'o}mez Maqueo Chew}, {Chiappini}, {Choi}, {Chojnowski}, {Chuang},
  {Chung}, {Cirolini}, {Clerc}, {Cohen}, {Comparat}, {da Costa}, {Cousinou},
  {Covey}, {Crane}, {Croft}, {Cruz-Gonzalez}, {Garrido Cuadra}, {Cunha},
  {Damke}, {Darling}, {Davies}, {Dawson}, {de la Macorra}, {Dell'Agli}, {De
  Lee}, {Delubac}, {Di Mille}, {Diamond-Stanic}, {Cano-D{\'\i}az}, {Donor},
  {Downes}, {Drory}, {du Mas des Bourboux}, {Duckworth}, {Dwelly}, {Dyer},
  {Ebelke}, {Eigenbrot}, {Eisenstein}, {Emsellem}, {Eracleous}, {Escoffier},
  {Evans}, {Fan}, {Fern{\'a}ndez-Alvar}, {Fernandez-Trincado}, {Feuillet},
  {Finoguenov}, {Fleming}, {Font-Ribera}, {Fredrickson}, {Freischlad},
  {Frinchaboy}, {Fuentes}, {Galbany}, {Garcia-Dias},
  {Garc{\'\i}a-Hern{\'a}ndez}, {Gaulme}, {Geisler}, {Gelfand},
  {Gil-Mar{\'\i}n}, {Gillespie}, {Goddard}, {Gonzalez-Perez}, {Grabowski},
  {Green}, {Grier}, {Gunn}, {Guo}, {Guy}, {Hagen}, {Hahn}, {Hall}, {Harding},
  {Hasselquist}, {Hawley}, {Hearty}, {Gonzalez Hern{\'a}ndez}, {Ho}, {Hogg},
  {Holley-Bockelmann}, {Holtzman}, {Holzer}, {Huehnerhoff}, {Hutchinson},
  {Hwang}, {Ibarra-Medel}, {da Silva Ilha}, {Ivans}, {Ivory}, {Jackson},
  {Jensen}, {Johnson}, {Jones}, {J{\"o}nsson}, {Jullo}, {Kamble}, {Kinemuchi},
  {Kirkby}, {Kitaura}, {Klaene}, {Knapp}, {Kneib}, {Kollmeier}, {Lacerna},
  {Lane}, {Lang}, {Law}, {Lazarz}, {Lee}, {Le Goff}, {Liang}, {Li}, {Li},
  {Lian}, {Lima}, {Lin}, {Lin}, {Bertran de Lis}, {Liu}, {de Icaza Lizaola},
  {Long}, {Lucatello}, {Lundgren}, {MacDonald}, {Deconto Machado}, {MacLeod},
  {Mahadevan}, {Geimba Maia}, {Maiolino}, {Majewski}, {Malanushenko},
  {Malanushenko}, {Manchado}, {Mao}, {Maraston}, {Marques-Chaves}, {Masseron},
  {Masters}, {McBride}, {McDermid}, {McGrath}, {McGreer}, {Medina Pe{\~n}a},
  {Melendez}, {Merloni}, {Merrifield}, {Meszaros}, {Meza}, {Minchev},
  {Minniti}, {Miyaji}, {More}, {Mulchaey}, {M{\"u}ller-S{\'a}nchez}, {Muna},
  {Munoz}, {Myers}, {Nair}, {Nandra}, {Correa do Nascimento}, {Negrete},
  {Ness}, {Newman}, {Nichol}, {Nidever}, {Nitschelm}, {Ntelis}, {O'Connell},
  {Oelkers}, {Oravetz}, {Oravetz}, {Pace}, {Padilla}, {Palanque-Delabrouille},
  {Alonso Palicio}, {Pan}, {Parejko}, {Parikh}, {P{\^a}ris}, {Park}, {Patten},
  {Peirani}, {Pellejero-Ibanez}, {Penny}, {Percival}, {Perez-Fournon},
  {Petitjean}, {Pieri}, {Pinsonneault}, {Pisani}, {Poleski}, {Prada},
  {Prakash}, {Queiroz}, {Raddick}, {Raichoor}, {Barboza Rembold}, {Richstein},
  {Riffel}, {Riffel}, {Rix}, {Robin}, {Rockosi}, {Rodr{\'\i}guez-Torres},
  {Roman-Lopes}, {Rom{\'a}n-Z{\'u}{\~n}iga}, {Rosado}, {Ross}, {Rossi}, {Ruan},
  {Ruggeri}, {Rykoff}, {Salazar-Albornoz}, {Salvato}, {S{\'a}nchez}, {Aguado},
  {S{\'a}nchez-Gallego}, {Santana}, {Santiago}, {Sayres}, {Schiavon}, {da Silva
  Schimoia}, {Schlafly}, {Schlegel}, {Schneider}, {Schultheis}, {Schuster},
  {Schwope}, {Seo}, {Shao}, {Shen}, {Shetrone}, {Shull}, {Simon}, {Skinner},
  {Skrutskie}, {Slosar}, {Smith}, {Sobeck}, {Sobreira}, {Somers}, {Souto},
  {Stark}, {Stassun}, {Stauffer}, {Steinmetz}, {Storchi-Bergmann},
  {Streblyanska}, {Stringfellow}, {Su{\'a}rez}, {Sun}, {Suzuki}, {Szigeti},
  {Taghizadeh-Popp}, {Tang}, {Tao}, {Tayar}, {Tembe}, {Teske}, {Thakar},
  {Thomas}, {Thompson}, {Tinker}, {Tissera}, {Tojeiro}, {Hernandez Toledo}, {de
  la Torre}, {Tremonti}, {Troup}, {Valenzuela}, {Martinez Valpuesta},
  {Vargas-Gonz{\'a}lez}, {Vargas-Maga{\~n}a}, {Vazquez}, {Villanova}, {Vivek},
  {Vogt}, {Wake}, {Walterbos}, {Wang}, {Weaver}, {Weijmans}, {Weinberg},
  {Westfall}, {Whelan}, {Wild}, {Wilson}, {Wood-Vasey}, {Wylezalek}, {Xiao},
  {Yan}, {Yang}, {Ybarra}, {Y{\`e}che}, {Zakamska}, {Zamora}, {Zarrouk},
  {Zasowski}, {Zhang}, {Zhao}, {Zheng}, {Zheng}, {Zhou}, {Zhou}, {Zhu},
  {Zoccali}, \& {Zou}}]{BlantonAJ2017}
{Blanton}, M.~R., {Bershady}, M.~A., {Abolfathi}, B., {et~al.} 2017, \aj, 154,
  28, \dodoi{10.3847/1538-3881/aa7567}

\bibitem[{{Bluck} {et~al.}(2020){Bluck}, {Maiolino}, {Piotrowska}, {Trussler},
  {Ellison}, {S{\'a}nchez}, {Thorp}, {Teimoorinia}, {Moreno}, \&
  {Conselice}}]{BluckMNRAS2020}
{Bluck}, A. F.~L., {Maiolino}, R., {Piotrowska}, J.~M., {et~al.} 2020, \mnras,
  499, 230, \dodoi{10.1093/mnras/staa2806}

\bibitem[{{Bradley} {et~al.}(2023){Bradley}, {Sip{\H{o}}cz}, {Robitaille},
  {Tollerud}, {Vin{\'\i}cius}, {Deil}, {Barbary}, {Wilson}, {Busko}, {Donath},
  {G{\"u}nther}, {Cara}, {Lim}, {Me{\ss}linger}, {Conseil}, {Burnett},
  {Bostroem}, {Droettboom}, {Bray}, {Andersen Bratholm}, {Jamieson},
  {Ginsburg}, {Barentsen}, {Craig}, {Morris}, {Perrin}, {Rathi}, {Pascual},
  {Perren}, \& {Georgiev}}]{Bradley2023}
{Bradley}, L., {Sip{\H{o}}cz}, B., {Robitaille}, T., {et~al.} 2023,
  {astropy/photutils: 1.10.0}, 1.10.0,  Zenodo, \dodoi{10.5281/zenodo.596036}

\bibitem[{{Bruzual} \& {Charlot}(2003)}]{BruzualMNRAS2003}
{Bruzual}, G., \& {Charlot}, S. 2003, \mnras, 344, 1000,
  \dodoi{10.1046/j.1365-8711.2003.06897.x}

\bibitem[{{Bryant} {et~al.}(2015){Bryant}, {Owers}, {Robotham}, {Croom},
  {Driver}, {Drinkwater}, {Lorente}, {Cortese}, {Scott}, {Colless}, {Schaefer},
  {Taylor}, {Konstantopoulos}, {Allen}, {Baldry}, {Barnes}, {Bauer},
  {Bland-Hawthorn}, {Bloom}, {Brooks}, {Brough}, {Cecil}, {Couch}, {Croton},
  {Davies}, {Ellis}, {Fogarty}, {Foster}, {Glazebrook}, {Goodwin}, {Green},
  {Gunawardhana}, {Hampton}, {Ho}, {Hopkins}, {Kewley}, {Lawrence},
  {Leon-Saval}, {Leslie}, {McElroy}, {Lewis}, {Liske}, {L{\'o}pez-S{\'a}nchez},
  {Mahajan}, {Medling}, {Metcalfe}, {Meyer}, {Mould}, {Obreschkow}, {O'Toole},
  {Pracy}, {Richards}, {Shanks}, {Sharp}, {Sweet}, {Thomas}, {Tonini}, \&
  {Walcher}}]{BryantMNRAS2015}
{Bryant}, J.~J., {Owers}, M.~S., {Robotham}, A.~S.~G., {et~al.} 2015, \mnras,
  447, 2857, \dodoi{10.1093/mnras/stu2635}

\bibitem[{{Bryant} {et~al.}(2016){Bryant}, {Bland-Hawthorn}, {Lawrence},
  {Croom}, {Brown}, {Venkatesan}, {Gillingham}, {Zhelem}, {Content},
  {Saunders}, {Staszak}, {van de Sande}, {Couch}, {Leon-Saval}, {Tims},
  {McDermid}, \& {Schaefer}}]{Bryant2016}
{Bryant}, J.~J., {Bland-Hawthorn}, J., {Lawrence}, J., {et~al.} 2016, in
  Society of Photo-Optical Instrumentation Engineers (SPIE) Conference Series,
  Vol. 9908, Ground-based and Airborne Instrumentation for Astronomy VI, ed.
  C.~J. {Evans}, L.~{Simard}, \& H.~{Takami}, 99081F,
  \dodoi{10.1117/12.2230740}

\bibitem[{{Bundy} {et~al.}(2015){Bundy}, {Bershady}, {Law}, {Yan}, {Drory},
  {MacDonald}, {Wake}, {Cherinka}, {S{\'a}nchez-Gallego}, {Weijmans}, {Thomas},
  {Tremonti}, {Masters}, {Coccato}, {Diamond-Stanic}, {Arag{\'o}n-Salamanca},
  {Avila-Reese}, {Badenes}, {Falc{\'o}n-Barroso}, {Belfiore}, {Bizyaev},
  {Blanc}, {Bland-Hawthorn}, {Blanton}, {Brownstein}, {Byler}, {Cappellari},
  {Conroy}, {Dutton}, {Emsellem}, {Etherington}, {Frinchaboy}, {Fu}, {Gunn},
  {Harding}, {Johnston}, {Kauffmann}, {Kinemuchi}, {Klaene}, {Knapen},
  {Leauthaud}, {Li}, {Lin}, {Maiolino}, {Malanushenko}, {Malanushenko}, {Mao},
  {Maraston}, {McDermid}, {Merrifield}, {Nichol}, {Oravetz}, {Pan}, {Parejko},
  {Sanchez}, {Schlegel}, {Simmons}, {Steele}, {Steinmetz}, {Thanjavur},
  {Thompson}, {Tinker}, {van den Bosch}, {Westfall}, {Wilkinson}, {Wright},
  {Xiao}, \& {Zhang}}]{BundyAPJ2015}
{Bundy}, K., {Bershady}, M.~A., {Law}, D.~R., {et~al.} 2015, \apj, 798, 7,
  \dodoi{10.1088/0004-637X/798/1/7}

\bibitem[{{Cappellari} {et~al.}(2013){Cappellari}, {McDermid}, {Alatalo},
  {Blitz}, {Bois}, {Bournaud}, {Bureau}, {Crocker}, {Davies}, {Davis}, {de
  Zeeuw}, {Duc}, {Emsellem}, {Khochfar}, {Krajnovi{\'c}}, {Kuntschner},
  {Morganti}, {Naab}, {Oosterloo}, {Sarzi}, {Scott}, {Serra}, {Weijmans}, \&
  {Young}}]{CappellariMNRAS2013}
{Cappellari}, M., {McDermid}, R.~M., {Alatalo}, K., {et~al.} 2013, \mnras, 432,
  1862, \dodoi{10.1093/mnras/stt644}

\bibitem[{{Cenarro} {et~al.}(2003){Cenarro}, {Gorgas}, {Vazdekis}, {Cardiel},
  \& {Peletier}}]{CenarroMNRAS2003}
{Cenarro}, A.~J., {Gorgas}, J., {Vazdekis}, A., {Cardiel}, N., \& {Peletier},
  R.~F. 2003, \mnras, 339, L12, \dodoi{10.1046/j.1365-8711.2003.06360.x}

\bibitem[{{Cheng} {et~al.}(2024){Cheng}, {Kriek}, {Beverage}, {van der Wel},
  {Bezanson}, {D'Eugenio}, {Franx}, {Mancera Pi{\~n}a}, {Nersesian}, {Slob},
  {Suess}, {van Dokkum}, {Wu}, {Gallazzi}, \& {Zibetti}}]{ChengMNRAS2024}
{Cheng}, C.~M., {Kriek}, M., {Beverage}, A.~G., {et~al.} 2024, \mnras, 532,
  3604, \dodoi{10.1093/mnras/stae1739}

\bibitem[{{Chilingarian} {et~al.}(2010){Chilingarian}, {Melchior}, \&
  {Zolotukhin}}]{ChilingarianMNRAS2010}
{Chilingarian}, I.~V., {Melchior}, A.-L., \& {Zolotukhin}, I.~Y. 2010, \mnras,
  405, 1409, \dodoi{10.1111/j.1365-2966.2010.16506.x}

\bibitem[{{Chilingarian} \& {Zolotukhin}(2012)}]{ChilingarianMNRAS2012}
{Chilingarian}, I.~V., \& {Zolotukhin}, I.~Y. 2012, \mnras, 419, 1727,
  \dodoi{10.1111/j.1365-2966.2011.19837.x}

\bibitem[{{Choi} {et~al.}(2016){Choi}, {Dotter}, {Conroy}, {Cantiello},
  {Paxton}, \& {Johnson}}]{ChoiAPJ2016}
{Choi}, J., {Dotter}, A., {Conroy}, C., {et~al.} 2016, \apj, 823, 102,
  \dodoi{10.3847/0004-637X/823/2/102}

\bibitem[{{Conroy} {et~al.}(2014){Conroy}, {Graves}, \& {van
  Dokkum}}]{ConroyAPJ2014}
{Conroy}, C., {Graves}, G.~J., \& {van Dokkum}, P.~G. 2014, \apj, 780, 33,
  \dodoi{10.1088/0004-637X/780/1/33}

\bibitem[{{Conroy} \& {van Dokkum}(2012{\natexlab{a}})}]{ConroyAPJ2012a}
{Conroy}, C., \& {van Dokkum}, P. 2012{\natexlab{a}}, \apj, 747, 69,
  \dodoi{10.1088/0004-637X/747/1/69}

\bibitem[{{Conroy} \& {van Dokkum}(2012{\natexlab{b}})}]{ConroyAPJ2012}
{Conroy}, C., \& {van Dokkum}, P.~G. 2012{\natexlab{b}}, \apj, 760, 71,
  \dodoi{10.1088/0004-637X/760/1/71}

\bibitem[{{Conroy} {et~al.}(2017){Conroy}, {van Dokkum}, \&
  {Villaume}}]{ConroyAPJ2017}
{Conroy}, C., {van Dokkum}, P.~G., \& {Villaume}, A. 2017, \apj, 837, 166,
  \dodoi{10.3847/1538-4357/aa6190}

\bibitem[{{Conroy} {et~al.}(2018){Conroy}, {Villaume}, {van Dokkum}, \&
  {Lind}}]{ConroyAPJ2018}
{Conroy}, C., {Villaume}, A., {van Dokkum}, P.~G., \& {Lind}, K. 2018, \apj,
  854, 139, \dodoi{10.3847/1538-4357/aaab49}

\bibitem[{{Croom} {et~al.}(2012){Croom}, {Lawrence}, {Bland-Hawthorn},
  {Bryant}, {Fogarty}, {Richards}, {Goodwin}, {Farrell}, {Miziarski}, {Heald},
  {Jones}, {Lee}, {Colless}, {Brough}, {Hopkins}, {Bauer}, {Birchall}, {Ellis},
  {Horton}, {Leon-Saval}, {Lewis}, {L{\'o}pez-S{\'a}nchez}, {Min}, {Trinh}, \&
  {Trowland}}]{CroomMNRAS2012}
{Croom}, S.~M., {Lawrence}, J.~S., {Bland-Hawthorn}, J., {et~al.} 2012, \mnras,
  421, 872, \dodoi{10.1111/j.1365-2966.2011.20365.x}

\bibitem[{{DESI Collaboration} {et~al.}(2016){DESI Collaboration}, {Aghamousa},
  {Aguilar}, {Ahlen}, {Alam}, {Allen}, {Allende Prieto}, {Annis}, {Bailey},
  {Balland}, {Ballester}, {Baltay}, {Beaufore}, {Bebek}, {Beers}, {Bell},
  {Bernal}, {Besuner}, {Beutler}, {Blake}, {Bleuler}, {Blomqvist}, {Blum},
  {Bolton}, {Briceno}, {Brooks}, {Brownstein}, {Buckley-Geer}, {Burden},
  {Burtin}, {Busca}, {Cahn}, {Cai}, {Cardiel-Sas}, {Carlberg}, {Carton},
  {Casas}, {Castander}, {Cervantes-Cota}, {Claybaugh}, {Close}, {Coker},
  {Cole}, {Comparat}, {Cooper}, {Cousinou}, {Crocce}, {Cuby}, {Cunningham},
  {Davis}, {Dawson}, {de la Macorra}, {De Vicente}, {Delubac}, {Derwent},
  {Dey}, {Dhungana}, {Ding}, {Doel}, {Duan}, {Ealet}, {Edelstein},
  {Eftekharzadeh}, {Eisenstein}, {Elliott}, {Escoffier}, {Evatt}, {Fagrelius},
  {Fan}, {Fanning}, {Farahi}, {Farihi}, {Favole}, {Feng}, {Fernandez},
  {Findlay}, {Finkbeiner}, {Fitzpatrick}, {Flaugher}, {Flender}, {Font-Ribera},
  {Forero-Romero}, {Fosalba}, {Frenk}, {Fumagalli}, {Gaensicke}, {Gallo},
  {Garcia-Bellido}, {Gaztanaga}, {Pietro Gentile Fusillo}, {Gerard},
  {Gershkovich}, {Giannantonio}, {Gillet}, {Gonzalez-de-Rivera},
  {Gonzalez-Perez}, {Gott}, {Graur}, {Gutierrez}, {Guy}, {Habib}, {Heetderks},
  {Heetderks}, {Heitmann}, {Hellwing}, {Herrera}, {Ho}, {Holland}, {Honscheid},
  {Huff}, {Hutchinson}, {Huterer}, {Hwang}, {Illa Laguna}, {Ishikawa},
  {Jacobs}, {Jeffrey}, {Jelinsky}, {Jennings}, {Jiang}, {Jimenez}, {Johnson},
  {Joyce}, {Jullo}, {Juneau}, {Kama}, {Karcher}, {Karkar}, {Kehoe}, {Kennamer},
  {Kent}, {Kilbinger}, {Kim}, {Kirkby}, {Kisner}, {Kitanidis}, {Kneib},
  {Koposov}, {Kovacs}, {Koyama}, {Kremin}, {Kron}, {Kronig}, {Kueter-Young},
  {Lacey}, {Lafever}, {Lahav}, {Lambert}, {Lampton}, {Landriau}, {Lang},
  {Lauer}, {Le Goff}, {Le Guillou}, {Le Van Suu}, {Lee}, {Lee}, {Leitner},
  {Lesser}, {Levi}, {L'Huillier}, {Li}, {Liang}, {Lin}, {Linder}, {Loebman},
  {Luki{\'c}}, {Ma}, {MacCrann}, {Magneville}, {Makarem}, {Manera}, {Manser},
  {Marshall}, {Martini}, {Massey}, {Matheson}, {McCauley}, {McDonald},
  {McGreer}, {Meisner}, {Metcalfe}, {Miller}, {Miquel}, {Moustakas}, {Myers},
  {Naik}, {Newman}, {Nichol}, {Nicola}, {Nicolati da Costa}, {Nie}, {Niz},
  {Norberg}, {Nord}, {Norman}, {Nugent}, {O'Brien}, {Oh}, {Olsen}, {Padilla},
  {Padmanabhan}, {Padmanabhan}, {Palanque-Delabrouille}, {Palmese},
  {Pappalardo}, {P{\^a}ris}, {Park}, {Patej}, {Peacock}, {Peiris}, {Peng},
  {Percival}, {Perruchot}, {Pieri}, {Pogge}, {Pollack}, {Poppett}, {Prada},
  {Prakash}, {Probst}, {Rabinowitz}, {Raichoor}, {Ree}, {Refregier}, {Regal},
  {Reid}, {Reil}, {Rezaie}, {Rockosi}, {Roe}, {Ronayette}, {Roodman}, {Ross},
  {Ross}, {Rossi}, {Rozo}, {Ruhlmann-Kleider}, {Rykoff}, {Sabiu}, {Samushia},
  {Sanchez}, {Sanchez}, {Schlegel}, {Schneider}, {Schubnell}, {Secroun},
  {Seljak}, {Seo}, {Serrano}, {Shafieloo}, {Shan}, {Sharples}, {Sholl},
  {Shourt}, {Silber}, {Silva}, {Sirk}, {Slosar}, {Smith}, {Smoot}, {Som},
  {Song}, {Sprayberry}, {Staten}, {Stefanik}, {Tarle}, {Sien Tie}, {Tinker},
  {Tojeiro}, {Valdes}, {Valenzuela}, {Valluri}, {Vargas-Magana}, {Verde},
  {Walker}, {Wang}, {Wang}, {Weaver}, {Weaverdyck}, {Wechsler}, {Weinberg},
  {White}, {Yang}, {Yeche}, {Zhang}, {Zhao}, {Zheng}, {Zhou}, {Zhou}, {Zhu},
  {Zou}, \& {Zu}}]{DESICollaborationARXIVEPRINTS2016}
{DESI Collaboration}, {Aghamousa}, A., {Aguilar}, J., {et~al.} 2016, arXiv
  e-prints, arXiv:1611.00036, \dodoi{10.48550/arXiv.1611.00036}

\bibitem[{{DESI Collaboration} {et~al.}(2022){DESI Collaboration}, {Abareshi},
  {Aguilar}, {Ahlen}, {Alam}, {Alexander}, {Alfarsy}, {Allen}, {Allende
  Prieto}, {Alves}, {Ameel}, {Armengaud}, {Asorey}, {Aviles}, {Bailey},
  {Balaguera-Antol{\'\i}nez}, {Ballester}, {Baltay}, {Bault}, {Beltran},
  {Benavides}, {BenZvi}, {Berti}, {Besuner}, {Beutler}, {Bianchi}, {Blake},
  {Blanc}, {Blum}, {Bolton}, {Bose}, {Bramall}, {Brieden}, {Brodzeller},
  {Brooks}, {Brownewell}, {Buckley-Geer}, {Cahn}, {Cai}, {Canning}, {Capasso},
  {Carnero Rosell}, {Carton}, {Casas}, {Castander}, {Cervantes-Cota},
  {Chabanier}, {Chaussidon}, {Chuang}, {Circosta}, {Cole}, {Cooper}, {da
  Costa}, {Cousinou}, {Cuceu}, {Davis}, {Dawson}, {de la Cruz-Noriega}, {de la
  Macorra}, {de Mattia}, {Della Costa}, {Demmer}, {Derwent}, {Dey}, {Dey},
  {Dhungana}, {Ding}, {Dobson}, {Doel}, {Donald-McCann}, {Donaldson},
  {Douglass}, {Duan}, {Dunlop}, {Edelstein}, {Eftekharzadeh}, {Eisenstein},
  {Enriquez-Vargas}, {Escoffier}, {Evatt}, {Fagrelius}, {Fan}, {Fanning},
  {Fawcett}, {Ferraro}, {Ereza}, {Flaugher}, {Font-Ribera}, {Forero-Romero},
  {Frenk}, {Fromenteau}, {G{\"a}nsicke}, {Garcia-Quintero}, {Garrison},
  {Gazta{\~n}aga}, {Gerardi}, {Gil-Mar{\'\i}n}, {Gontcho a Gontcho},
  {Gonzalez-Morales}, {Gonzalez-de-Rivera}, {Gonzalez-Perez}, {Gordon},
  {Graur}, {Green}, {Grove}, {Gruen}, {Gutierrez}, {Guy}, {Hahn}, {Harris},
  {Herrera}, {Herrera-Alcantar}, {Honscheid}, {Howlett}, {Huterer},
  {Ir{\v{s}}i{\v{c}}}, {Ishak}, {Jelinsky}, {Jiang}, {Jimenez}, {Jing},
  {Joyce}, {Jullo}, {Juneau}, {Kara{\c{c}}ayl{\i}}, {Karamanis}, {Karcher},
  {Karim}, {Kehoe}, {Kent}, {Kirkby}, {Kisner}, {Kitaura}, {Koposov},
  {Kov{\'a}cs}, {Kremin}, {Krolewski}, {L'Huillier}, {Lahav}, {Lambert},
  {Lamman}, {Lan}, {Landriau}, {Lane}, {Lang}, {Lange}, {Lasker}, {Le Guillou},
  {Leauthaud}, {Le Van Suu}, {Levi}, {Li}, {Magneville}, {Manera}, {Manser},
  {Marshall}, {Martini}, {McCollam}, {McDonald}, {Meisner},
  {Mena-Fern{\'a}ndez}, {Meneses-Rizo}, {Mezcua}, {Miller}, {Miquel},
  {Montero-Camacho}, {Moon}, {Moustakas}, {Mueller}, {Mu{\~n}oz-Guti{\'e}rrez},
  {Myers}, {Nadathur}, {Najita}, {Napolitano}, {Neilsen}, {Newman}, {Nie},
  {Ning}, {Niz}, {Norberg}, {Noriega}, {O'Brien}, {Obuljen},
  {Palanque-Delabrouille}, {Palmese}, {Zhiwei}, {Pappalardo}, {PENG},
  {Percival}, {Perruchot}, {Pogge}, {Poppett}, {Porredon}, {Prada},
  {Prochaska}, {Pucha}, {P{\'e}rez-Fern{\'a}ndez}, {P{\'e}rez-R{\`a}fols},
  {Rabinowitz}, {Raichoor}, {Ramirez-Solano}, {Ram{\'\i}rez-P{\'e}rez},
  {Ravoux}, {Reil}, {Rezaie}, {Rocher}, {Rockosi}, {Roe}, {Roodman}, {Ross},
  {Rossi}, {Ruggeri}, {Ruhlmann-Kleider}, {Sabiu}, {Safonova}, {Said},
  {Saintonge}, {Salas Catonga}, {Samushia}, {Sanchez}, {Saulder}, {Schaan},
  {Schlafly}, {Schlegel}, {Schmoll}, {Scholte}, {Schubnell}, {Secroun}, {Seo},
  {Serrano}, {Sharples}, {Sholl}, {Silber}, {Silva}, {Sirk}, {Siudek}, {Smith},
  {Sprayberry}, {Staten}, {Stupak}, {Tan}, {Tarl{\'e}}, {Tie}, {Tojeiro},
  {Ure{\~n}a-L{\'o}pez}, {Valdes}, {Valenzuela}, {Valluri},
  {Vargas-Maga{\~n}a}, {Verde}, {Walther}, {Wang}, {Wang}, {Weaver},
  {Weaverdyck}, {Wechsler}, {Wilson}, {Yang}, {Yu}, {Yuan}, {Y{\`e}che},
  {Zhang}, {Zhang}, {Zhao}, {Zhou}, {Zhou}, {Zou}, {Zou}, {Zou}, {Zu}, \& {DESI
  Collaboration}}]{DESICollaborationAJ2022}
{DESI Collaboration}, {Abareshi}, B., {Aguilar}, J., {et~al.} 2022, \aj, 164,
  207, \dodoi{10.3847/1538-3881/ac882b}

\bibitem[{{Dey} {et~al.}(2019){Dey}, {Schlegel}, {Lang}, {Blum}, {Burleigh},
  {Fan}, {Findlay}, {Finkbeiner}, {Herrera}, {Juneau}, {Landriau}, {Levi},
  {McGreer}, {Meisner}, {Myers}, {Moustakas}, {Nugent}, {Patej}, {Schlafly},
  {Walker}, {Valdes}, {Weaver}, {Y{\`e}che}, {Zou}, {Zhou}, {Abareshi},
  {Abbott}, {Abolfathi}, {Aguilera}, {Alam}, {Allen}, {Alvarez}, {Annis},
  {Ansarinejad}, {Aubert}, {Beechert}, {Bell}, {BenZvi}, {Beutler}, {Bielby},
  {Bolton}, {Brice{\~n}o}, {Buckley-Geer}, {Butler}, {Calamida}, {Carlberg},
  {Carter}, {Casas}, {Castander}, {Choi}, {Comparat}, {Cukanovaite}, {Delubac},
  {DeVries}, {Dey}, {Dhungana}, {Dickinson}, {Ding}, {Donaldson}, {Duan},
  {Duckworth}, {Eftekharzadeh}, {Eisenstein}, {Etourneau}, {Fagrelius},
  {Farihi}, {Fitzpatrick}, {Font-Ribera}, {Fulmer}, {G{\"a}nsicke},
  {Gaztanaga}, {George}, {Gerdes}, {Gontcho}, {Gorgoni}, {Green}, {Guy},
  {Harmer}, {Hernandez}, {Honscheid}, {Huang}, {James}, {Jannuzi}, {Jiang},
  {Joyce}, {Karcher}, {Karkar}, {Kehoe}, {Kneib}, {Kueter-Young}, {Lan},
  {Lauer}, {Le Guillou}, {Le Van Suu}, {Lee}, {Lesser}, {Perreault Levasseur},
  {Li}, {Mann}, {Marshall}, {Mart{\'\i}nez-V{\'a}zquez}, {Martini}, {du Mas des
  Bourboux}, {McManus}, {Meier}, {M{\'e}nard}, {Metcalfe},
  {Mu{\~n}oz-Guti{\'e}rrez}, {Najita}, {Napier}, {Narayan}, {Newman}, {Nie},
  {Nord}, {Norman}, {Olsen}, {Paat}, {Palanque-Delabrouille}, {Peng},
  {Poppett}, {Poremba}, {Prakash}, {Rabinowitz}, {Raichoor}, {Rezaie},
  {Robertson}, {Roe}, {Ross}, {Ross}, {Rudnick}, {Safonova}, {Saha},
  {S{\'a}nchez}, {Savary}, {Schweiker}, {Scott}, {Seo}, {Shan}, {Silva},
  {Slepian}, {Soto}, {Sprayberry}, {Staten}, {Stillman}, {Stupak}, {Summers},
  {Sien Tie}, {Tirado}, {Vargas-Maga{\~n}a}, {Vivas}, {Wechsler}, {Williams},
  {Yang}, {Yang}, {Yapici}, {Zaritsky}, {Zenteno}, {Zhang}, {Zhang}, {Zhou}, \&
  {Zhou}}]{DeyAJ2019}
{Dey}, A., {Schlegel}, D.~J., {Lang}, D., {et~al.} 2019, \aj, 157, 168,
  \dodoi{10.3847/1538-3881/ab089d}

\bibitem[{{Dom{\'\i}nguez S{\'a}nchez} {et~al.}(2022){Dom{\'\i}nguez
  S{\'a}nchez}, {Margalef}, {Bernardi}, \&
  {Huertas-Company}}]{SanchezMNRAS2022}
{Dom{\'\i}nguez S{\'a}nchez}, H., {Margalef}, B., {Bernardi}, M., \&
  {Huertas-Company}, M. 2022, \mnras, 509, 4024, \dodoi{10.1093/mnras/stab3089}

\bibitem[{{Faber}(1973)}]{FaberApJ1973}
{Faber}, S.~M. 1973, \apj, 179, 423, \dodoi{10.1086/151881}

\bibitem[{{Feldmeier-Krause} {et~al.}(2021){Feldmeier-Krause}, {Lonoce}, \&
  {Freedman}}]{FeldmeierKrauseAPJ2021}
{Feldmeier-Krause}, A., {Lonoce}, I., \& {Freedman}, W.~L. 2021, \apj, 923, 65,
  \dodoi{10.3847/1538-4357/ac281e}

\bibitem[{{Ferr{\'e}-Mateu} {et~al.}(2017){Ferr{\'e}-Mateu}, {Trujillo},
  {Mart{\'\i}n-Navarro}, {Vazdekis}, {Mezcua}, {Balcells}, \&
  {Dom{\'\i}nguez}}]{Ferre-MateuMNRAS2017}
{Ferr{\'e}-Mateu}, A., {Trujillo}, I., {Mart{\'\i}n-Navarro}, I., {et~al.}
  2017, \mnras, 467, 1929, \dodoi{10.1093/mnras/stx171}

\bibitem[{{Ferreras} {et~al.}(2019){Ferreras}, {Scott}, {La Barbera}, {Croom},
  {van de Sande}, {Hopkins}, {Colless}, {Barone}, {d'Eugenio},
  {Bland-Hawthorn}, {Brough}, {Bryant}, {Konstantopoulos}, {Lagos}, {Lawrence},
  {L{\'o}pez-S{\'a}nchez}, {Medling}, {Owers}, \&
  {Richards}}]{FerrerasMNRAS2019}
{Ferreras}, I., {Scott}, N., {La Barbera}, F., {et~al.} 2019, \mnras, 489, 608,
  \dodoi{10.1093/mnras/stz2095}

\bibitem[{{Fischer} {et~al.}(2017){Fischer}, {Bernardi}, \&
  {Meert}}]{FischerMNRAS2017}
{Fischer}, J.~L., {Bernardi}, M., \& {Meert}, A. 2017, \mnras, 467, 490,
  \dodoi{10.1093/mnras/stx136}

\bibitem[{{Flaugher} {et~al.}(2015){Flaugher}, {Diehl}, {Honscheid}, {Abbott},
  {Alvarez}, {Angstadt}, {Annis}, {Antonik}, {Ballester}, {Beaufore},
  {Bernstein}, {Bernstein}, {Bigelow}, {Bonati}, {Boprie}, {Brooks},
  {Buckley-Geer}, {Campa}, {Cardiel-Sas}, {Castander}, {Castilla}, {Cease},
  {Cela-Ruiz}, {Chappa}, {Chi}, {Cooper}, {da Costa}, {Dede}, {Derylo},
  {DePoy}, {de Vicente}, {Doel}, {Drlica-Wagner}, {Eiting}, {Elliott}, {Emes},
  {Estrada}, {Fausti Neto}, {Finley}, {Flores}, {Frieman}, {Gerdes},
  {Gladders}, {Gregory}, {Gutierrez}, {Hao}, {Holland}, {Holm}, {Huffman},
  {Jackson}, {James}, {Jonas}, {Karcher}, {Karliner}, {Kent}, {Kessler},
  {Kozlovsky}, {Kron}, {Kubik}, {Kuehn}, {Kuhlmann}, {Kuk}, {Lahav}, {Lathrop},
  {Lee}, {Levi}, {Lewis}, {Li}, {Mandrichenko}, {Marshall}, {Martinez},
  {Merritt}, {Miquel}, {Mu{\~n}oz}, {Neilsen}, {Nichol}, {Nord}, {Ogando},
  {Olsen}, {Palaio}, {Patton}, {Peoples}, {Plazas}, {Rauch}, {Reil}, {Rheault},
  {Roe}, {Rogers}, {Roodman}, {Sanchez}, {Scarpine}, {Schindler}, {Schmidt},
  {Schmitt}, {Schubnell}, {Schultz}, {Schurter}, {Scott}, {Serrano}, {Shaw},
  {Smith}, {Soares-Santos}, {Stefanik}, {Stuermer}, {Suchyta}, {Sypniewski},
  {Tarle}, {Thaler}, {Tighe}, {Tran}, {Tucker}, {Walker}, {Wang}, {Watson},
  {Weaverdyck}, {Wester}, {Woods}, {Yanny}, \& {DES
  Collaboration}}]{FlaugherAJ2015}
{Flaugher}, B., {Diehl}, H.~T., {Honscheid}, K., {et~al.} 2015, \aj, 150, 150,
  \dodoi{10.1088/0004-6256/150/5/150}

\bibitem[{{Foreman-Mackey} {et~al.}(2013){Foreman-Mackey}, {Hogg}, {Lang}, \&
  {Goodman}}]{ForemanMackeyPASP2013}
{Foreman-Mackey}, D., {Hogg}, D.~W., {Lang}, D., \& {Goodman}, J. 2013, \pasp,
  125, 306, \dodoi{10.1086/670067}

\bibitem[{{Fouesneau}(2022)}]{Fouesneau2022}
{Fouesneau}, M. 2022, {pyphot}, 1.4.3,  Zenodo, \dodoi{10.5281/zenodo.7016775}

\bibitem[{{Gallazzi} {et~al.}(2006){Gallazzi}, {Charlot}, {Brinchmann}, \&
  {White}}]{GallazziMNRAS2006}
{Gallazzi}, A., {Charlot}, S., {Brinchmann}, J., \& {White}, S. D.~M. 2006,
  \mnras, 370, 1106, \dodoi{10.1111/j.1365-2966.2006.10548.x}

\bibitem[{{Gardner} {et~al.}(2023){Gardner}, {Mather}, {Abbott}, {Abell},
  {Abernathy}, {Abney}, {Abraham}, {Abraham}, {Abul-Huda}, {Acton}, {Adams},
  {Adams}, {Adler}, {Adriaensen}, {Aguilar}, {Ahmed}, {Ahmed}, {Ahmed},
  {Albat}, {Albert}, {Alberts}, {Aldridge}, {Allen}, {Allen}, {Altenburg},
  {Altunc}, {Alvarez}, {{\'A}lvarez-M{\'a}rquez}, {Alves de Oliveira},
  {Ambrose}, {Anandakrishnan}, {Andersen}, {Anderson}, {Anderson}, {Anderson},
  {Anderson}, {Aprea}, {Archer}, {Arenberg}, {Argyriou}, {Arribas}, {Artigau},
  {Arvai}, {Atcheson}, {Atkinson}, {Averbukh}, {Aymergen}, {Bacinski},
  {Baggett}, {Bagnasco}, {Baker}, {Balzano}, {Banks}, {Baran}, {Barker},
  {Barrett}, {Barringer}, {Barto}, {Bast}, {Baudoz}, {Baum}, {Beatty},
  {Beaulieu}, {Bechtold}, {Beck}, {Beddard}, {Beichman}, {Bellagama}, {Bely},
  {Berger}, {Bergeron}, {Bernier}, {Bertch}, {Beskow}, {Betz}, {Biagetti},
  {Birkmann}, {Bjorklund}, {Blackwood}, {Blazek}, {Blossfeld}, {Bluth},
  {Boccaletti}, {Boegner}, {Bohlin}, {Boia}, {B{\"o}ker}, {Bonaventura},
  {Bond}, {Bosley}, {Boucarut}, {Bouchet}, {Bouwman}, {Bower}, {Bowers},
  {Bowers}, {Boyce}, {Boyer}, {Boyer}, {Boyer}, {Boyer}, {Bradley}, {Brady},
  {Brandl}, {Brannen}, {Breda}, {Bremmer}, {Brennan}, {Bresnahan}, {Bright},
  {Broiles}, {Bromenschenkel}, {Brooks}, {Brooks}, {Brown}, {Brown}, {Brown},
  {Bruce}, {Bryson}, {Bujanda}, {Bullock}, {Bunker}, {Bureo}, {Burt}, {Bush},
  {Bushouse}, {Bussman}, {Cabaud}, {Cale}, {Calhoon}, {Calvani}, {Canipe},
  {Caputo}, {Cara}, {Carey}, {Case}, {Cesari}, {Cetorelli}, {Chance},
  {Chandler}, {Chaney}, {Chapman}, {Charlot}, {Chayer}, {Cheezum}, {Chen},
  {Chen}, {Cherinka}, {Chichester}, {Chilton}, {Chittiraibalan}, {Clampin},
  {Clark}, {Clark}, {Clark}, {Claybrooks}, {Cleveland}, {Cohen}, {Cohen},
  {Col{\'o}n}, {Coleman}, {Colina}, {Comber}, {Comeau}, {Comer}, {Conde Reis},
  {Connolly}, {Conroy}, {Contos}, {Contreras}, {Cook}, {Cooper}, {Cooper},
  {Correia}, {Correnti}, {Cossou}, {Costanza}, {Coulais}, {Cox}, {Coyle},
  {Cracraft}, {Crew}, {Curtis}, {Cusveller}, {Da Costa Maciel}, {Dailey},
  {Daugeron}, {Davidson}, {Davies}, {Davis}, {Davis}, {Day}, {de Chambure}, {de
  Jong}, {De Marchi}, {Dean}, {Decker}, {Delisa}, {Dell}, {Dellagatta},
  {Dembinska}, {Demosthenes}, {Dencheva}, {Deneu}, {DePriest}, {Deschenes},
  {Dethienne}, {Detre}, {Diaz}, {Dicken}, {DiFelice}, {Dillman}, {Disharoon},
  {Dixon}, {Doggett}, {Dominguez}, {Donaldson}, {Doria-Warner}, {Santos},
  {Doty}, {Douglas}, {Doyon}, {Dressler}, {Driggers}, {Driggers}, {Dunn},
  {DuPrie}, {Dupuis}, {Durning}, {Dutta}, {Earl}, {Eccleston}, {Ecobichon},
  {Egami}, {Ehrenwinkler}, {Eisenhamer}, {Eisenhower}, {Eisenstein}, {El
  Hamel}, {Elie}, {Elliott}, {Elliott}, {Engesser}, {Espinoza}, {Etienne},
  {Etxaluze}, {Evans}, {Fabreguettes}, {Falcolini}, {Falini}, {Fatig},
  {Feeney}, {Feinberg}, {Fels}, {Ferdous}, {Ferguson}, {Ferrarese}, {Ferreira},
  {Ferruit}, {Ferry}, {Filippazzo}, {Firre}, {Fix}, {Flagey}, {Flanagan},
  {Fleming}, {Florian}, {Flynn}, {Foiadelli}, {Fontaine}, {Fontanella},
  {Forshay}, {Fortner}, {Fox}, {Framarini}, {Francisco}, {Franck}, {Franx},
  {Franz}, {Friedman}, {Friend}, {Frost}, {Fu}, {Fullerton}, {Gaillard},
  {Galkin}, {Gallagher}, {Galyer}, {Garc{\'\i}a Mar{\'\i}n}, {Gardner},
  {Garland}, {Garrett}, {Gasman}, {G{\'a}sp{\'a}r}, {Gastaud}, {Gaudreau},
  {Gauthier}, {Geers}, {Geithner}, {Gennaro}, {Gerber}, {Gereau}, {Giampaoli},
  {Giardino}, {Gibbons}, {Gilbert}, {Gilman}, {Girard}, {Giuliano}, {Gkountis},
  {Glasse}, {Glassmire}, {Glauser}, {Glazer}, {Goldberg}, {Golimowski},
  {Gonzaga}, {Gordon}, {Gordon}, {Goudfrooij}, {Gough}, {Graham}, {Grau},
  {Green}, {Greene}, {Greene}, {Greenfield}, {Greenhouse}, {Greve}, {Greville},
  {Grimaldi}, {Groe}, {Groebner}, {Grumm}, {Grundy}, {G{\"u}del}, {Guillard},
  {Guldalian}, {Gunn}, {Gurule}, {Gutman}, {Guy}, {Guyot}, {Hack}, {Haderlein},
  {Hagan}, {Hagedorn}, {Hainline}, {Haley}, {Hami}, {Hamilton}, {Hammann},
  {Hammel}, {Hanley}, {Hansen}, {Hardy}, {Harnisch}, {Harr}, {Harris}, {Hart},
  {Hartig}, {Hasan}, {Hashim}, {Hashimoto}, {Haskins}, {Hawkins}, {Hayden},
  {Hayden}, {Healy}, {Hecht}, {Heeg}, {Hejal}, {Helm}, {Hengemihle}, {Henning},
  {Henry}, {Henry}, {Henshaw}, {Hernandez}, {Herrington}, {Heske}, {Hesman},
  {Hickey}, {Hilbert}, {Hines}, {Hinz}, {Hirsch}, {Hitcho}, {Hodapp}, {Hodge},
  {Hoffman}, {Holfeltz}, {Holler}, {Hoppa}, {Horner}, {Howard}, {Howard},
  {Huber}, {Hunkeler}, {Hunter}, {Hunter}, {Hurd}, {Hurst}, {Hutchings},
  {Hylan}, {Ignat}, {Illingworth}, {Irish}, {Isaacs}, {Jackson}, {Jaffe},
  {Jahic}, {Jahromi}, {Jakobsen}, {James}, {James}, {James}, {Jamieson},
  {Jandra}, {Jayawardhana}, {Jedrzejewski}, {Jeffers}, {Jensen}, {Joanne},
  {Johns}, {Johnson}, {Johnson}, {Johnson}, {Johnson}, {Johnson}, {Johnson},
  {Johnstone}, {Jollet}, {Jones}, {Jones}, {Jones}, {Jones}, {Jones}, {Jordan},
  {Jordan}, {Jue}, {Jurkowski}, {Justis}, {Justtanont}, {Kaleida}, {Kalirai},
  {Kalmanson}, {Kaltenegger}, {Kammerer}, {Kan}, {Kanarek}, {Kao}, {Karakla},
  {Karl}, {Kassin}, {Kauffman}, {Kavanagh}, {Kelley}, {Kelly}, {Kendrew},
  {Kennedy}, {Kenny}, {Keski-Kuha}, {Keyes}, {Khan}, {Kidwell}, {Kimble},
  {King}, {King}, {Kinzel}, {Kirk}, {Kirkpatrick}, {Klaassen}, {Klingemann},
  {Klintworth}, {Knapp}, {Knight}, {Knollenberg}, {Knutsen}, {Koehler},
  {Koekemoer}, {Kofler}, {Kontson}, {Kovacs}, {Kozhurina-Platais}, {Krause},
  {Kriss}, {Krist}, {Kristoffersen}, {Krogel}, {Krueger}, {Kulp}, {Kumari},
  {Kwan}, {Kyprianou}, {Labador}, {Labiano}, {Lafreni{\`e}re}, {Lagage},
  {Laidler}, {Laine}, {Laird}, {Lajoie}, {Lallo}, {Lam}, {LaMassa}, {Lambros},
  {Lampenfield}, {Lander}, {Langston}, {Larson}, {Larson}, {LaVerghetta},
  {Law}, {Lawrence}, {Lee}, {Lee}, {Lee}, {Leisenring}, {Leveille}, {Levenson},
  {Levi}, {Levine}, {Lewis}, {Lewis}, {Lewis}, {Libralato}, {Lidon},
  {Liebrecht}, {Lightsey}, {Lilly}, {Lim}, {Lim}, {Ling}, {Link}, {Link},
  {Lipinski}, {Liu}, {Lo}, {Lobmeyer}, {Logue}, {Long}, {Long}, {Long}, {Long},
  {L{\'o}pez-Caniego}, {Lotz}, {Love-Pruitt}, {Lubskiy}, {Luers}, {Luetgens},
  {Luevano}, {Lui}, {Lund}, {Lundquist}, {Lunine}, {L{\"u}tzgendorf}, {Lynch},
  {MacDonald}, {MacDonald}, {Macias}, {Macklis}, {Maghami}, {Maharaja},
  {Maiolino}, {Makrygiannis}, {Malla}, {Malumuth}, {Manjavacas}, {Marini},
  {Marrione}, {Marston}, {Martel}, {Martin}, {Martin}, {Martinez}, {Maschmann},
  {Masci}, {Masetti}, {Maszkiewicz}, {Matthews}, {Matuskey}, {McBrayer},
  {McCarthy}, {McCaughrean}, {McClare}, {McClare}, {McCloskey}, {McClurg},
  {McCoy}, {McElwain}, {McGregor}, {McGuffey}, {McKay}, {McKenzie}, {McLean},
  {McMaster}, {McNeil}, {De Meester}, {Mehalick}, {Meixner}, {Mel{\'e}ndez},
  {Menzel}, {Menzel}, {Merz}, {Mesterharm}, {Meyer}, {Meyett}, {Meza},
  {Midwinter}, {Milam}, {Miller}, {Miller}, {Miskey}, {Misselt}, {Mitchell},
  {Mohan}, {Montoya}, {Moran}, {Morishita}, {Moro-Mart{\'\i}n}, {Morrison},
  {Morrison}, {Morse}, {Moschos}, {Moseley}, {Mosier}, {Mosner}, {Mountain},
  {Muckenthaler}, {Mueller}, {Mueller}, {Muhiem}, {M{\"u}hlmann}, {Mullally},
  {Mullen}, {Munger}, {Murphy}, {Murray}, {Muzerolle}, {Mycroft}, {Myers},
  {Myers}, {Myers}, {Myers}, {Myrick}, {Nagle}, {Nayak}, {Naylor}, {Neff},
  {Nelan}, {Nella}, {Nguyen}, {Nguyen}, {Nickson}, {Nidhiry}, {Niedner},
  {Nieto-Santisteban}, {Nikolov}, {Nishisaka}, {Noriega-Crespo}, {Nota},
  {O'Mara}, {Oboryshko}, {O'Brien}, {Ochs}, {Offenberg}, {Ogle}, {Ohl},
  {Olmsted}, {Osborne}, {O'Shaughnessy}, {{\"O}stlin}, {O'Sullivan}, {Otor},
  {Ottens}, {Ouellette}, {Outlaw}, {Owens}, {Pacifici}, {Page}, {Paranilam},
  {Park}, {Parrish}, {Paschal}, {Patapis}, {Patel}, {Patrick}, {Pattishall},
  {Paul}, {Paul}, {Pauly}, {Pavlovsky}, {Pe{\~n}a-Guerrero}, {Pedder}, {Peek},
  {Pelham}, {Penanen}, {Perriello}, {Perrin}, {Perrine}, {Perrygo}, {Peslier},
  {Petach}, {Peterson}, {Pfarr}, {Pierson}, {Pietraszkiewicz}, {Pilchen},
  {Pipher}, {Pirzkal}, {Pitman}, {Player}, {Plesha}, {Plitzke}, {Pohner},
  {Poletis}, {Pollizzi}, {Polster}, {Pontius}, {Pontoppidan}, {Porges},
  {Potter}, {Prescott}, {Proffitt}, {Pueyo}, {Quispe Neira}, {Radich}, {Rager},
  {Rameau}, {Ramey}, {Ramos Alarcon}, {Rampini}, {Rapp}, {Rashford},
  {Rauscher}, {Ravindranath}, {Rawle}, {Rawlings}, {Ray}, {Regan}, {Rehm},
  {Rehm}, {Reid}, {Reis}, {Renk}, {Reoch}, {Ressler}, {Rest}, {Reynolds},
  {Richon}, {Richon}, {Ridgaway}, {Riedel}, {Rieke}, {Rieke}, {Rifelli},
  {Rigby}, {Riggs}, {Ringel}, {Ritchie}, {Rix}, {Robberto}, {Robinson},
  {Robinson}, {Robinson}, {Rock}, {Rodriguez}, {Rodr{\'\i}guez del Pino},
  {Roellig}, {Rohrbach}, {Roman}, {Romelfanger}, {Romo}, {Rosales}, {Rose},
  {Roteliuk}, {Roth}, {Rothwell}, {Rouzaud}, {Rowe}, {Rowlands}, {Roy},
  {Royer}, {Rui}, {Rumler}, {Rumpl}, {Russ}, {Ryan}, {Ryan}, {Saad}, {Sabata},
  {Sabatino}, {Sabbi}, {Sabelhaus}, {Sabia}, {Sahu}, {Saif}, {Salvignol},
  {Samara-Ratna}, {Samuelson}, {Sanders}, {Sappington}, {Sargent}, {Sauer},
  {Savadkin}, {Sawicki}, {Schappell}, {Scheffer}, {Scheithauer}, {Scherer},
  {Schiff}, {Schlawin}, {Schmeitzky}, {Schmitz}, {Schmude}, {Schneider},
  {Schreiber}, {Schroeven-Deceuninck}, {Schultz}, {Schwab}, {Schwartz},
  {Scoccimarro}, {Scott}, {Scott}, {Seaton}, {Seely}, {Seery}, {Seidleck},
  {Sembach}, {Shanahan}, {Shaughnessy}, {Shaw}, {Shay}, {Sheehan}, {Sheth},
  {Shih}, {Shivaei}, {Siegel}, {Sienkiewicz}, {Simmons}, {Simon}, {Sirianni},
  {Sivaramakrishnan}, {Slade}, {Sloan}, {Slocum}, {Slowinski}, {Smith},
  {Smith}, {Smith}, {Smith}, {Smith}, {Smith}, {Smolik}, {Soderblom}, {Sohn},
  {Sokol}, {Sonneborn}, {Sontag}, {Sooy}, {Soummer}, {Southwood}, {Spain},
  {Sparmo}, {Speer}, {Spencer}, {Sprofera}, {Stallcup}, {Stanley},
  {Stansberry}, {Stark}, {Starr}, {Stassi}, {Steck}, {Steeley}, {Stephens},
  {Stephenson}, {Stewart}, {Stiavelli}, {}, {Strada}, {Straughn}, {Streetman},
  {Strickland}, {Strobele}, {Stuhlinger}, {Stys}, {Such}, {Sukhatme},
  {Sullivan}, {Sullivan}, {Sumner}, {Sun}, {Sunnquist}, {Swade}, {Swam},
  {Swenton}, {Swoish}, {Tam Litten}, {Tamas}, {Tao}, {Taylor}, {Taylor}, {te
  Plate}, {Van Tea}, {Teague}, {Telfer}, {Temim}, {Texter}, {Thatte},
  {Thompson}, {Thompson}, {Thomson}, {Thronson}, {Tierney}, {Tikkanen},
  {Tinnin}, {Tippet}, {Todd}, {Tran}, {Trauger}, {Trejo}, {Vinh Truong},
  {Tsukamoto}, {Tufail}, {Tumlinson}, {Tustain}, {Tyra}, {Ubeda}, {Underwood},
  {Uzzo}, {Vaclavik}, {Valenduc}, {Valenti}, {Van Campen}, {van de Wetering},
  {Van Der Marel}, {van Haarlem}, {Vandenbussche}, {van Dishoeck},
  {Vanterpool}, {Vernoy}, {Vila Costas}, {Volk}, {Voorzaat}, {Voyton}, {Vydra},
  {Waddy}, {Waelkens}, {Wahlgren}, {Walker}, {Wander}, {Warfield}, {Warner},
  {Wasiak}, {Wasiak}, {Wehner}, {Weiler}, {Weilert}, {Weiss}, {Wells}, {Welty},
  {Wheate}, {Wheeler}, {White}, {Whitehouse}, {Whiteleather}, {Whitman},
  {Williams}, {Willmer}, {Willott}, {Willoughby}, {Wilson}, {Wilson}, {Wilson},
  {Windhorst}, {Wislowski}, {Wolfe}, {Wolfe}, {Wolff}, {Wondel}, {Woo},
  {Woods}, {Worden}, {Workman}, {Wright}, {Wu}, {Wu}, {Wun}, {Wymer},
  {Yadetie}, {Yan}, {Yang}, {Yates}, {Yeager}, {Yerger}, {Young}, {Young},
  {Yu}, {Yu}, {Zak}, {Zeidler}, {Zepp}, {Zhou}, {Zincke}, {Zonak}, \&
  {Zondag}}]{GardnerPASP2023}
{Gardner}, J.~P., {Mather}, J.~C., {Abbott}, R., {et~al.} 2023, \pasp, 135,
  068001, \dodoi{10.1088/1538-3873/acd1b5}

\bibitem[{{Gonz{\'a}lez}(1993)}]{Gonzalez1993}
{Gonz{\'a}lez}, J.~J. 1993, PhD thesis, -

\bibitem[{{Graves} {et~al.}(2009){Graves}, {Faber}, \&
  {Schiavon}}]{GravesApJ2009}
{Graves}, G.~J., {Faber}, S.~M., \& {Schiavon}, R.~P. 2009, \apj, 698, 1590,
  \dodoi{10.1088/0004-637X/698/2/1590}

\bibitem[{{Greene} {et~al.}(2015){Greene}, {Janish}, {Ma}, {McConnell},
  {Blakeslee}, {Thomas}, \& {Murphy}}]{GreeneAPJ2015}
{Greene}, J.~E., {Janish}, R., {Ma}, C.-P., {et~al.} 2015, \apj, 807, 11,
  \dodoi{10.1088/0004-637X/807/1/11}

\bibitem[{{Gu} {et~al.}(2018){Gu}, {Conroy}, \& {Brammer}}]{GuApJ2018}
{Gu}, M., {Conroy}, C., \& {Brammer}, G. 2018, \apjl, 862, L18,
  \dodoi{10.3847/2041-8213/aad336}

\bibitem[{{Gu} {et~al.}(2022){Gu}, {Greene}, {Newman}, {Kreisch},
  {Quenneville}, {Ma}, \& {Blakeslee}}]{GuAPJ2022}
{Gu}, M., {Greene}, J.~E., {Newman}, A.~B., {et~al.} 2022, \apj, 932, 103,
  \dodoi{10.3847/1538-4357/ac69ea}

\bibitem[{{Gunn} {et~al.}(2006){Gunn}, {Siegmund}, {Mannery}, {Owen}, {Hull},
  {Leger}, {Carey}, {Knapp}, {York}, {Boroski}, {Kent}, {Lupton}, {Rockosi},
  {Evans}, {Waddell}, {Anderson}, {Annis}, {Barentine}, {Bartoszek}, {Bastian},
  {Bracker}, {Brewington}, {Briegel}, {Brinkmann}, {Brown}, {Carr},
  {Czarapata}, {Drennan}, {Dombeck}, {Federwitz}, {Gillespie}, {Gonzales},
  {Hansen}, {Harvanek}, {Hayes}, {Jordan}, {Kinney}, {Klaene}, {Kleinman},
  {Kron}, {Kresinski}, {Lee}, {Limmongkol}, {Lindenmeyer}, {Long}, {Loomis},
  {McGehee}, {Mantsch}, {Neilsen}, {Neswold}, {Newman}, {Nitta}, {Peoples},
  {Pier}, {Prieto}, {Prosapio}, {Rivetta}, {Schneider}, {Snedden}, \&
  {Wang}}]{GunnAJ2006}
{Gunn}, J.~E., {Siegmund}, W.~A., {Mannery}, E.~J., {et~al.} 2006, \aj, 131,
  2332, \dodoi{10.1086/500975}

\bibitem[{{Harris} {et~al.}(2020){Harris}, {Millman}, {van der Walt},
  {Gommers}, {Virtanen}, {Cournapeau}, {Wieser}, {Taylor}, {Berg}, {Smith},
  {Kern}, {Picus}, {Hoyer}, {van Kerkwijk}, {Brett}, {Haldane}, {del R{\'\i}o},
  {Wiebe}, {Peterson}, {G{\'e}rard-Marchant}, {Sheppard}, {Reddy}, {Weckesser},
  {Abbasi}, {Gohlke}, \& {Oliphant}}]{HarrisNAT2020}
{Harris}, C.~R., {Millman}, K.~J., {van der Walt}, S.~J., {et~al.} 2020, \nat,
  585, 357, \dodoi{10.1038/s41586-020-2649-2}

\bibitem[{{Hirschmann} {et~al.}(2015){Hirschmann}, {Naab}, {Ostriker},
  {Forbes}, {Duc}, {Dav{\'e}}, {Oser}, \& {Karabal}}]{HirschmannMNRAS2015}
{Hirschmann}, M., {Naab}, T., {Ostriker}, J.~P., {et~al.} 2015, \mnras, 449,
  528, \dodoi{10.1093/mnras/stv274}

\bibitem[{{Huang} {et~al.}(2013){Huang}, {Ho}, {Peng}, {Li}, \&
  {Barth}}]{HuangAPJL2013}
{Huang}, S., {Ho}, L.~C., {Peng}, C.~Y., {Li}, Z.-Y., \& {Barth}, A.~J. 2013,
  \apjl, 768, L28, \dodoi{10.1088/2041-8205/768/2/L28}

\bibitem[{{Huang} {et~al.}(2018){Huang}, {Leauthaud}, {Greene}, {Bundy}, {Lin},
  {Tanaka}, {Mandelbaum}, {Miyazaki}, \& {Komiyama}}]{HuangMNRAS2018}
{Huang}, S., {Leauthaud}, A., {Greene}, J., {et~al.} 2018, \mnras, 480, 521,
  \dodoi{10.1093/mnras/sty1136}

\bibitem[{{Huang} {et~al.}(2020){Huang}, {Leauthaud}, {Hearin}, {Behroozi},
  {Bradshaw}, {Ardila}, {Speagle}, {Tenneti}, {Bundy}, {Greene}, {Sif{\'o}n},
  \& {Bahcall}}]{Huang2020ASAP}
{Huang}, S., {Leauthaud}, A., {Hearin}, A., {et~al.} 2020, \mnras, 492, 3685,
  \dodoi{10.1093/mnras/stz3314}

\bibitem[{{Huang} {et~al.}(2022){Huang}, {Leauthaud}, {Bradshaw}, {Hearin},
  {Behroozi}, {Lange}, {Greene}, {DeRose}, {Speagle}, \&
  {Xhakaj}}]{HuangMNRAS2022}
{Huang}, S., {Leauthaud}, A., {Bradshaw}, C., {et~al.} 2022, \mnras, 515, 4722,
  \dodoi{10.1093/mnras/stac1680}

\bibitem[{{Hunter}(2007)}]{HunterCOMPUTINGINSCIENCEANDENGINEERING2007}
{Hunter}, J.~D. 2007, Computing in Science and Engineering, 9, 90,
  \dodoi{10.1109/MCSE.2007.55}

\bibitem[{{Husemann} {et~al.}(2013){Husemann}, {Jahnke}, {S{\'a}nchez},
  {Barrado}, {Bekerait{\.{e}}}, {Bomans}, {Castillo-Morales},
  {Catal{\'a}n-Torrecilla}, {Cid Fernandes}, {Falc{\'o}n-Barroso},
  {Garc{\'\i}a-Benito}, {Gonz{\'a}lez Delgado}, {Iglesias-P{\'a}ramo},
  {Johnson}, {Kupko}, {L{\'o}pez-Fernandez}, {Lyubenova}, {Marino}, {Mast},
  {Miskolczi}, {Monreal-Ibero}, {Gil de Paz}, {P{\'e}rez}, {P{\'e}rez},
  {Rosales-Ortega}, {Ruiz-Lara}, {Schilling}, {van de Ven}, {Walcher}, {Alves},
  {de Amorim}, {Backsmann}, {Barrera-Ballesteros}, {Bland-Hawthorn}, {Cortijo},
  {Dettmar}, {Demleitner}, {D{\'\i}az}, {Enke}, {Florido}, {Flores}, {Galbany},
  {Gallazzi}, {Garc{\'\i}a-Lorenzo}, {Gomes}, {Gruel}, {Haines}, {Holmes},
  {Jungwiert}, {Kalinova}, {Kehrig}, {Kennicutt}, {Klar}, {Lehnert},
  {L{\'o}pez-S{\'a}nchez}, {de Lorenzo-C{\'a}ceres}, {M{\'a}rmol-Queralt{\'o}},
  {M{\'a}rquez}, {Mendez-Abreu}, {Moll{\'a}}, {del Olmo}, {Meidt}, {Papaderos},
  {Puschnig}, {Quirrenbach}, {Roth}, {S{\'a}nchez-Bl{\'a}zquez}, {Spekkens},
  {Singh}, {Stanishev}, {Trager}, {Vilchez}, {Wild}, {Wisotzki}, {Zibetti}, \&
  {Ziegler}}]{HusemannAAP2013}
{Husemann}, B., {Jahnke}, K., {S{\'a}nchez}, S.~F., {et~al.} 2013, \aap, 549,
  A87, \dodoi{10.1051/0004-6361/201220582}

\bibitem[{{Johansson} {et~al.}(2012{\natexlab{a}}){Johansson}, {Thomas}, \&
  {Maraston}}]{JohanssonMNRAS2012}
{Johansson}, J., {Thomas}, D., \& {Maraston}, C. 2012{\natexlab{a}}, \mnras,
  421, 1908, \dodoi{10.1111/j.1365-2966.2011.20316.x}

\bibitem[{{Johansson} {et~al.}(2012{\natexlab{b}}){Johansson}, {Naab}, \&
  {Ostriker}}]{JohanssonAPJ2012}
{Johansson}, P.~H., {Naab}, T., \& {Ostriker}, J.~P. 2012{\natexlab{b}}, \apj,
  754, 115, \dodoi{10.1088/0004-637X/754/2/115}

\bibitem[{{Johnson} {et~al.}(2021){Johnson}, {Leja}, {Conroy}, \&
  {Speagle}}]{JohnsonApJS2021}
{Johnson}, B.~D., {Leja}, J., {Conroy}, C., \& {Speagle}, J.~S. 2021, \apjs,
  254, 22, \dodoi{10.3847/1538-4365/abef67}

\bibitem[{{Knowles} {et~al.}(2019){Knowles}, {Sansom}, {Coelho}, {Allende
  Prieto}, {Conroy}, \& {Vazdekis}}]{KnowlesMNRAS2019}
{Knowles}, A.~T., {Sansom}, A.~E., {Coelho}, P.~R.~T., {et~al.} 2019, \mnras,
  486, 1814, \dodoi{10.1093/mnras/stz754}

\bibitem[{{Knowles} {et~al.}(2023){Knowles}, {Sansom}, {Vazdekis}, \& {Allende
  Prieto}}]{KnowlesMNRAS2023}
{Knowles}, A.~T., {Sansom}, A.~E., {Vazdekis}, A., \& {Allende Prieto}, C.
  2023, \mnras, 523, 3450, \dodoi{10.1093/mnras/stad1647}

\bibitem[{{Korn} {et~al.}(2005){Korn}, {Maraston}, \& {Thomas}}]{KornA&A2005}
{Korn}, A.~J., {Maraston}, C., \& {Thomas}, D. 2005, \aap, 438, 685,
  \dodoi{10.1051/0004-6361:20042126}

\bibitem[{{Kroupa}(2001)}]{KroupaMNRAS2001}
{Kroupa}, P. 2001, \mnras, 322, 231, \dodoi{10.1046/j.1365-8711.2001.04022.x}

\bibitem[{{Kuntschner} {et~al.}(2001){Kuntschner}, {Lucey}, {Smith}, {Hudson},
  \& {Davies}}]{KuntschnerMNRAS2001}
{Kuntschner}, H., {Lucey}, J.~R., {Smith}, R.~J., {Hudson}, M.~J., \& {Davies},
  R.~L. 2001, \mnras, 323, 615, \dodoi{10.1046/j.1365-8711.2001.04263.x}

\bibitem[{{La Barbera} {et~al.}(2014){La Barbera}, {Pasquali}, {Ferreras},
  {Gallazzi}, {de Carvalho}, \& {de la Rosa}}]{LaBarbera2014MNRAS}
{La Barbera}, F., {Pasquali}, A., {Ferreras}, I., {et~al.} 2014, \mnras, 445,
  1977, \dodoi{10.1093/mnras/stu1626}

\bibitem[{{La Barbera} {et~al.}(2016){La Barbera}, {Vazdekis}, {Ferreras},
  {Pasquali}, {Cappellari}, {Mart{\'\i}n-Navarro}, {Sch{\"o}nebeck}, \&
  {Falc{\'o}n-Barroso}}]{LaBarberaMNRAS2016}
{La Barbera}, F., {Vazdekis}, A., {Ferreras}, I., {et~al.} 2016, \mnras, 457,
  1468, \dodoi{10.1093/mnras/stv2996}

\bibitem[{{La Barbera} {et~al.}(2019){La Barbera}, {Vazdekis}, {Ferreras},
  {Pasquali}, {Allende Prieto}, {Mart{\'\i}n-Navarro}, {Aguado}, {de Carvalho},
  {Rembold}, {Falc{\'o}n-Barroso}, \& {van de Ven}}]{LaBarberaMNRAS2019}
---. 2019, \mnras, 489, 4090, \dodoi{10.1093/mnras/stz2192}

\bibitem[{{Law} {et~al.}(2016){Law}, {Cherinka}, {Yan}, {Andrews}, {Bershady},
  {Bizyaev}, {Blanc}, {Blanton}, {Bolton}, {Brownstein}, {Bundy}, {Chen},
  {Drory}, {D'Souza}, {Fu}, {Jones}, {Kauffmann}, {MacDonald}, {Masters},
  {Newman}, {Parejko}, {S{\'a}nchez-Gallego}, {S{\'a}nchez}, {Schlegel},
  {Thomas}, {Wake}, {Weijmans}, {Westfall}, \& {Zhang}}]{LawAJ2016}
{Law}, D.~R., {Cherinka}, B., {Yan}, R., {et~al.} 2016, \aj, 152, 83,
  \dodoi{10.3847/0004-6256/152/4/83}

\bibitem[{{Li} {et~al.}(2018){Li}, {Mao}, {Cappellari}, {Ge}, {Long}, {Li},
  {Mo}, {Li}, {Zheng}, {Bundy}, {Thomas}, {Brownstein}, {Roman Lopes}, {Law},
  \& {Drory}}]{LiMNRAS2018}
{Li}, H., {Mao}, S., {Cappellari}, M., {et~al.} 2018, \mnras, 476, 1765,
  \dodoi{10.1093/mnras/sty334}

\bibitem[{{Li} {et~al.}(2022){Li}, {Huang}, {Leauthaud}, {Moustakas},
  {Danieli}, {Greene}, {Abraham}, {Ardila}, {Kado-Fong}, {Lokhorst}, {Lupton},
  \& {Price}}]{LiMNRAS2022}
{Li}, J., {Huang}, S., {Leauthaud}, A., {et~al.} 2022, \mnras, 515, 5335,
  \dodoi{10.1093/mnras/stac2121}

\bibitem[{{Liu}(2020)}]{LiuMNRAS2020}
{Liu}, Y. 2020, \mnras, 497, 3011, \dodoi{10.1093/mnras/staa2012}

\bibitem[{{Liu} {et~al.}(2016){Liu}, {Ho}, \& {Peng}}]{LiuAPJL2016}
{Liu}, Y., {Ho}, L.~C., \& {Peng}, E. 2016, \apjl, 829, L26,
  \dodoi{10.3847/2041-8205/829/2/L26}

\bibitem[{{Lu} {et~al.}(2023){Lu}, {Zhu}, {Cappellari}, {Li}, {Mao}, \&
  {Xu}}]{LuMNRAS2023}
{Lu}, S., {Zhu}, K., {Cappellari}, M., {et~al.} 2023, \mnras, 526, 1022,
  \dodoi{10.1093/mnras/stad2732}

\bibitem[{{Mart{\'\i}n-Navarro}
  {et~al.}(2015{\natexlab{a}}){Mart{\'\i}n-Navarro}, {La Barbera}, {Vazdekis},
  {Falc{\'o}n-Barroso}, \& {Ferreras}}]{Martin-NavarroMNRAS2015}
{Mart{\'\i}n-Navarro}, I., {La Barbera}, F., {Vazdekis}, A.,
  {Falc{\'o}n-Barroso}, J., \& {Ferreras}, I. 2015{\natexlab{a}}, \mnras, 447,
  1033, \dodoi{10.1093/mnras/stu2480}

\bibitem[{{Mart{\'\i}n-Navarro} {et~al.}(2018){Mart{\'\i}n-Navarro},
  {Vazdekis}, {Falc{\'o}n-Barroso}, {La Barbera}, {Y{\i}ld{\i}r{\i}m}, \& {van
  de Ven}}]{Martin-NavarroMNRAS2018}
{Mart{\'\i}n-Navarro}, I., {Vazdekis}, A., {Falc{\'o}n-Barroso}, J., {et~al.}
  2018, \mnras, 475, 3700, \dodoi{10.1093/mnras/stx3346}

\bibitem[{{Mart{\'\i}n-Navarro}
  {et~al.}(2015{\natexlab{b}}){Mart{\'\i}n-Navarro}, {Vazdekis}, {La Barbera},
  {Falc{\'o}n-Barroso}, {Lyubenova}, {van de Ven}, {Ferreras}, {S{\'a}nchez},
  {Trager}, {Garc{\'\i}a-Benito}, {Mast}, {Mendoza},
  {S{\'a}nchez-Bl{\'a}zquez}, {Gonz{\'a}lez Delgado}, {Walcher}, \& {CALIFA
  Team}}]{Martin-NavarroApJ2015}
{Mart{\'\i}n-Navarro}, I., {Vazdekis}, A., {La Barbera}, F., {et~al.}
  2015{\natexlab{b}}, \apjl, 806, L31, \dodoi{10.1088/2041-8205/806/2/L31}

\bibitem[{{McDermid} {et~al.}(2015){McDermid}, {Alatalo}, {Blitz}, {Bournaud},
  {Bureau}, {Cappellari}, {Crocker}, {Davies}, {Davis}, {de Zeeuw}, {Duc},
  {Emsellem}, {Khochfar}, {Krajnovi{\'c}}, {Kuntschner}, {Morganti}, {Naab},
  {Oosterloo}, {Sarzi}, {Scott}, {Serra}, {Weijmans}, \&
  {Young}}]{McDermidMNRAS2015}
{McDermid}, R.~M., {Alatalo}, K., {Blitz}, L., {et~al.} 2015, \mnras, 448,
  3484, \dodoi{10.1093/mnras/stv105}

\bibitem[{{Milone} {et~al.}(2011){Milone}, {Sansom}, \&
  {S{\'a}nchez-Bl{\'a}zquez}}]{MiloneMNRAS2011}
{Milone}, A. D.~C., {Sansom}, A.~E., \& {S{\'a}nchez-Bl{\'a}zquez}, P. 2011,
  \mnras, 414, 1227, \dodoi{10.1111/j.1365-2966.2011.18457.x}

\bibitem[{{Moustakas} {et~al.}(2023){Moustakas}, {Lang}, {Dey}, {Juneau},
  {Meisner}, {Myers}, {Schlafly}, {Schlegel}, {Valdes}, {Weaver}, \&
  {Zhou}}]{MoustakasAPJS2023}
{Moustakas}, J., {Lang}, D., {Dey}, A., {et~al.} 2023, \apjs, 269, 3,
  \dodoi{10.3847/1538-4365/acfaa2}

\bibitem[{{Naab} {et~al.}(2009){Naab}, {Johansson}, \&
  {Ostriker}}]{NaabAPJL2009}
{Naab}, T., {Johansson}, P.~H., \& {Ostriker}, J.~P. 2009, \apjl, 699, L178,
  \dodoi{10.1088/0004-637X/699/2/L178}

\bibitem[{{Newman} {et~al.}(2012){Newman}, {Ellis}, {Bundy}, \&
  {Treu}}]{NewmanAPJ2012}
{Newman}, A.~B., {Ellis}, R.~S., {Bundy}, K., \& {Treu}, T. 2012, \apj, 746,
  162, \dodoi{10.1088/0004-637X/746/2/162}

\bibitem[{{Noll} {et~al.}(2014){Noll}, {Kausch}, {Kimeswenger}, {Barden},
  {Jones}, {Modigliani}, {Szyszka}, \& {Taylor}}]{NollAAP2014}
{Noll}, S., {Kausch}, W., {Kimeswenger}, S., {et~al.} 2014, \aap, 567, A25,
  \dodoi{10.1051/0004-6361/201423908}

\bibitem[{{Oke} \& {Gunn}(1983)}]{Oke1983}
{Oke}, J.~B., \& {Gunn}, J.~E. 1983, \apj, 266, 713, \dodoi{10.1086/160817}

\bibitem[{{Oogi} \& {Habe}(2013)}]{OogiMNRAS2013}
{Oogi}, T., \& {Habe}, A. 2013, \mnras, 428, 641, \dodoi{10.1093/mnras/sts047}

\bibitem[{{Oser} {et~al.}(2012){Oser}, {Naab}, {Ostriker}, \&
  {Johansson}}]{OserAPJ2012}
{Oser}, L., {Naab}, T., {Ostriker}, J.~P., \& {Johansson}, P.~H. 2012, \apj,
  744, 63, \dodoi{10.1088/0004-637X/744/1/63}

\bibitem[{{Oser} {et~al.}(2010){Oser}, {Ostriker}, {Naab}, {Johansson}, \&
  {Burkert}}]{OserAPJ2010}
{Oser}, L., {Ostriker}, J.~P., {Naab}, T., {Johansson}, P.~H., \& {Burkert}, A.
  2010, \apj, 725, 2312, \dodoi{10.1088/0004-637X/725/2/2312}

\bibitem[{{Oyarz{\'u}n} {et~al.}(2023){Oyarz{\'u}n}, {Bundy}, {Westfall},
  {Lacerna}, {Yan}, {Brownstein}, {Drory}, \& {Lane}}]{OyarzunAPJ2023}
{Oyarz{\'u}n}, G.~A., {Bundy}, K., {Westfall}, K.~B., {et~al.} 2023, \apj, 947,
  13, \dodoi{10.3847/1538-4357/acbbca}

\bibitem[{{Oyarz{\'u}n} {et~al.}(2022){Oyarz{\'u}n}, {Bundy}, {Westfall},
  {Tinker}, {Belfiore}, {Argudo-Fern{\'a}ndez}, {Zheng}, {Conroy}, {Masters},
  {Wake}, {Law}, {McDermid}, {Arag{\'o}n-Salamanca}, {Parikh}, {Yan},
  {Bershady}, {S{\'a}nchez}, {Andrews}, {Fern{\'a}ndez-Trincado}, {Lane},
  {Bizyaev}, {Boardman}, {Lacerna}, {Brownstein}, {Drory}, \&
  {Zhang}}]{OyarzunAPJ2022}
---. 2022, \apj, 933, 88, \dodoi{10.3847/1538-4357/ac7048}

\bibitem[{{Parikh} {et~al.}(2021){Parikh}, {Thomas}, {Maraston}, {Westfall},
  {Andrews}, {Boardman}, {Drory}, \& {Oyarzun}}]{ParikhMNRAS2021}
{Parikh}, T., {Thomas}, D., {Maraston}, C., {et~al.} 2021, \mnras, 502, 5508,
  \dodoi{10.1093/mnras/stab449}

\bibitem[{{Parikh} {et~al.}(2018){Parikh}, {Thomas}, {Maraston}, {Westfall},
  {Goddard}, {Lian}, {Meneses-Goytia}, {Jones}, {Vaughan}, {Andrews},
  {Bershady}, {Bizyaev}, {Brinkmann}, {Brownstein}, {Bundy}, {Drory},
  {Emsellem}, {Law}, {Newman}, {Roman-Lopes}, {Wake}, {Yan}, \&
  {Zheng}}]{ParikhMNRAS2018}
---. 2018, \mnras, 477, 3954, \dodoi{10.1093/mnras/sty785}

\bibitem[{{Parikh} {et~al.}(2019){Parikh}, {Thomas}, {Maraston}, {Westfall},
  {Lian}, {Fraser-McKelvie}, {Andrews}, {Drory}, \&
  {Meneses-Goytia}}]{ParikhMNRAS2019}
---. 2019, \mnras, 483, 3420, \dodoi{10.1093/mnras/sty3339}

\bibitem[{{Peng} {et~al.}(2012){Peng}, {Lilly}, {Renzini}, \&
  {Carollo}}]{PengAPJ2012}
{Peng}, Y.-j., {Lilly}, S.~J., {Renzini}, A., \& {Carollo}, M. 2012, \apj, 757,
  4, \dodoi{10.1088/0004-637X/757/1/4}

\bibitem[{{Price} {et~al.}(2011){Price}, {Phillipps}, {Huxor}, {Smith}, \&
  {Lucey}}]{PriceMNRAS2011}
{Price}, J., {Phillipps}, S., {Huxor}, A., {Smith}, R.~J., \& {Lucey}, J.~R.
  2011, \mnras, 411, 2558, \dodoi{10.1111/j.1365-2966.2010.17862.x}

\bibitem[{{Qu} {et~al.}(2017){Qu}, {Helly}, {Bower}, {Theuns}, {Crain},
  {Frenk}, {Furlong}, {McAlpine}, {Schaller}, {Schaye}, \&
  {White}}]{QuMNRAS2017}
{Qu}, Y., {Helly}, J.~C., {Bower}, R.~G., {et~al.} 2017, \mnras, 464, 1659,
  \dodoi{10.1093/mnras/stw2437}

\bibitem[{{Remus} \& {Forbes}(2022)}]{RemusApJ2022}
{Remus}, R.-S., \& {Forbes}, D.~A. 2022, \apj, 935, 37,
  \dodoi{10.3847/1538-4357/ac7b30}

\bibitem[{{Robotham} {et~al.}(2020){Robotham}, {Bellstedt}, {Lagos}, {Thorne},
  {Davies}, {Driver}, \& {Bravo}}]{RobothamMNRAS2020}
{Robotham}, A.~S.~G., {Bellstedt}, S., {Lagos}, C. d.~P., {et~al.} 2020,
  \mnras, 495, 905, \dodoi{10.1093/mnras/staa1116}

\bibitem[{{Rodriguez-Gomez} {et~al.}(2016){Rodriguez-Gomez}, {Pillepich},
  {Sales}, {Genel}, {Vogelsberger}, {Zhu}, {Wellons}, {Nelson}, {Torrey},
  {Springel}, {Ma}, \& {Hernquist}}]{Rodriguez-GomezMNRAS2016}
{Rodriguez-Gomez}, V., {Pillepich}, A., {Sales}, L.~V., {et~al.} 2016, \mnras,
  458, 2371, \dodoi{10.1093/mnras/stw456}

\bibitem[{{Roediger} \& {Courteau}(2015)}]{RoedigerMNRAS2015}
{Roediger}, J.~C., \& {Courteau}, S. 2015, \mnras, 452, 3209,
  \dodoi{10.1093/mnras/stv1499}

\bibitem[{{Romero-G{\'o}mez} {et~al.}(2023){Romero-G{\'o}mez}, {Peletier},
  {Aguerri}, {Mieske}, {Scott}, {Bland-Hawthorn}, {Bryant}, {Croom},
  {Eftekhari}, {Falc{\'o}n-Barroso}, {Hilker}, {van de Ven}, \&
  {Venhola}}]{RomeroGomezMNRAS2023}
{Romero-G{\'o}mez}, J., {Peletier}, R.~F., {Aguerri}, J.~A.~L., {et~al.} 2023,
  \mnras, 522, 130, \dodoi{10.1093/mnras/stad953}

\bibitem[{{S{\'a}nchez} {et~al.}(2012){S{\'a}nchez}, {Kennicutt}, {Gil de Paz},
  {van de Ven}, {V{\'\i}lchez}, {Wisotzki}, {Walcher}, {Mast}, {Aguerri},
  {Albiol-P{\'e}rez}, {Alonso-Herrero}, {Alves}, {Bakos}, {Bart{\'a}kov{\'a}},
  {Bland-Hawthorn}, {Boselli}, {Bomans}, {Castillo-Morales}, {Cortijo-Ferrero},
  {de Lorenzo-C{\'a}ceres}, {Del Olmo}, {Dettmar}, {D{\'\i}az}, {Ellis},
  {Falc{\'o}n-Barroso}, {Flores}, {Gallazzi}, {Garc{\'\i}a-Lorenzo},
  {Gonz{\'a}lez Delgado}, {Gruel}, {Haines}, {Hao}, {Husemann},
  {Igl{\'e}sias-P{\'a}ramo}, {Jahnke}, {Johnson}, {Jungwiert}, {Kalinova},
  {Kehrig}, {Kupko}, {L{\'o}pez-S{\'a}nchez}, {Lyubenova}, {Marino},
  {M{\'a}rmol-Queralt{\'o}}, {M{\'a}rquez}, {Masegosa}, {Meidt},
  {Mendez-Abreu}, {Monreal-Ibero}, {Montijo}, {Mour{\~a}o}, {Palacios-Navarro},
  {Papaderos}, {Pasquali}, {Peletier}, {P{\'e}rez}, {P{\'e}rez}, {Quirrenbach},
  {Rela{\~n}o}, {Rosales-Ortega}, {Roth}, {Ruiz-Lara},
  {S{\'a}nchez-Bl{\'a}zquez}, {Sengupta}, {Singh}, {Stanishev}, {Trager},
  {Vazdekis}, {Viironen}, {Wild}, {Zibetti}, \& {Ziegler}}]{SanchezAAP2012}
{S{\'a}nchez}, S.~F., {Kennicutt}, R.~C., {Gil de Paz}, A., {et~al.} 2012,
  \aap, 538, A8, \dodoi{10.1051/0004-6361/201117353}

\bibitem[{{Sarzi} {et~al.}(2018){Sarzi}, {Spiniello}, {La Barbera},
  {Krajnovi{\'c}}, \& {van den Bosch}}]{SarziMNRAS2018}
{Sarzi}, M., {Spiniello}, C., {La Barbera}, F., {Krajnovi{\'c}}, D., \& {van
  den Bosch}, R. 2018, \mnras, 478, 4084, \dodoi{10.1093/mnras/sty1092}

\bibitem[{{Schlafly} \& {Finkbeiner}(2011)}]{SchlaflyApJ2011}
{Schlafly}, E.~F., \& {Finkbeiner}, D.~P. 2011, \apj, 737, 103,
  \dodoi{10.1088/0004-637X/737/2/103}

\bibitem[{{Schlegel} {et~al.}(2021){Schlegel}, {Dey}, {Herrera}, {Juneau},
  {Landriau}, {Lang}, {Meisner}, {Moustakas}, {Myers}, {Schlafly}, {Valdes},
  {Weaver}, {Zhang}, {Zhou}, \& {DESI Legacy Imaging Surveys
  Team}}]{SchlegelAAS2021}
{Schlegel}, D., {Dey}, A., {Herrera}, D., {et~al.} 2021, in American
  Astronomical Society Meeting Abstracts, Vol. 237, American Astronomical
  Society Meeting Abstracts, 235.03

\bibitem[{{Schlegel} {et~al.}(1998){Schlegel}, {Finkbeiner}, \&
  {Davis}}]{Schlegel1998}
{Schlegel}, D.~J., {Finkbeiner}, D.~P., \& {Davis}, M. 1998, \apj, 500, 525,
  \dodoi{10.1086/305772}

\bibitem[{{Scholz-D{\'\i}az} {et~al.}(2022){Scholz-D{\'\i}az},
  {Mart{\'\i}n-Navarro}, \& {Falc{\'o}n-Barroso}}]{ScholzDiaz2022MNRAS}
{Scholz-D{\'\i}az}, L., {Mart{\'\i}n-Navarro}, I., \& {Falc{\'o}n-Barroso}, J.
  2022, \mnras, 511, 4900, \dodoi{10.1093/mnras/stac361}

\bibitem[{{Scott} {et~al.}(2017{\natexlab{a}}){Scott}, {Brough}, {Croom},
  {Davies}, {van de Sande}, {Allen}, {Bland-Hawthorn}, {Bryant}, {Cortese},
  {D'Eugenio}, {Federrath}, {Ferreras}, {Goodwin}, {Groves}, {Konstantopoulos},
  {Lawrence}, {Medling}, {Moffett}, {Owers}, {Richards}, {Robotham}, {Tonini},
  \& {Yi}}]{ScottMNRAS2017}
{Scott}, N., {Brough}, S., {Croom}, S.~M., {et~al.} 2017{\natexlab{a}}, \mnras,
  472, 2833, \dodoi{10.1093/mnras/stx2166}

\bibitem[{{Scott} {et~al.}(2017{\natexlab{b}}){Scott}, {Brough}, {Croom},
  {Davies}, {van de Sande}, {Allen}, {Bland-Hawthorn}, {Bryant}, {Cortese},
  {D'Eugenio}, {Federrath}, {Ferreras}, {Goodwin}, {Groves}, {Konstantopoulos},
  {Lawrence}, {Medling}, {Moffett}, {Owers}, {Richards}, {Robotham}, {Tonini},
  \& {Yi}}]{Scott2017MNRAS}
---. 2017{\natexlab{b}}, \mnras, 472, 2833, \dodoi{10.1093/mnras/stx2166}

\bibitem[{{Sohn} {et~al.}(2020){Sohn}, {Geller}, {Diaferio}, \&
  {Rines}}]{Sohn2020}
{Sohn}, J., {Geller}, M.~J., {Diaferio}, A., \& {Rines}, K.~J. 2020, \apj, 891,
  129, \dodoi{10.3847/1538-4357/ab6e6a}

\bibitem[{{Spiniello} {et~al.}(2012){Spiniello}, {Trager}, {Koopmans}, \&
  {Chen}}]{SpinielloApJ2012}
{Spiniello}, C., {Trager}, S.~C., {Koopmans}, L.~V.~E., \& {Chen}, Y.~P. 2012,
  \apjl, 753, L32, \dodoi{10.1088/2041-8205/753/2/L32}

\bibitem[{{Thomas} {et~al.}(2003){Thomas}, {Maraston}, \&
  {Bender}}]{ThomasMNRAS2003}
{Thomas}, D., {Maraston}, C., \& {Bender}, R. 2003, \mnras, 339, 897,
  \dodoi{10.1046/j.1365-8711.2003.06248.x}

\bibitem[{{Thomas} {et~al.}(2005){Thomas}, {Maraston}, {Bender}, \& {Mendes de
  Oliveira}}]{ThomasAPJ2005}
{Thomas}, D., {Maraston}, C., {Bender}, R., \& {Mendes de Oliveira}, C. 2005,
  \apj, 621, 673, \dodoi{10.1086/426932}

\bibitem[{{Tinker}(2021)}]{Tinker2021}
{Tinker}, J.~L. 2021, \apj, 923, 154, \dodoi{10.3847/1538-4357/ac2aaa}

\bibitem[{{Trager} {et~al.}(2000){Trager}, {Faber}, {Worthey}, \&
  {Gonz{\'a}lez}}]{TragerAJ2000}
{Trager}, S.~C., {Faber}, S.~M., {Worthey}, G., \& {Gonz{\'a}lez}, J.~J. 2000,
  \aj, 120, 165, \dodoi{10.1086/301442}

\bibitem[{{Trujillo} {et~al.}(2009){Trujillo}, {Cenarro}, {de
  Lorenzo-C{\'a}ceres}, {Vazdekis}, {de la Rosa}, \& {Cava}}]{TrujilloApJ2009}
{Trujillo}, I., {Cenarro}, A.~J., {de Lorenzo-C{\'a}ceres}, A., {et~al.} 2009,
  \apjl, 692, L118, \dodoi{10.1088/0004-637X/692/2/L118}

\bibitem[{{Trujillo} {et~al.}(2014){Trujillo}, {Ferr{\'e}-Mateu}, {Balcells},
  {Vazdekis}, \& {S{\'a}nchez-Bl{\'a}zquez}}]{TrujilloApJ2014}
{Trujillo}, I., {Ferr{\'e}-Mateu}, A., {Balcells}, M., {Vazdekis}, A., \&
  {S{\'a}nchez-Bl{\'a}zquez}, P. 2014, \apjl, 780, L20,
  \dodoi{10.1088/2041-8205/780/2/L20}

\bibitem[{{Utsumi} {et~al.}(2020){Utsumi}, {Geller}, {Zahid}, {Sohn},
  {Dell'Antonio}, {Kawanomoto}, {Komiyama}, {Koshida}, \&
  {Miyazaki}}]{Utsumi2020}
{Utsumi}, Y., {Geller}, M.~J., {Zahid}, H.~J., {et~al.} 2020, \apj, 900, 50,
  \dodoi{10.3847/1538-4357/aba61c}

\bibitem[{{van der Wel} {et~al.}(2014){van der Wel}, {Franx}, {van Dokkum},
  {Skelton}, {Momcheva}, {Whitaker}, {Brammer}, {Bell}, {Rix}, {Wuyts},
  {Ferguson}, {Holden}, {Barro}, {Koekemoer}, {Chang}, {McGrath},
  {H{\"a}ussler}, {Dekel}, {Behroozi}, {Fumagalli}, {Leja}, {Lundgren},
  {Maseda}, {Nelson}, {Wake}, {Patel}, {Labb{\'e}}, {Faber}, {Grogin}, \&
  {Kocevski}}]{vanderWelAPJ2014}
{van der Wel}, A., {Franx}, M., {van Dokkum}, P.~G., {et~al.} 2014, \apj, 788,
  28, \dodoi{10.1088/0004-637X/788/1/28}

\bibitem[{{van Dokkum} {et~al.}(2017){van Dokkum}, {Conroy}, {Villaume},
  {Brodie}, \& {Romanowsky}}]{vanDokkumAPJ2017}
{van Dokkum}, P., {Conroy}, C., {Villaume}, A., {Brodie}, J., \& {Romanowsky},
  A.~J. 2017, \apj, 841, 68, \dodoi{10.3847/1538-4357/aa7135}

\bibitem[{{van Dokkum} \& {Conroy}(2010)}]{vanDokkumNatur2010}
{van Dokkum}, P.~G., \& {Conroy}, C. 2010, \nat, 468, 940,
  \dodoi{10.1038/nature09578}

\bibitem[{{van Dokkum} {et~al.}(2010){van Dokkum}, {Whitaker}, {Brammer},
  {Franx}, {Kriek}, {Labb{\'e}}, {Marchesini}, {Quadri}, {Bezanson},
  {Illingworth}, {Muzzin}, {Rudnick}, {Tal}, \& {Wake}}]{vanDokkumAPJ2010}
{van Dokkum}, P.~G., {Whitaker}, K.~E., {Brammer}, G., {et~al.} 2010, \apj,
  709, 1018, \dodoi{10.1088/0004-637X/709/2/1018}

\bibitem[{{Vazdekis} {et~al.}(1996){Vazdekis}, {Casuso}, {Peletier}, \&
  {Beckman}}]{VazdekisAPJS1996}
{Vazdekis}, A., {Casuso}, E., {Peletier}, R.~F., \& {Beckman}, J.~E. 1996,
  \apjs, 106, 307, \dodoi{10.1086/192340}

\bibitem[{{Vazdekis} {et~al.}(2010){Vazdekis}, {S{\'a}nchez-Bl{\'a}zquez},
  {Falc{\'o}n-Barroso}, {Cenarro}, {Beasley}, {Cardiel}, {Gorgas}, \&
  {Peletier}}]{VazdekisMNRAS2010}
{Vazdekis}, A., {S{\'a}nchez-Bl{\'a}zquez}, P., {Falc{\'o}n-Barroso}, J.,
  {et~al.} 2010, \mnras, 404, 1639, \dodoi{10.1111/j.1365-2966.2010.16407.x}

\bibitem[{{Villaume} {et~al.}(2017){Villaume}, {Conroy}, {Johnson}, {Rayner},
  {Mann}, \& {van Dokkum}}]{VillaumeAPJS2017}
{Villaume}, A., {Conroy}, C., {Johnson}, B., {et~al.} 2017, \apjs, 230, 23,
  \dodoi{10.3847/1538-4365/aa72ed}

\bibitem[{{Virtanen} {et~al.}(2020){Virtanen}, {Gommers}, {Oliphant},
  {Haberland}, {Reddy}, {Cournapeau}, {Burovski}, {Peterson}, {Weckesser},
  {Bright}, {van der Walt}, {Brett}, {Wilson}, {Millman}, {Mayorov}, {Nelson},
  {Jones}, {Kern}, {Larson}, {Carey}, {Polat}, {Feng}, {Moore}, {VanderPlas},
  {Laxalde}, {Perktold}, {Cimrman}, {Henriksen}, {Quintero}, {Harris},
  {Archibald}, {Ribeiro}, {Pedregosa}, {van Mulbregt}, \& {SciPy 1. 0
  Contributors}}]{VirtanenNATUREMETHODS2020}
{Virtanen}, P., {Gommers}, R., {Oliphant}, T.~E., {et~al.} 2020, Nature
  Methods, 17, 261, \dodoi{10.1038/s41592-019-0686-2}

\bibitem[{{Wake} {et~al.}(2017){Wake}, {Bundy}, {Diamond-Stanic}, {Yan},
  {Blanton}, {Bershady}, {S{\'a}nchez-Gallego}, {Drory}, {Jones}, {Kauffmann},
  {Law}, {Li}, {MacDonald}, {Masters}, {Thomas}, {Tinker}, {Weijmans}, \&
  {Brownstein}}]{WakeAJ2017}
{Wake}, D.~A., {Bundy}, K., {Diamond-Stanic}, A.~M., {et~al.} 2017, \aj, 154,
  86, \dodoi{10.3847/1538-3881/aa7ecc}

\bibitem[{{Walcher} {et~al.}(2015){Walcher}, {Coelho}, {Gallazzi}, {Bruzual},
  {Charlot}, \& {Chiappini}}]{WalcherA&A2015}
{Walcher}, C.~J., {Coelho}, P.~R.~T., {Gallazzi}, A., {et~al.} 2015, \aap, 582,
  A46, \dodoi{10.1051/0004-6361/201525924}

\bibitem[{{Watson} {et~al.}(2022){Watson}, {Davies}, {Brough}, {Croom},
  {D'Eugenio}, {Glazebrook}, {Groves}, {L{\'o}pez-S{\'a}nchez}, {van de Sande},
  {Scott}, {Vaughan}, {Walcher}, {Bland-Hawthorn}, {Bryant}, {Goodwin},
  {Lawrence}, {Lorente}, {Owers}, \& {Richards}}]{WatsonMNRAS2022}
{Watson}, P.~J., {Davies}, R.~L., {Brough}, S., {et~al.} 2022, \mnras, 510,
  1541, \dodoi{10.1093/mnras/stab3477}

\bibitem[{{Willmer}(2018)}]{WillmerApJS2018}
{Willmer}, C. N.~A. 2018, \apjs, 236, 47, \dodoi{10.3847/1538-4365/aabfdf}

\bibitem[{{Worthey}(1999)}]{WortheyASPC1999}
{Worthey}, G. 1999, in Astronomical Society of the Pacific Conference Series,
  Vol. 192, Spectrophotometric Dating of Stars and Galaxies, ed. I.~{Hubeny},
  S.~{Heap}, \& R.~{Cornett}, 283

\bibitem[{{Worthey} {et~al.}(1994){Worthey}, {Faber}, {Gonzalez}, \&
  {Burstein}}]{WortheyAPJS1994}
{Worthey}, G., {Faber}, S.~M., {Gonzalez}, J.~J., \& {Burstein}, D. 1994,
  \apjs, 94, 687, \dodoi{10.1086/192087}

\bibitem[{{Worthey} \& {Ottaviani}(1997)}]{WortheyAPJS1997}
{Worthey}, G., \& {Ottaviani}, D.~L. 1997, \apjs, 111, 377,
  \dodoi{10.1086/313021}

\bibitem[{{Xu} {et~al.}(2024){Xu}, {Huang}, {Leauthaud}, {Diemer}, {Leidig},
  {Cannarozzo}, \& {Zhou}}]{Xu2024}
{Xu}, S., {Huang}, S., {Leauthaud}, A., {et~al.} 2024, arXiv e-prints,
  arXiv:2412.03406, \dodoi{10.48550/arXiv.2412.03406}

\bibitem[{{Yan} {et~al.}(2016{\natexlab{a}}){Yan}, {Bundy}, {Law}, {Bershady},
  {Andrews}, {Cherinka}, {Diamond-Stanic}, {Drory}, {MacDonald},
  {S{\'a}nchez-Gallego}, {Thomas}, {Wake}, {Weijmans}, {Westfall}, {Zhang},
  {Arag{\'o}n-Salamanca}, {Belfiore}, {Bizyaev}, {Blanc}, {Blanton},
  {Brownstein}, {Cappellari}, {D'Souza}, {Emsellem}, {Fu}, {Gaulme}, {Graham},
  {Goddard}, {Gunn}, {Harding}, {Jones}, {Kinemuchi}, {Li}, {Li}, {Maiolino},
  {Mao}, {Maraston}, {Masters}, {Merrifield}, {Oravetz}, {Pan}, {Parejko},
  {Sanchez}, {Schlegel}, {Simmons}, {Thanjavur}, {Tinker}, {Tremonti}, {van den
  Bosch}, \& {Zheng}}]{YanAJ2016}
{Yan}, R., {Bundy}, K., {Law}, D.~R., {et~al.} 2016{\natexlab{a}}, \aj, 152,
  197, \dodoi{10.3847/0004-6256/152/6/197}

\bibitem[{{Yan} {et~al.}(2016{\natexlab{b}}){Yan}, {Tremonti}, {Bershady},
  {Law}, {Schlegel}, {Bundy}, {Drory}, {MacDonald}, {Bizyaev}, {Blanc},
  {Blanton}, {Cherinka}, {Eigenbrot}, {Gunn}, {Harding}, {Hogg},
  {S{\'a}nchez-Gallego}, {S{\'a}nchez}, {Wake}, {Weijmans}, {Xiao}, \&
  {Zhang}}]{YanAJ2016a}
{Yan}, R., {Tremonti}, C., {Bershady}, M.~A., {et~al.} 2016{\natexlab{b}}, \aj,
  151, 8, \dodoi{10.3847/0004-6256/151/1/8}

\bibitem[{{York} {et~al.}(2000){York}, {Adelman}, {Anderson}, {Anderson},
  {Annis}, {Bahcall}, {Bakken}, {Barkhouser}, {Bastian}, {Berman}, {Boroski},
  {Bracker}, {Briegel}, {Briggs}, {Brinkmann}, {Brunner}, {Burles}, {Carey},
  {Carr}, {Castander}, {Chen}, {Colestock}, {Connolly}, {Crocker}, {Csabai},
  {Czarapata}, {Davis}, {Doi}, {Dombeck}, {Eisenstein}, {Ellman}, {Elms},
  {Evans}, {Fan}, {Federwitz}, {Fiscelli}, {Friedman}, {Frieman}, {Fukugita},
  {Gillespie}, {Gunn}, {Gurbani}, {de Haas}, {Haldeman}, {Harris}, {Hayes},
  {Heckman}, {Hennessy}, {Hindsley}, {Holm}, {Holmgren}, {Huang}, {Hull},
  {Husby}, {Ichikawa}, {Ichikawa}, {Ivezi{\'c}}, {Kent}, {Kim}, {Kinney},
  {Klaene}, {Kleinman}, {Kleinman}, {Knapp}, {Korienek}, {Kron}, {Kunszt},
  {Lamb}, {Lee}, {Leger}, {Limmongkol}, {Lindenmeyer}, {Long}, {Loomis},
  {Loveday}, {Lucinio}, {Lupton}, {MacKinnon}, {Mannery}, {Mantsch}, {Margon},
  {McGehee}, {McKay}, {Meiksin}, {Merelli}, {Monet}, {Munn}, {Narayanan},
  {Nash}, {Neilsen}, {Neswold}, {Newberg}, {Nichol}, {Nicinski}, {Nonino},
  {Okada}, {Okamura}, {Ostriker}, {Owen}, {Pauls}, {Peoples}, {Peterson},
  {Petravick}, {Pier}, {Pope}, {Pordes}, {Prosapio}, {Rechenmacher}, {Quinn},
  {Richards}, {Richmond}, {Rivetta}, {Rockosi}, {Ruthmansdorfer}, {Sandford},
  {Schlegel}, {Schneider}, {Sekiguchi}, {Sergey}, {Shimasaku}, {Siegmund},
  {Smee}, {Smith}, {Snedden}, {Stone}, {Stoughton}, {Strauss}, {Stubbs},
  {SubbaRao}, {Szalay}, {Szapudi}, {Szokoly}, {Thakar}, {Tremonti}, {Tucker},
  {Uomoto}, {Vanden Berk}, {Vogeley}, {Waddell}, {Wang}, {Watanabe},
  {Weinberg}, {Yanny}, {Yasuda}, \& {SDSS Collaboration}}]{YorkAJ2000}
{York}, D.~G., {Adelman}, J., {Anderson}, John~E., J., {et~al.} 2000, \aj, 120,
  1579, \dodoi{10.1086/301513}

\bibitem[{{Zahid} \& {Geller}(2017)}]{ZahidApJ2017}
{Zahid}, H.~J., \& {Geller}, M.~J. 2017, \apj, 841, 32,
  \dodoi{10.3847/1538-4357/aa7056}

\bibitem[{{Zahid} {et~al.}(2018){Zahid}, {Sohn}, \& {Geller}}]{Zahid2018}
{Zahid}, H.~J., {Sohn}, J., \& {Geller}, M.~J. 2018, \apj, 859, 96,
  \dodoi{10.3847/1538-4357/aabe31}

\bibitem[{{Zou} {et~al.}(2017){Zou}, {Zhou}, {Fan}, {Zhang}, {Zhou}, {Nie},
  {Peng}, {McGreer}, {Jiang}, {Dey}, {Fan}, {He}, {Jiang}, {Lang}, {Lesser},
  {Ma}, {Mao}, {Schlegel}, \& {Wang}}]{ZouPASP2017}
{Zou}, H., {Zhou}, X., {Fan}, X., {et~al.} 2017, \pasp, 129, 064101,
  \dodoi{10.1088/1538-3873/aa65ba}

\end{thebibliography}
\bibliographystyle{aasjournal}

%\listofchanges
\end{CJK*}
\end{document}